\documentclass{ws-ijmpd}
\usepackage[super,compress]{cite}

\pdfoutput=1
\usepackage{graphicx}
\usepackage{float}
\usepackage{epstopdf,cancel}
\usepackage{epsf,latexsym,bbm,euscript,eufrak}
\usepackage{amssymb,amsmath}
\usepackage{mathtools}
\usepackage{times,graphics}
\usepackage{soul,xcolor}
\usepackage{mathtools}
\usepackage{braket}
\usepackage{enumitem}

\usepackage{mathrsfs}
\usepackage{ulem}

\usepackage[russian,english]{babel}
\usepackage{csquotes}

\usepackage{tabularx}
\newcolumntype{C}{>{\centering\arraybackslash}X}

\usepackage{tensor}
\newcommand{\RNum}[1]{\uppercase\expandafter{\romannumeral #1\relax}}

\usepackage[caption=false]{subfig}
\newcommand{\subfigimg}[3][,]{%
  \setbox1=\hbox{\includegraphics[#1]{#3}}%
  \leavevmode\rlap{\usebox1}%
  \rlap{\hspace*{-5pt}\raisebox{.5\baselineskip}{\small{#2}}}%
  \phantom{\usebox1}%
}

\usepackage{array}
\usepackage{booktabs}
\usepackage{xcolor}

\def\6{{\langle}}
\def\9{{\rangle}}
\newcommand{\be}{\begin{equation}}
\newcommand{\ee}{\end{equation}}
\newcommand{\ba}{\begin{eqnarray}}
\newcommand{\ea}{\end{eqnarray}}
\def\half{{\tfrac{1}{2}}}
\def\pad{{\partial}}

\def\vK{\mathfrak{k}}
\def\vKo{\mathbbm{k}}

\def\sg{\textsl{g}}

\def\eh{\mathrm{eh}}

\def\ha{{\hat{a}}}

\def\cO{\mathcal{O}}

\def\rA{\mathrm{A}}
\def\rE{\mathrm{E}}
\def\rH{\mathrm{H}}
\def\rP{\mathrm{P}}
\def\rS{\mathrm{S}}
\def\vS{v_\mathrm{S}}

\def\rin{\mathrm{in}}
\newcommand{\ah}{\mathrm{AH}}
\def\vS{v_\mathrm{S}}

\def\mb{\mathcal{B}}
\def\mf{\mathcal{F}}
\def\mh{\mathcal{H}}
\def\mh{\mathcal{H}}
\def\mh{\mathcal{H}}

\def\cO{\mathcal{O}}
\def\ms{\mathcal{S}}
\def\mt{\mathcal{T}}

\def\sB{\mathscr{B}}
\def\sH{\mathscr{H}}

\def\eD{\EuScript{D}}
\def\eE{\EuScript{E}}
\def\eF{\EuScript{F}}
\def\eG{\EuScript{G}}
\def\eH{\EuScript{H}}
\def\eK{\EuScript{K}}
\def\eL{\EuScript{L}}
\def\eM{\EuScript{M}}
\def\eR{\EuScript{R}}
\def\eS{\EuScript{S}}
\def\eV{\EuScript{V}}

\def\maf{\mathfrak{f}}
\def\maT{\mathfrak{T}}

\def\maf{\mathfrak{f}}
\def\maT{\mathfrak{T}}

\def\rA{\mathrm{A}}
\def\rB{\mathrm{B}}
\def\rE{\mathrm{E}}
\def\rK{\mathrm{K}}
\def\rP{\mathrm{P}}
\def\rS{\mathrm{S}}
\def\rT{\mathrm{T}}

\makeatletter
\newcommand*{\defeq}{\mathrel{\rlap{%
	\raisebox{0.3ex}{$\m@th\cdot$}}%
	\raisebox{-0.3ex}{$\m@th\cdot$}}%
	=}
\newcommand*{\eqdef}{=\mathrel{\rlap{%
	\raisebox{0.3ex}{$\m@th\cdot$}}%
	\raisebox{-0.3ex}{$\m@th\cdot$}}%
	}
\makeatother

\newcommand{\sumj}{\sum\limits_{j \geqslant \frac{1}{2}}^\infty}
\newcommand{\overbar}[1]{\mkern 1.5mu\overline{\mkern-1.5mu#1\mkern-1.5mu}\mkern 1.5mu}

\usepackage{pifont}
\newcommand{\cmark}{\ding{51}}
\newcommand{\xmark}{\ding{55}}

\usepackage{url,hyperref}
\hypersetup{colorlinks,linkcolor={blue!55!black},citecolor={red!45!black},urlcolor={blue!45!black},breaklinks=true}

\begin{document}

\markboth{R.\ B.\ Mann, S.\ Murk, D.\ R.\ Terno}
{Black holes and their horizons in semiclassical and modified theories of gravity}

\title{Black holes and their horizons in semiclassical and modified theories of gravity}

\author{Robert B.\ Mann}
\address{Department of Physics and Astronomy, University of Waterloo, \\ Waterloo, Ontario N2L 3G1, Canada \\
and \\
Perimeter Institute for Theoretical Physics, \\ Waterloo, Ontario N2L 6B9, Canada \\
\href{mailto:rbmann@uwaterloo.ca}{rbmann@uwaterloo.ca}}

\author{Sebastian Murk}
\address{Department of Physics and Astronomy, Macquarie University, \\ Sydney, NSW 2109, Australia \\
and \\
Sydney Quantum Academy, \\ Sydney, NSW 2006, Australia \\
\href{mailto:sebastian.murk@mq.edu.au}{sebastian.murk@mq.edu.au}}

\author{Daniel R.\ Terno}
\address{Department of Physics and Astronomy, Macquarie University, \\ Sydney, NSW 2109, Australia \\
\href{mailto:daniel.terno@mq.edu.au}{daniel.terno@mq.edu.au}}

\maketitle

\begin{abstract}
For distant observers black holes are trapped spacetime domains bounded by apparent horizons. We review properties of the near-horizon geometry emphasizing the consequences of two common implicit assumptions of semiclassical physics. The first is a consequence of the cosmic censorship conjecture, namely that curvature scalars are finite at apparent horizons. The second is that horizons form in finite asymptotic time (i.e. according to distant observers), a property implicitly assumed in conventional descriptions of black hole formation and evaporation. Taking these as the only requirements within the semiclassical framework, we find that in spherical symmetry only two classes of dynamic solutions are admissible, both describing evaporating black holes and expanding white holes. We review their properties and present the implications. The null energy condition is violated in the vicinity of the outer horizon and satisfied in the vicinity of the inner apparent/anti-trapping horizon. Apparent and anti-trapping horizons are timelike surfaces of intermediately singular behavior, which manifests itself in negative energy density firewalls. These and other properties are also present in axially symmetric solutions. Different generalizations of surface gravity to dynamic spacetimes are discordant and do not match the semiclassical results. We conclude by discussing signatures of these models and implications for the identification of observed ultra-compact objects.
\end{abstract}


\thispagestyle{empty}
\newpage
\tableofcontents

\section{Introduction}
Recent observations from the LIGO, Virgo, KAGRA \cite{LIGO.Virgo:19,LIGOScientific:2020kqk,LIGOScientific:2021qlt}, and Event Horizon Telescope (EHT) \cite{EHT:19} Collaborations have opened up a new avenue for exploring our universe. So far, over 90 gravitational wave events from compact binary coalescences of black holes have been reported, and the first image of a black hole in the galaxy Messier 87 has been captured in unprecedented detail.

These observations indicate that we can no longer regard black holes as strange mathematical solutions to the equations of general relativity (GR) \cite{Schwarzschild:1916uq}, but rather must come to grips with their physical character. Despite the wealth of new observations, this task is not as simple as it may sound. To invoke the theoretical concept of a black hole to explain these observations is to implicitly accept the singularities that inevitably exist at their cores and the strange causal structures they introduce into spacetime at large \cite{Penrose:1964wq}. It is generally assumed that some mechanism will resolve the former issue without doing too much damage to
the notion of a black hole and its horizon, and that departures from the standard relativistic description will be small. This assumption has been challenged recently by a demonstration that the singularity resolution has important physical consequences that must be addressed \cite{CFLV:20c}.

Regarding the latter issue, ``the birth of a black hole signifies the formation of a non-trivial causal structure in spacetime" \cite{FN:98}, and the collective task of the physics community is now to interpret the physical implications of this statement. The notion of a horizon has been particularly perplexing ever since Wheeler asked how black holes could be consistent with the second law of thermodynamics \cite{Almeida:2021wpl}. The problem was that a black hole, emitting nothing, must be an object of zero temperature, and so any hot object thrown into it would not heat up the black hole, in violation of the second law. The introduction of black hole entropy \cite{Bekenstein:1973ur} and subsequently temperature \cite{H:75} ameliorated this paradox and allowed black holes to be understood as thermodynamic objects \cite{Bardeen:1973gs}. However, the use of quantum physics to introduce temperature gave birth to a new paradox \cite{H:76}: the eventual evaporation of a black hole that is permitted due to its thermodynamic character apparently violates the unitarity of quantum physics that allows this process to occur in the first place.

Much study has been devoted to this problem \cite{Mathur:2009hf,Ydri:2016,Mann:2015kpl}, yielding mounting theoretical evidence that a quantum black hole is likely to be significantly different from its classical counterpart. These modifications need not be only due
to non-perturbative quantum gravitational effects \cite{Lunin:2001jy,Braunstein:2009my,Almheiri:2012rt}, but could perhaps be present semiclassically \cite{Abedi:2013xua,Abedi:2015yga}. This in turn has generated a growing number of investigations into exotic compact objects (ECOs) as the culprits behind the EHT and LIGO/Virgo observations. Their characteristic feature is their compactness: their radius is very close to that of a black hole with the same mass whilst lacking an event horizon. Explicit examples include wormholes \cite{Morris:1988tu,Visser}, gravastars \cite{Mazur:2001fv}, boson stars \cite{Schunck:2003kk}, and 2-2 holes \cite{Holdom:2016nek,Holdom:2019ouz,Ren:2019afg,Holdom:2019bdv,CP:19,BCNS:19}. The chief concern is in seeing if such objects yield distinctive signatures in postmerger gravitational wave echoes \cite{Holdom:2019ouz,Ren:2019afg,Holdom:2019bdv,AAOW:20} that arise when a wave falling inside the gravitational potential barrier  travels to a reflecting boundary and after some time delay  returns to the barrier.

\subsection{What is a black hole?} \label{sec:bh}
Ever since their conceptualization in 1783 by the Reverend John Michell \cite{Michell:1783}, black holes have presented us with paradoxes. Michell was originally seeking a method for determining stellar magnitudes and distances, reasoning that a star's gravitational pull would reduce the speed of the light leaving it. At that time light was thought to be corpuscular, and therefore subject to gravity. The maximal effect measurable is limited by the escape velocity from the star, which would be 301,000 km/s (speed of light $c$ as measured at that time \cite{Bradley:1728}). Any star more massive than this upper bound (assuming the same average density, about 497 ``in round numbers'' times the mass of the sun) would make the emitted light ``to return towards it, by its own proper gravity'' \cite{Michell:1783}. No theoretical constraints for objects moving faster than $c$ were known at that time \cite{I:86}, nor were there any empirical measurements indicating the existence of such `dark stars'. Invisible to an outside observer, the existence of such objects paradoxically could be indirectly inferred from their gravitational influence on nearby luminous objects. The relationship between their mass and radius  is given by the same relativistic value $r_\sg=2GM/c^2$ as for Schwarzschild black holes.

Today, we are well aware of a variety of objects that are the result of the gravitational collapse of matter, the best known being brown dwarfs, white dwarfs, and neutron stars. The pioneering work of Oppenheimer and Snyder \cite{OS:939} indicated that a collapsing ball of dust (a form of stress-energy with density but no pressure), matched to Schwarzschild's spherically symmetric solution \cite{Schwarzschild:1916uq} to Einstein's equations, will yield a spacetime that results in what we regard as black hole. Many other collapse solutions have been obtained since then \cite{I:86,exact:03}.

According to a distant observer (Bob), gravitational collapse beyond the density of a neutron star can have three possible outcomes:
\begin{enumerate}
	\item Perpetual ongoing collapse, whose progress is quantified by some closeness parameter $\epsilon$, with a suitable horizon as an asymptotic ($t \to \infty$) concept.
 	\item Formation of a transient or an asymptotic object, where the closeness parameter reaches a minimal value $\epsilon_\text{min}$ at some finite time $t = t_{\text{min}}$ as measured by Bob (some finite asymptotic time).
	\item Formation of an apparent horizon  (see \ref{a:hor} for the formal definition)  in finite asymptotic time $t_\rS$; in other words, Bob determines that a horizon has formed in finite time according to his clock.   	
\end{enumerate}

All three scenarios lead to formation of dark, compact and massive objects. Observational differences between them are subtle and are still beyond the resolving capacity of our technology. The black hole paradigm provides the simplest explanation for all current observations of astrophysical black hole (ABH) candidates. Known physics all but excludes BH alternatives \cite{CP:19}. The success of the paradigm is pitted against  conceptual difficulties that are inherent in the mathematical description of black holes, difficulties  that are absent by design in alternative models of ultra-compact objects without an event horizon\cite{BCNS:19,CP:19}. This is the background for the question ``is there any observational evidence for the existence of black holes?"\cite{CP:17}

There is no unanimously agreed upon definition of a black hole \cite{eC:19,L:21}, and the different alternatives for gravitational collapse leave us with a variety of notions for what is meant by this concept. In our review we organize this ambiguity by presenting clear mathematical differences that should serve as a starting point for the extraction of observable properties. For objects with a horizon it is useful to adapt the terminology of Ref.~\refcite{F:14}, according to which a mathematical black hole (MBH) is a solution of the Einstein equations of classical GR, and the source of our ideas about what features are necessary in order for an object to be considered a black hole. The most well-known feature is the event horizon $r_\mathrm{eh}$ (located at the gravitational radius $r_\sg=2GM/c^2$ for the Schwarzschild black hole) that separates an interior spacetime containing a singularity that is forever inaccessible to observers remaining in our outside world. Due to their simplicity, MBHs are regarded as ``the most perfect macroscopic objects there are in the universe: the only elements in their construction are our concepts of space and time''.\cite{C:92}.  It should be noted that all current observational data can be explained (and in fact, is often predicted) within the MBH paradigm.

An event horizon is a global teleological construct and as such is not observable by local mortal observers \cite{H:00,mV:14,J:14}.   Its location can be known only after the entire evolution of the Universe is recorded, with this record resulting in a giant Carter-Penrose diagram. Moreover,  quantum effects may prevent   formation of the event horizon altogether \cite{H:14}.  On the other hand, a trapped spacetime region from which currently nothing, not even light, can escape --- a crucial black hole property \cite{Michell:1783,I:86,eC:19} --- constitutes what one would reasonably regard as a physical black hole (PBH)\footnote{This notation should be distinguished from that of a primordial black hole, which we denote by pBH.} \cite{F:14}. A detailed definition of a PBH $\sB$ that formalizes ``now'' as a spacelike surface of simultaneity $\Sigma$ that allows the introduction of a (2D) apparent horizon $\sH_\Sigma$ and the PBH boundary (dynamical (3D) apparent horizon $\sH$) is given in \ref{a:hor}. Bounded by a potentially observable apparent horizon at $r_\sg$, a PBH may be a regular black hole (RBH) without an event horizon or singularity, or may overlap or be contained in an MBH.  We note than in classical physics (more precisely, if the classical energy conditions are satisfied), formation of a PBH implies existence of an MBH that actually contains it.

\begin{figure}[!htbp] \centering
	\includegraphics[scale=0.825]{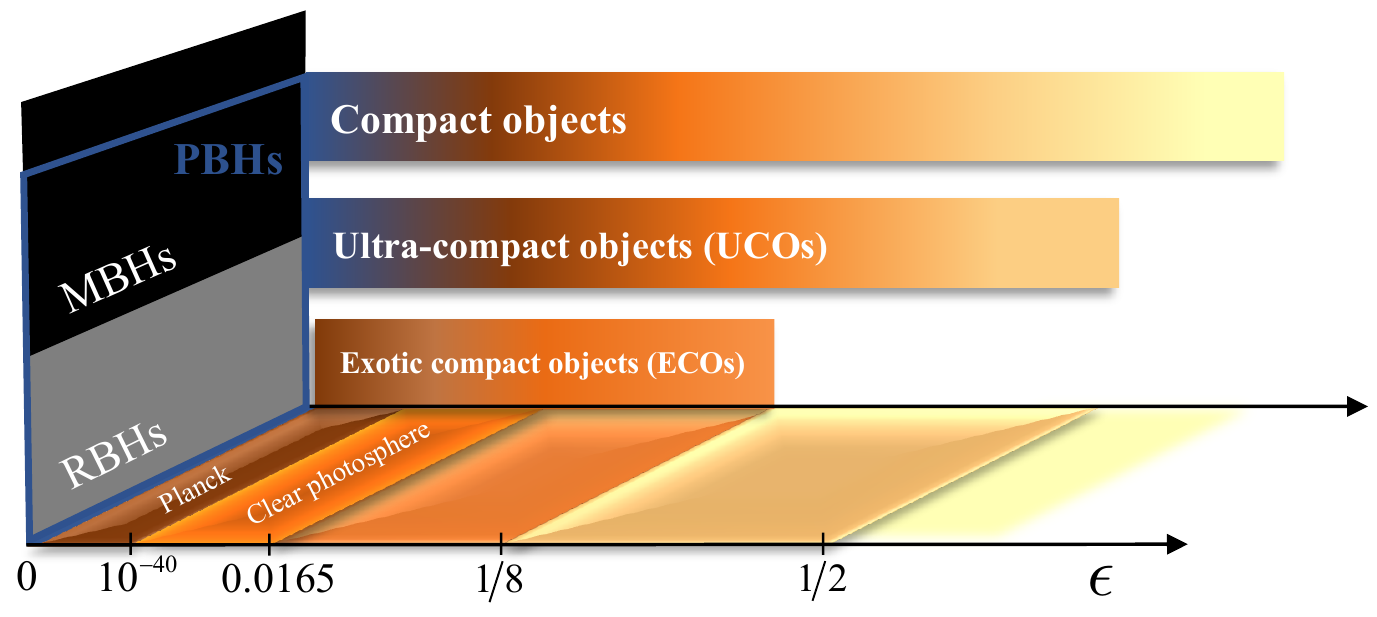}
	\caption{Classification of ultra-compact objects (UCOs) by their compactness $r_\sg/r_0=1/(1+\epsilon)$. UCOs are compact objects that can have a light sphere, i.e.\ $r_0<\tfrac{3}{2}r_\sg$. Black holes, whether mathematical (MBH) or regular (RBH), correspond to $\epsilon=0$. UCOs with $r_0<\tfrac{9}{8}r_\sg$ are excluded by the Buchdahl theorem \cite{B:59}. Therefore the hypothetical horizonless alternatives to black holes whose models violate one or more assumptions of the theorem are exotic compact objects (ECOs). Their subset with $\epsilon \lesssim 0.0165$ are clear photosphere objects \cite{CP:19}, and various quantum structures are the basis of ECOs with $\epsilon \lesssim 10^{-40}$. A more refined classification and numerous examples can be found in Refs.~\citenum{CP:19,BCNS:19}. Unlike the classical case, where the apparent horizon is inside of the event horizon and approaches it (at some finite evolution parameter or asymptotically), \cite{FN:98,HE:73,J:14} and is used in numerical relativity \cite{RZ:13,BS:10} as a proxy for the latter, the spacetime domains of a physical black hole (PBH) and a MBH overlap, but neither is contained within the other. Fig.~\ref{fig:time-g}(a) provides an example.}
	\label{fig:scheme}
\end{figure}
Black holes are part of the hierarchy of ultra-compact objects (UCOs). Fig.~\ref{fig:scheme} presents an overview of spherically symmetric UCOs. In brief, a UCO is a compact object, with or without a horizon, that has a photosphere \cite{CP:17,CP:19}. The key classification parameter is the compactness $r_\sg/r_0$, where $r_\sg$ is the gravitational radius of a MBH of the same mass as the object and $r_0=r_\sg(1+\epsilon)$ is the radius of its effective surface. UCOs are compact objects that null particles (idealizations of high-frequency electromagnetic or gravitational waves) can orbit in circular motion. The unstable circular orbit of massless particles \cite{C:92,MTW} is situated at $r_\mathrm{ph}=\tfrac{3}{2}r_\sg$, defining a surface  that is variously known as  the photon (or light) sphere,  or the photosphere\footnote{This should be distinguished from the homonymous surface in a stellar atmosphere to which the observed radiation is ascribed.}.  It governs the appearance of UCOs when illuminated by accretion disks or stars, thus defining their so-called shadow \cite{CP:17,CP:19,BCNS:19,sbG:17}.

The next important scale is set by the Buchdahl limit $\epsilon_\rB=1/8$. Buchdahl's theorem \cite{B:59} states that, under certain assumptions, the maximum compactness of a spherical self-gravitating object is bounded by $1/(1+\epsilon_\rB)=8/9$. Models that predict higher compactness approaching that of black holes violate one or more assumptions of this theorem, either by admitting modifications of GR, anisotropic fluids, negative density and pressure, or something else \cite{CP:19}. They are referred to as exotic compact objects (ECOs).

The first scenario (perpetually ongoing collapse) corresponds to the gravitational collapse of classical matter whose emissions (and $\epsilon$) as detected by Bob approach zero exponentially with $t$, i.e.\ $t \propto - \log \epsilon$. The second scenario (asymptotic object formation) is realized by various ECO models \cite{CP:19,BCNS:19}. Quasi-normal modes and, as a result, the late-time parts of the ringdown signals, are different in these two cases. From Bob's viewpoint, a compact object is a PBH only if the apparent horizon has formed prior to emission of the signals that he detects. Hence, regardless of whether or not they contain singularities, PBHs represent the third scenario.

\subsection{Semiclassical horizon physics} \label{s:shori}
EHT observations \cite{EHT:19} confirm that UCOs exist, but do not identify them. Whether they are horizonless objects or PBHs is an open question. Understanding the true nature of the objects giving rise to these observations is one of the most important open problems in physics, with implications ranging from understanding properties of exotic matter, to characterizing information, to thermodynamics in gravitational fields, and to the formulation of theories of quantum gravity \cite{Mann:2015kpl,W:01,CK:07,GAC:13,PT:04}. This is why it is important to understand the differences between MBHs and PBHs.

For these reasons we are interested in what might be called {\it horizon physics}: understanding the physical implications of the formation of an object whose escape velocity is greater than the speed of light. We earlier noted that an apparent horizon is a more appropriate definition for connecting this idea to that of a PBH. It is an observer-dependent\footnote{More preciesly, it is a folitation-dependent notion. However, particular families of observers can be related to specific spacetime foliations.} notion \cite{J:14,S:11,vF:15,FEFHM:17}, which can be both an advantage --- a perspective that can be in tune with that of a distant observer, and a disadvantage --- making it difficult to  identify invariant objects. In \ref{a:hor} we describe this definition and that of other horizons in more detail, and in Sec.~\ref{scope} we discuss it in relation to other approaches.

\begin{figure*}[!htbp]
	\begin{tabular}{@{\hspace*{0.05\linewidth}}p{0.45\linewidth}@{\hspace*{0.025\linewidth}}p{0.45\linewidth}@{}}
   		\subfigimg[scale=0.625]{(a)}{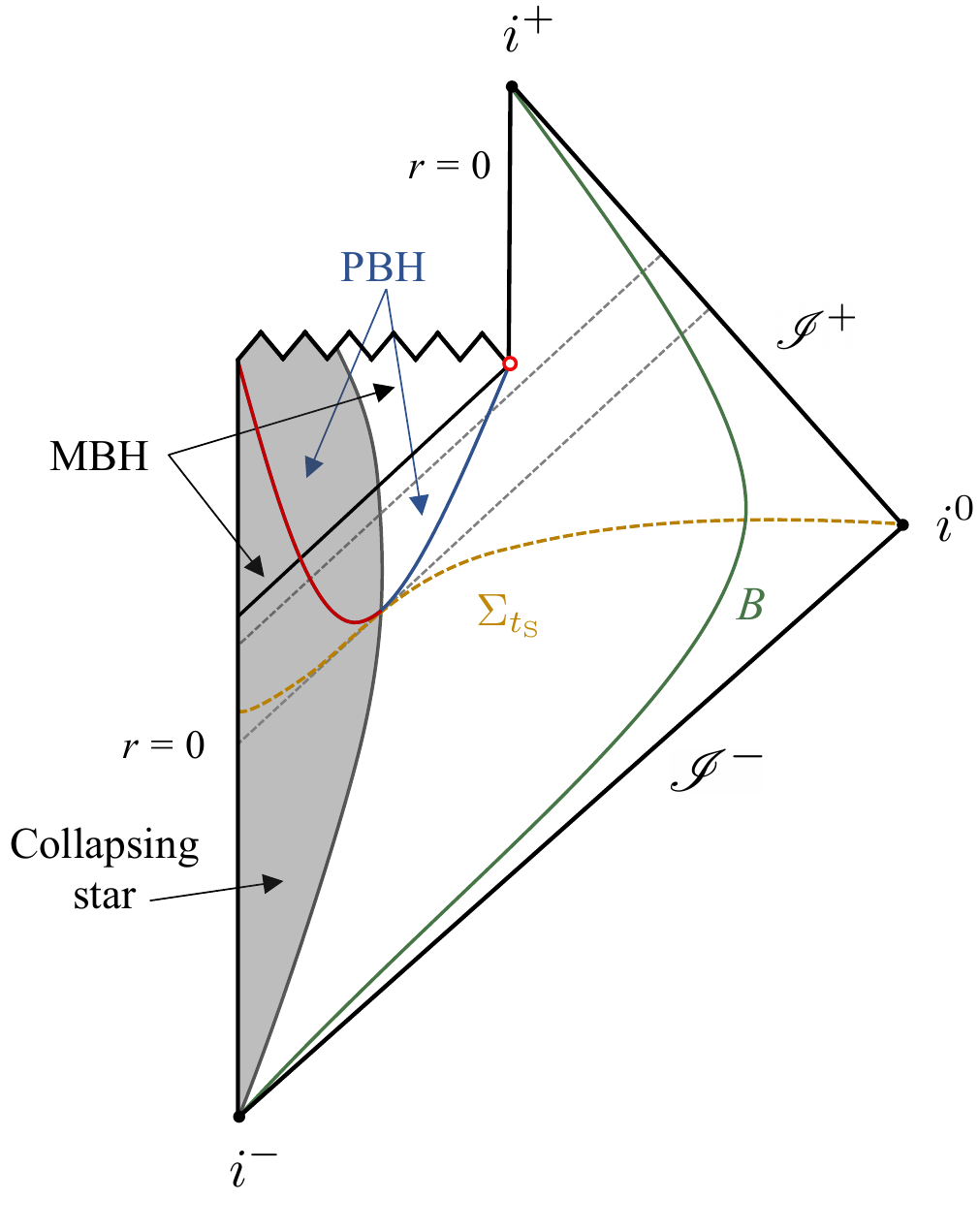} &
		\subfigimg[scale=0.575]{(b)}{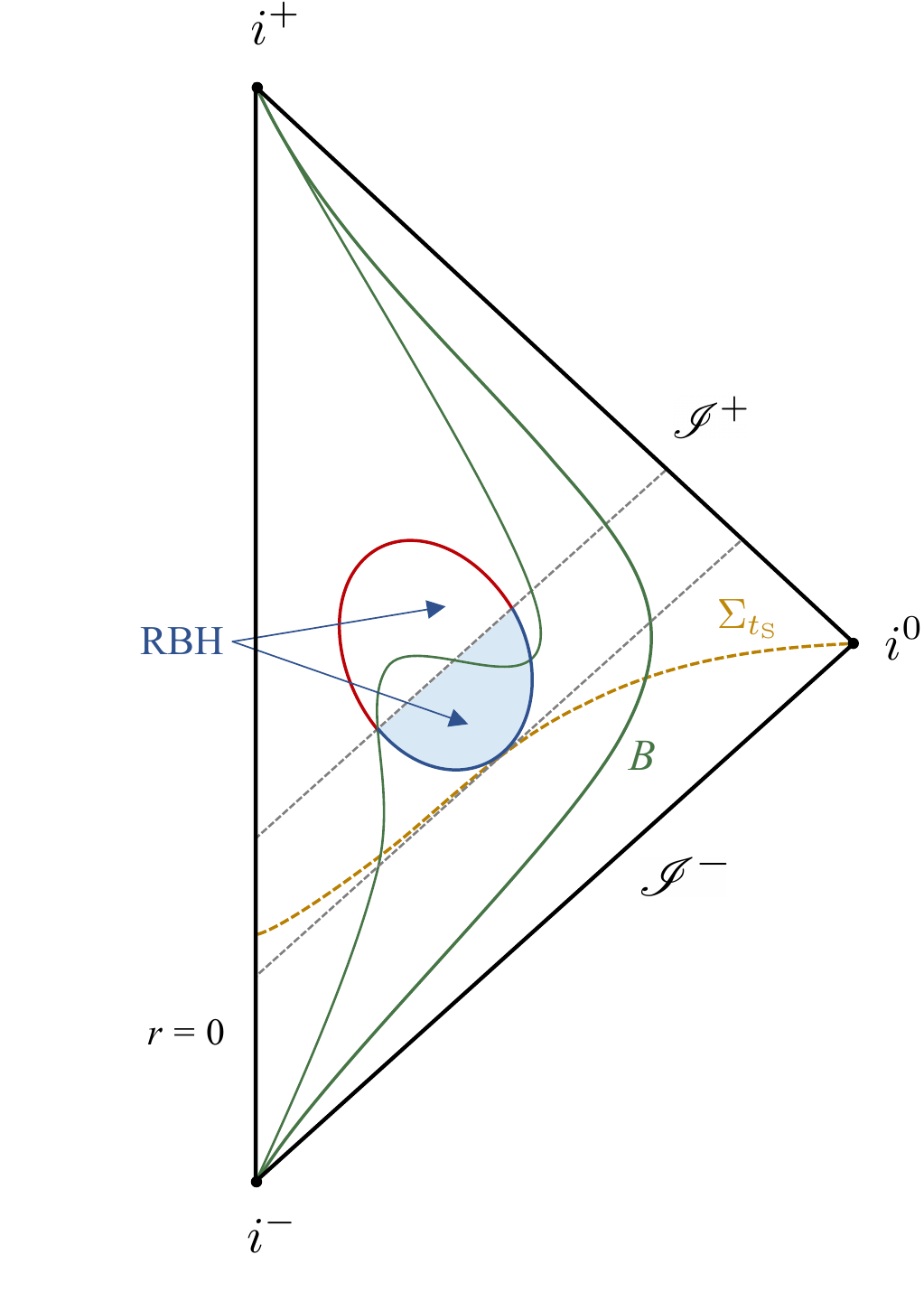}
  \end{tabular}
  \caption{Schematic Carter\textendash{}Penrose diagram for the conventional description of the formation and evaporation of (a) a black hole, and (b) a RBH.
The equal time surface $\Sigma_{t_\mathrm{S}}$  {(see Sec.~\ref{s:formation}) is shown} as a dashed orange line. The trajectory of a distant observer Bob is indicated in green  {and marked by the initial \textit{B}.} The dashed grey lines correspond to outgoing radial null geodesics,  {i.e.\ lines along which the retarded Penrose coordinate $\bar u$ is constant}. As illustrated, they can pass through the quantum ergosphere (see Sec.~\ref{sb:ergo}) and reach  {future null infinity} $\mathscr{I}^+$.
(a) Diagrams of this type are elaborations of the original sketch by Hawking (Fig.~5 of Ref.~\citenum{H:75}).  {The outer $r_\sg(t)$ (dark blue) and inner $r_\mathrm{in}(t)$ (dark red) components of the apparent horizon are identified as the roots of $f(t,r)=0$.} The precise significance of the end point of evaporation (indicated by the red circle) is unclear \cite{H:75,BD:82}. If taken literally, it must be considered as a naked singularity, raising questions about the applicability of this diagram for analyzing the unitarity of black hole formation and evaporation processes \cite{SAK:20}. Spacetime regions corresponding to PBH (MBH) solutions are indicated by blue (black) arrows. The light from any event that lies outside of the  {event} horizon reaches Bob at some finite proper time. The collapsing matter and its surface are shown as in conventional depictions of the collapse. However, the matter in the vicinity of the apparent horizon $\big(t,r_\sg(t)\big)$ violates the NEC for $t\geqslant t_\mathrm{S}$. Moreover, the energy density, pressure, and flux as seen by an infalling observer Alice vary continuously across it, and the equation of state dramatically differs from that of normal matter that may have been used to model the initial EMT of the collapse  {(see Secs.~\ref{k=0} and \ref{s:formation})}. The quantum ergosphere is the region between the event and the apparent horizon.
(b) The asymptotic structure of a simple RBH spacetime \cite{CDLV:20} coincides with that of Minkowski spacetime. An immediate neighborhood of $r=0$ never belongs to the trapped region. The outer (dark blue) and inner (dark red) apparent horizon are indicated according to the invariant definition of Eq.~\eqref{in-out} that correspond to the largest and smallest root of $f(v,r)=0$. This RBH has a smoothly joined inner and outer horizon \cite{BHL:18}. The quantum ergosphere is indicated by the light blue shading. For a PBH both sections of the apparent horizon are non-spacelike (see Sec.~\ref{k=0}).}
	\label{fig:time-g}
\end{figure*}

Notwithstanding these difficulties, the purpose of this article is to review  the physical implications of horizon formation based on the following principles of semiclassical gravity \cite{CK:07,PP:09,BMT:18,HV:20}:
\begin{enumerate}
\item Pseudo-Riemannian geometry on a manifold $\eM$ provides an effective  description of spacetime.  We can use classical notions such as horizons, trajectories, etc.
\item Solutions to the field equations of GR  and/or modified theories of gravity (MTG)
\begin{align}
	\tensor{G}{_\mu_\nu} = 8 \pi \tensor{T}{_\mu_\nu} \equiv 8 \pi \6 \tensor{\hat{T}}{_\mu_\nu} \9_\omega,
	\label{eq=1}
\end{align}
provide an adequate description of spacetime, including the vicinity of an apparent horizon. Here the right-hand side contains the expectation value of the renormalized energy-momentum tensor (EMT) $\tensor{T}{_\mu_\nu}$, understood as an operator in quantum field theory. Treating the entire matter content jointly is the key feature of the self-consistent approach in understanding semiclassical apparent horizon formation \cite{BMsMT:19}.
\item No assumptions about the global structure of the spacetime manifold $\eM$ (event horizon, global hyperbolicity, geodesic completeness, structure of infinity) are made.
\item The state $\omega$ and EMT are not specified a priori. In particular, we do not assume the presence of Hawking-like radiation.
\end{enumerate}
The key elements of this approach are two natural and almost unavoidable assumptions: $\quad$ (i) regularity of the apparent horizon and (ii) its finite-time formation according to the clock of a distant observer. We now outline the motivation and justification for these requirements, which we illustrate in Fig.~\ref{fig:time-g}.
\begin{enumerate}[label=(\roman*)]
	\item In classical GR, non-spacelike singularities destroy predictability. The weak cosmic censorship conjecture (which is the idea that we essentially follow here) \cite{SG:15,P:79,C-B:09} is the statement that spacetime singularities are obscured by event horizons. The original idea is due to Penrose \cite{P:79} and follows the study of spherical gravitational collapse scenarios that succeed in hiding the singularity from external observers. We formulate our first criterion as the absence of curvature singularities at the apparent horizon. A precise formulation is provided in Sec.~\ref{s:Ein}.
	\item A philosophical justification of the second assumption is based on the principle \cite{L:21,E:04} that ``no effect can be counted as a genuine physical effect if it disappears when the idealizations are removed''. Consequently, in order for a horizon to be considered a genuine physical object rather than merely a useful mathematical tool, it must form in finite time according to a distant observer, and there should be some potentially observable consequences of this formation. Thus, if Hawking radiation is real and we accept a finite evaporation time, then the formulation of the information loss problem necessarily requires the formation of an event horizon at some finite time \cite{W:01,mV:14}, which in turn implies the formation of an apparent horizon at some finite time $t_\rS$ as measured by a distant observer \cite{MsMT:21}.
\end{enumerate}
Moreover, if the schematics of the spacetime structure of Fig.~\ref{fig:time-g}(a) are accepted, then all signals that are emitted from the so-called quantum ergosphere \cite{Y:83,BHL:18} --- part of the trapped region that lies outside the event horizon (Sec.~\ref{sb:ergo}) --- reach future null infinity  $\mathscr{I}^+$. Models of transient (even if long-lived) RBHs imply the same \cite{CDLV:20}, as illustrated in Fig.~\ref{fig:time-g}(b). In fact, if one assumes the formation of a transient trapped region, one has to accept our second requirement: the finite-time formation of its apparent horizon according to a distant observer.

\subsection{Scope and relations} \label{scope}
The remainder of this review is dedicated to derivations of mathematical consequences that follow from these principles and exploration of their properties. We are agnostic in our approach: we do not take a stand on the question of the nature of UCOs. Instead, we derive the consequences of modeling them as trapped spacetime regions that have been formed ``by now'', are adequately described by semiclassical physics, and are sufficiently regular. Since we are primarily concerned with apparent horizons, we also do not take a stance on the type of PBHs in question, but follow through with the consequences of their existence.

In Sec.~\ref{s:Ein} we present the near-horizon geometry of PBHs using GR as the underlying theory. This is done mostly in spherical symmetry, where it is easier to extract metamathematical consequences of our assumptions. There are many points of agreement with the usual semiclassical results, particularly on consistency grounds, such as necessity of the violation of the null energy condition (NEC). More interestingly, the PBH solutions show some dramatic differences compared to the semiclassical results on fixed backgrounds (or evolving backgrounds without taking into account the details of backreaction). Sec.~\ref{physcs01} describes their properties and contrasts them with results that are obtained on a fixed background. We extend the analysis to some modified theories of gravity in Sec.~\ref{s:MTG}. Useful background information and additional technical details are summarized in the appendices.

There are several fruitful approaches to the study of the near-horizon physics. The formalism of dynamical horizons \cite{AK:04,K:14,vF:15} produced a number of important results. However, it is based on imposing additional assumptions and a causal structure on a dynamical apparent horizon, whereas we are more interested in following the development of the initial assumptions without additional input to whatever end it brings them.

We are not using the membrane paradigm \cite{TPM:86} as our starting point. Instead, having a timelike  outer apparent horizon as the only admissible solution makes it a good candidate for the role of a stretched horizon. If one is not interested in details of the immediate vicinity of the horizon,   it is possible to summarize its influence on exterior physics by appropriate boundary conditions. Effectively this is a membrane, which can be penetrated by fields and particles from the exterior, but practically nothing can go through it from the inside. Its position is chosen to coincide with an equipotential surface of the lapse function at some unspecified but sufficiently small value. Our scale function $\xi$ may play a similar role.

A recently developed formalism of geometric horizons \cite{CMS:17,CM:18,MJGMBC:21} defines a quasi-local horizon as a hypersurface on which the curvature tensor is algebraically special. Indeed, known black hole horizons are more algebraically special than other regions of spacetime. By its very definition the geometric horizon is invariantly defined. Studies of general spacetimes are ongoing and the exact definition of a geometric horizon has not been fully determined in terms of a particular set of curvature invariants \cite{MJGMBC:21}. However, in the case of spherically symmetric black holes, a spherically symmetric apparent horizon is a geometric horizon \cite{CM:18}.

\subsection{Housekeeping: selection and conventions}
We close this section by commenting on the choice of the selected literature. It is not comprehensive. White holes are mentioned only in passing, as well as scenarios of black-to-white hole transitions. Our selection is biased towards providing a description of the near-horizon geometry in as much detail as possible, and showing how these results mesh with black hole physics in general. We are theorists, so observational investigations are mentioned only in their relation to particular theoretical features or experimental signatures that are extracted from theory. We strive to appropriately give credit, and several times refer to papers that deserve more attention. However, many important works are not cited because the phenomena they first described are well-presented in texts, monographs, and reviews that we briefly present below. In some cases the most useful and succinct descriptions in our view are still contained in the original papers, and these works are explicitly mentioned, and occasionally directly quoted. We admit that despite our best efforts for objectivity the choice reflects our preferences and working habits, and thus some excellent references may have been unjustly omitted.

We use Refs.~\citenum{HE:73,MTW,C:92,exact:03,P:04,C-B:09,P:10,B:16,W:20} as references on GR and its technical aspects, and Refs.~\citenum{BD:82,PT:09} for quantum field theory on curved backgrounds. Our two most commonly used sources on black hole physics in general are Refs.~\citenum{C:92} and \citenum{FN:98}, and Refs.~\citenum{S:11} and \citenum{vF:15} for properties of various horizons. For specific properties of Hawking radiation we consulted Refs.~\citenum{BMPS:95,VAD:11}. Refs.~\citenum{W:01,P:05} are our main sources for  black hole thermodynamics, and Ref.~\citenum{KS:20} for energy conditions. The information loss problem and the final state of a black hole are the subject of a number of excellent reviews. Refs.~\citenum{M:15,COY:15,H:16,M:17,UW:17,info:21,dM:17} were particularly useful in our work.

Apart from a few exceptions we work in Planck units $G=\hbar=c=1$. We work in a four-dimensional spacetime with the metric signature $(-,+,+,+)$ and follow the conventions of Misner, Thorne, and Wheeler \cite{MTW}. To simplify the exposition we often refer to named observers: Alice crosses the apparent horizon, Eve is stationary in its vicinity, and Bob is a distant observer. The terminology that is used to describe varieties of horizons and singularities is not entirely consistent. We summarize it in \ref{a:hor} and \ref{singu}, respectively, where we also indicate the usage we adapt.

\section{Einstein Equations and their Solutions} \label{s:Ein}
Henceforth, we regard classical GR as the underlying physical theory: the standard Einstein tensor $\tensor{G}{_\mu_\nu} = \tensor{R}{_\mu_\nu} - \half \eR \tensor{\sg}{_\mu_\nu}$ (with $\eR \defeq \tensor{\sg}{^\mu^\nu} \tensor{R}{_\mu_\nu}$) appears on the left-hand side of Eq.~\eqref{eq=1}. In this section, we derive the necessary properties of the expectation value of the renormalized EMT and explore properties of the resulting solutions.

First, we briefly sketch the hierarchy of singularities and specify the minimal conditions that a physically acceptable solution should satisfy. A detailed account is presented in \ref{singu}. As singular points are excluded from the (sufficiently smooth) manifold representing the spacetime, they are identified by having incomplete geodesics in their vicinity. Incomplete geodesics are inextendible in at least one direction, but their generalized affine parameter has a finite (or bounded on one side) range \cite{HE:73,C-B:09,TCE:80,J:14}. We focus only on curvature singularities and formalize the regularity requirement as the demand that curvature scalars that are built from polynomials of components of the Riemann tensor are finite.

In other words, we require the absence of any essential scalar curvature singularity (also known as s.p.\ singularity \cite{HE:73,TCE:80}). We do not impose demands on the behavior of invariants that involve covariant derivatives of the Riemann tensor $\tensor{R}{^\mu_\nu_\lambda_\sigma}$, nor on the behavior of its individual components or their contractions in various orthonormal frames. Hence the intermediate (also known as whimper \cite{EK:74,TCE:80}) singularities are not excluded, nor are the matter singularities that are characterized by the divergence of some Ricci tensor components in some frames \cite{EK:74}. We will see that apparent horizons are singular in this particular sense.

To ensure that the trapped region forms in finite time according to the clock of a distant Bob, we use it as one of the coordinates. This coordinate system becomes singular at the apparent horizon. We extract information about the EMT and therefore about the near-horizon geometry by studying how various divergences cancel to produce finite curvature scalars. Once the basic features of the solutions are established we work in coordinates that are continuous across the horizon.

These two assumptions --- (i) curvature scalars are finite on the boundaries of trapped regions, and (ii) real-valued solutions that describe the trapping of light exist at finite time of a distant observer  --- allow for a complete characterization of the near-horizon geometry of a spherically symmetric black hole. This is the subject of the following Secs.~\ref{spher:a} and~\ref{spher:b}. We compare properties of PBHs with those of selected popular models in Sec.~\ref{models}. General issues arising in axial symmetry and specific properties of Kerr--Vaidya metrics are described in Sec.~\ref{axial}.

\subsection{Spherical symmetry: the setup} \label{spher:a}
A general spherically symmetric metric in Schwarzschild coordinates is given by \cite{C:92,vF:15}
\begin{align}
	ds^2 = -e^{2h(t,r)}f(t,r)dt^2+f(t,r)^{-1}dr^2+r^2d\Omega_2 , \label{eq:metric}
\end{align}
where $r$ is the circumferential radius and $d\Omega_2$ is the area element on a unit two-sphere.  These coordinates provide geometrically preferred foliations with respect to Kodama time, which is derived from a natural divergence-free vector field \cite{vF:15,hK:80,AV:10}. This field can be introduced in any spherically symmetric spacetime. In dynamical spacetimes it retains many useful properties of the Killing field to which, modulo possible rescaling, it reduces in the static case \cite{vF:15,AV:10}.

Some of the derivations become more transparent when they are expressed in radiative coordinates $w_\pm$, in the so-called Bondi gauge \cite{BBM:62}. Using the advanced null coordinate $w_+\equiv v$, the metric is written as
\begin{align}
	ds^2=-e^{2h_+}f_+(v,r)dv^2+2e^{h_+}dvdr +r^2d\Omega_2 , \label{eq:metric_vr}
\end{align}
while in $(u,r)$ coordinates, where $w_-\equiv u$ is the retarded null coordinate, the same geometry is described by
\begin{align}
	ds^2=-e^{2h_-}f_-(u,r)du^2-2e^{h_-}dudr +r^2d\Omega_2 . \label{metric_ur}
\end{align}
Gravitational radiation is absent in spherically symmetric spacetimes. This feature makes a privileged notion of energy possible --- the so-called Misner\textendash{}Sharp (MS) mass \cite{VAD:11,vF:15,MS:64} $C(t,r)/2$. It is invariantly defined via
\begin{align}
	f(t,r) \defeq 1 - C/r \defeq \partial_\mu r \partial^\mu r , \label{eq:MSmass}
\end{align}
and thus $C(t,r) \equiv C_+ \big( v(t,r),r \big)\equiv C_- \big( u(t,r),r \big)$. We will omit the subscripts $\pm$ from the function $f$ in what follows. The functions $h(t,r)$ and $h_\pm(w_\pm,r)$ play the role of integrating factors in coordinate transformations\cite{B:16}, such as
\begin{align}
	dt=e^{-h}(e^{h_+}dv- f^{-1}dr) . \label{eq:intfactor}
\end{align}
For the Schwarzschild metric $C=2M=\mathrm{const}$ and $h\equiv 0$, while the coordinates $w_\pm$ become the ingoing and outgoing Eddington--Finkelstein coordinates, respectively.

Similar to a Killing vector field $\vK$ that is defined in static spacetimes ($\vK_{(\mu;\nu)}=0$), the Kodama field is timelike outside of the apparent horizon and spacelike inside of it. It is most conveniently expressed in $(w_\pm,r)$ coordinates. For example, in $(v,r)$ coordinates
\begin{align}
	\vKo^\mu=(e^{-h_+},0,0,0).
\end{align}
It is covariantly conserved and generates the conserved current
\begin{align}
	&\nabla_\mu \vKo^\mu=0, \\
	& J^\mu\defeq G^{\mu\nu}\vKo_\nu, \qquad \nabla_\mu J^\mu=0,
\end{align}
thereby giving a natural geometric meaning \cite{AV:10} to the Schwarzschild coordinate time $t$. The MS mass is its Noether charge \cite{vF:15}.

The Schwarzschild radius $r_\sg(t)$ is the largest root of $f(t,r)=0$. Due to the invariance of $C$, it is invariant in the sense that $r_\sg(t) \equiv r_+\big(v(t,r_\sg))$, etc. Tangents to the congruences of ingoing and outgoing radial null geodesics\footnote{These designations make literal sense only in a space with simple topology.} are given in $(v,r)$ coordinates by
\begin{align}
	l_{\mathrm{in}}^\mu=(0,-e^{-h_+},0,0), \qquad l_{\mathrm{out}}^\mu=(1,\half e^{h_+}f,0,0), \label{null-v}
\end{align}
respectively. The vectors are normalized to satisfy $l_{\mathrm{in}}\cdot\l_{\mathrm{out}}=-1$. Their expansions \cite{HE:73,P:04,C-B:09} are
\begin{align}
	\vartheta_{{\mathrm{in}}} = - \frac{2e^{-h_+}}{r}, \qquad \vartheta_{{\mathrm{out}}}=\frac{e^{h_+}f}{r},
\end{align}
respectively. Hence the (outer) apparent horizon is located at the Schwarzschild radius $r_\sg$ \cite{vF:15,B:16,FEFHM:17}, justifying the definition of the black hole mass as \cite{jB:81} $2M(v)=r_+(v)$. Despite the fact that the apparent horizon is observer-dependent in general (see \ref{a:hor}), in spherically symmetric spacetimes it is invariantly defined in all foliations that respect this symmetry \cite{S:11,FEFHM:17}.

Working in $(u,r)$ coordinates allows us to identify white holes (see \ref{a:hor} and \ref{AForm} for details). In this case, both expansions are positive inside of the anti-trapped domain that is bounded by the anti-trapping horizon $r_-(u)=r_\sg\big(t(u,r_-)\big)$.

It is convenient to introduce the effective EMT components
\begin{align}
	\tensor{\tau}{_t} \defeq e^{-2h} \tensor{T}{_t_t}, \qquad \tensor{\tau}{^r} \defeq \tensor{T}{^r^r}, \qquad \tensor{\tau}{_t^r} \defeq e^{-h} \tensor{T}{_t^r} . \label{eq:taus}
\end{align}
In spherical symmetry, the three Einstein equations (for the components $\tensor{G}{_t_t}$, $\tensor{G}{_t^r}$, and $\tensor{G}{^r^r}$) are
\begin{align}
	\partial_r C &= 8 \pi r^2 \tensor{\tau}{_t} / f , \label{eq:Gtt} \\
	\partial_t C &= 8 \pi r^2 e^h \tensor{\tau}{_t^r} , \label{eq:Gtr} \\
	\partial_r h &= 4 \pi r \left( \tensor{\tau}{_t} + \tensor{\tau}{^r} \right) / f^2 . \label{eq:Grr}
\end{align}
A useful relationship between the EMT components in $(t,r)$ and $(v,r)$ coordinates is given by
\begin{align}
	&	\tensor{\theta}{_v} \defeq e^{-2h_+} \tensor{\Theta}{_v_v} = \tensor{\tau}{_t} , \label{eq:thev} \\
	&	\tensor{\theta}{_v_r} \defeq e^{-h_+} \tensor{\Theta}{_v_r} = \left( \tensor{\tau}{_t^r} - \tensor{\tau}{_t} \right) / f , \label{eq:thevr}\\
	&	\tensor{\theta}{_r} \defeq \tensor{\Theta}{_r_r} = \left( \tensor{\tau}{^r} + \tensor{\tau}{_t} - 2 \tensor{\tau}{_t^r} \right) / f^2 ,   \label{eq:ther}
\end{align}
where $\tensor{\Theta}{_\mu_\nu}$ denotes the EMT components in $(v,r)$ coordinates. We denote the limit of $\tensor{\theta}{_v}$ as $r \to r_+$ as $\theta^+_v$, etc. The Einstein equations are then given by
\begin{align}
	& e^{-h_+} \pad_v C_+ + f \pad_r C_+ = 8 \pi r^2 \tensor{\theta}{_v} , \label{eq:Gvv}\\
	& \pad_r C_+ = - 8 \pi r^2 \tensor{\theta}{_v_r} , \\
	& \pad_r h_+ = 4 \pi r \tensor{\theta}{_r} . \label{Grr}
\end{align}
Additional relations are collected in \ref{AForm}.

In a general four-dimensional spacetime, there are 14 algebraically independent scalar invariants (i.e.\ invariants not satisfying any polynomial relation) that can be constructed from the Riemann tensor \cite{exact:03}. A convenient system of polynomial invariants consists of the Ricci scalar and 15 additonal invariants \cite{CM:91}. In a general spacetime the 20 independent components of the Riemann tensor can be represented with the ten scalars that are particular contractions of the Weyl and Ricci tensors with the vectors of the Newman--Penrose null tetrad \cite{C:92,FN:98,exact:03,C-B:09}. Spherical symmetry reduces the number of independent scalars, as well as the number of independent components of the Weyl and Ricci tensors. As spherically symmetric spacetimes are of type-D in the Petrov classification, a natural tetrad (see \ref{AForm}) results in only four non-zero real values of the scalars $\Lambda$,  $\Psi_2$, and  $\Phi_{00}$, $\Phi_{11}$, and $\Phi_{22}$.

By virtue of the Gauss--Bonnet invariant \cite{C-B:09}
\begin{align}
	\eG \equiv \eR^2 - 4 R_{\mu\nu}R^{\mu\nu} + \eK = \mathrm{const} ,
\end{align}
where $\eK \defeq R^{\mu\nu\lambda\sigma} R_{\mu\nu\lambda\sigma}$ is the Kretschmann scalar, imposing the regularity conditions on ${\tilde{\maT}}$ (see below) is equivalent to imposing the same conditions on $\eK$.  In general, for a manifold of fixed topology in four spacetime dimensions, there are two independent quadratic curvature invariants\cite{L:938}.  We use two quantities that can be expressed directly from EMT components:
\begin{align}
	\tilde{\mathrm{T}}\defeq \tensor{T}{^\mu_\mu}, \qquad \tilde{\maT} \defeq \tensor{T}{_\mu_\nu} \tensor{T}{^\mu^\nu} .
\end{align}
The Einstein equations relate them to the curvature scalars as $\tilde{\mathrm{T}} \equiv - {\eR}/8\pi$ and $\tilde{\mathfrak{T}} \equiv \tensor{R}{^\mu^\nu}\tensor{R}{_\mu_\nu}/64\pi^2$, where $\tensor{R}{_\mu_\nu}$ and $\eR$ are the Ricci tensor and Ricci scalar, respectively.
We note in passing that, while the choice of $\tilde{\maT}$ as one of the curvature scalars is not very popular, it has been used in studies of various no-hair conjectures \cite{B:95,NQS:96} (see also Sec.~\ref{s:mMTG}).

An ostensibly stronger requirement is finiteness of all EMT components at the horizon, expressed in a local orthonormal frame (that is obtained from a regular coordinate system). This requirement was used in the study of the Hawking-radiation-induced EMT \cite{CF:77}, wherein the regularity considerations were used to identify and/or constrain various quantities. We will see that the PBH solutions exhibit the same type of finiteness of the EMT components.

In spherical symmetry $\tensor{T}{^\theta_\theta} \equiv \tensor{T}{^\varphi_\varphi}$, and the most general form of the EMT \cite{C:92,P:10} in an orthonormal basis attached to a fiducial static observer is given by
\begin{align}
	\tensor{T}{_{\hat{\mu}}_{\hat{\nu}}} = \begin{pmatrix}
		\rho$ $\,~ & \psi$ $~ & 0$ $~ & 0$ $~ \vspace{1mm}\\
		\psi & p & 0 & 0 \vspace{1mm}\\
		0 & 0 & { \mathfrak{p}} & 0 \vspace{1mm} \\
		0 & 0 & 0 & { \mathfrak{p}} \vspace{1mm} \\
	\end{pmatrix},
	\label{tspher}
\end{align}
where $\rho$, $p$, $\psi$, and $\mathfrak{p}$ are functions of $t$ and $r$.

 It can be shown that  if the curvature scalars are finite, then $\mathfrak{p}$ cannot diverge as fast as the three effective EMT components $\tensor{\tau}{_a}$, $a \in \lbrace \tensor{}{_t}, \tensor{}{_t^r}, \tensor{}{^r} \rbrace$, do when approaching the apparent horizon \cite{BMsMT:19,T:20}. Different leading rates of divergence among the four quantities $\tau_a$, $\mathfrak{p}$ are also excluded\cite{MT:21mg}.
  Hence the absence of a scalar curvature singularity at the apparent horizon requires that the quantities
\begin{align}
     \mathrm{T} \defeq (\tensor{\tau}{^r} - \tensor{\tau}{_t}) / f , \qquad \mathfrak{T} \defeq \big( (\tensor{\tau}{^r})^2 + (\tensor{\tau}{_t})^2 - 2 (\tensor{\tau}{_t^r})^2 \big) / f^2 ,
     \label{eq:TwoScalars}
\end{align}
are finite at $r=r_\sg$  {(and the subsequent verification that $\eR$ is finite\cite{MT:21mg})}.

\subsection{Spherical symmetry: consistent solutions} \label{spher:b}
To satisfy the regularity requirements of $\mathrm{T}$ and $\mathfrak{T}$, the effective EMT components $\tau_a$,  $a \in \lbrace \tensor{}{_t}, \tensor{}{_t^r}, \tensor{}{^r} \rbrace$, may behave in two qualitatively different ways as $r \to r_\sg$. One possibility is that $\tau_a \to 0$ sufficiently fast, such that the regularity of Eq.~\eqref{eq:TwoScalars} does not constrain the convergence any further. Alternatively, they may converge to zero slower than $f$, converge to a constant, or diverge in such a way that $\maT$ and $\mathrm{T}$ remain finite.

We apply the standard procedure for series solutions of ordinary differential equations near regular-singular points \cite{BO:78} to express the metric functions $C(t,r)$ and $h(t,r)$ given a particular mode of behavior of $\tau_a$. Since we have three partial differential equations for two functions [e.g.\ Eqs.~\eqref{eq:Gtt}--\eqref{eq:Grr} or Eqs.~\eqref{eq:Gvv}--\eqref{Grr}], consistency requirements lead to additional relations between the rate of change of $r_\sg$ and other parameters. The solution is constructed near the apparent horizon, and in $(t,r)$ coordinates is expressed in terms of powers and logarithms of $x(t,r)\defeq r-r_\sg(t)$. By checking the self-consistency of the resulting solution, we find that only two possibilities are viable for dynamical black holes.

Close to the Schwarzschild radius $r_\sg$, the effective EMT components may scale with $f(t,r)$ as
\begin{align}
	\tensor{\tau}{_t} \sim f^{k_t}, \qquad \tensor{\tau}{^r} \sim f^{k_r}, \qquad \tensor{\tau}{_t^r} \sim f^{k_{tr}} , \label{tauS}
\end{align}
for some powers $k_a$. However, only two scaling behaviors correspond to viable dynamic solutions: those with $k_a=0$ and $k_a=1$ $\forall a$.  More importantly, they correspond to $\theta^+_v\neq 0$ ($\theta^+_u\neq 0$) and  $\theta^+_v=0$ ($\theta^+_u=0)$, respectively.  We establish this fact in what follows. Solutions of the $k=0$ class are described in Sec.~\ref{k=0}, and $k=1$ solutions are presented in Sec~\ref{k=1}. The main features of both classes are summarized in Table~\ref{tab:PBHsol} that is adapted from Ref.~\refcite{sM:21}.

\begin{table}[!htpb]
	\tbl{Comparison of the two classes of dynamic solutions in spherical symmetry. Equations in the text associated with the relevant quantities are given in brackets. The metric functions $C$ and $h$ [cf.\ Eqs.~\eqref{eq:metric}--\eqref{eq:MSmass}] are obtained as the solutions of Eqs.~\eqref{eq:Gtt} and \eqref{eq:Grr} and are written together with the effective EMT components and Ricci scalar as series expansions in terms of the coordinate distance $x \defeq r - r_\sg$ from the apparent horizon $r_\sg$. The function $\Upsilon(t)>0$ parametrizes the leading contributions to the effective EMT components for $k=0$ solutions, and $\xi(t)$ is determined by the choice of time variable. The index $j \in \mathbb{Z} \frac{1}{2}$ labels half-integer and integer coefficients and powers of $x$. Since only the leading terms in each series are relevant, we simplify the notation by writing $c_{12}$ instead of $c_{1/2}$, and similarly for higher orders and coefficients of the EMT expansion and Ricci scalar. To remind us of their connection to physical quantities, the coefficients of the effective EMT components are denoted $e_j$ (energy density), $\phi_j$ (flux), and $p_j$ (pressure). Consistency of Eqs.~\eqref{eq:Gtr}--\eqref{eq:Grr} implies $E = - P = 1/(8 \pi r_\sg^2)$ and $\Phi=0$, i.e.\ energy density and pressure take on their maximal possible values at the horizon. This extreme-valued solution is the only self-consistent dynamic solution for $k=1$. (Static $k=1$ solutions can have $E \leqslant 1/(8 \pi r_\sg^2)$ and are described in Sec.~\ref{k=1}). The lower (upper) signature in Eqs.~\eqref{eq:k0rp}, \eqref{eq:k0tautr}, and \eqref{rder1} describes an evaporating PBH (an expanding white hole). The dynamic behavior of the horizon $r_\sg^\prime \defeq dr_\sg/dt$ is determined by Eq.~\eqref{eq:Gtr}. The Einstein equations \eqref{eq:Gtt}--\eqref{eq:Grr} hold order-by-order in powers of $x$. Accordingly, explicit expressions for higher-order terms in the metric functions are obtained by matching those of the same order in the EMT expansion \cite{sM:21,BsMT:19,sMT:21b}.\vspace*{3mm}}	
	{
	\centering
	\resizebox{\textwidth}{!}{
		\begin{tabular}{ >{\raggedright\arraybackslash}m{0.15\linewidth} | @{\hskip 0.025\linewidth} >{\raggedright\arraybackslash}m{0.385\linewidth} @{\hskip 0.05\linewidth} | @{\hskip 0.025\linewidth} >{\raggedright\arraybackslash}m{0.365\linewidth}}
			& $k=0$ solutions & $k=1$ solution
			\\ \toprule
			Metric functions &
			{\begin{flalign*}
				C &= r_\sg + c_{12} \sqrt{x} + \sum\limits_{j \geqslant 1}^\infty c_j x^j & \eqref{eq:k0C} \\
				h &= - \frac{1}{2} \ln \frac{x}{\xi} + \sumj h_j x^j & \eqref{eq:k0h}
			\end{flalign*}}
			\quad & \quad
			{\begin{flalign*}
				C &= r + c_{32} x^{3/2} + \sum\limits_{j \geqslant 2}^\infty c_j x^j & \eqref{fk1} \\
				h &= - \frac{3}{2} \ln \frac{x}{\xi} + \sumj h_j x^j & \eqref{hk1}
			\end{flalign*}}
			\\[-4mm]
			Leading coefficient &
			{\begin{flalign*}
				c_{12}&= -4 \sqrt{\pi} r_\sg^{3/2} \Upsilon & \eqref{eq:k0C}
			\end{flalign*}}
			\quad & \quad
			{\begin{flalign*}
				c_{32}&= - 4 r_\sg^{3/2} \sqrt{- \pi e_2 / 3} & \eqref{c32-1}
			\end{flalign*}}
			\\[-4mm]
			{\begin{flalign*}
				& f(t,r_\sg+x) & \\
				& \hspace*{0.1mm} x>0 &
			\end{flalign*}}
			&
			{\begin{flalign*}
				f&= |c_{12}|\sqrt{x}/r_\sg &
			\end{flalign*}}
			\quad & \quad
			{\begin{flalign*}
				f&=|c_{32}|x^{3/2}/r_\sg &
			\end{flalign*}}
			\\[-4mm]  			
			Horizon dynamics &
			{\begin{flalign*}
				r_\sg^\prime &= \pm4 \sqrt{\pi r_\sg\xi} \Upsilon & \eqref{eq:k0rp}
			\end{flalign*}}
			\quad & \quad
			{\begin{flalign*}
				r_\sg^\prime &= \pm |c_{32}| \xi^{3/2} / r_\sg & \eqref{rder1}
			\end{flalign*}}
			\\[-4mm]
			Effective EMT &
			{\begin{flalign*}
				\tensor{\tau}{_t} &= - \Upsilon^2 + \sum\limits_{j \geqslant \frac{1}{2}}^\infty e_j x^j & \eqref{eq:k0taut} \\
				\tensor{\tau}{_t^r} &= \pm \Upsilon^2 + \sum\limits_{j \geqslant \frac{1}{2}}^\infty \phi_j x^j & \eqref{eq:k0tautr} \\
				\tensor{\tau}{^r} &= - \Upsilon^2 + \sum\limits_{j \geqslant \frac{1}{2}}^\infty p_j x^j & \eqref{eq:k0taur}
			\end{flalign*}}
			\quad & \quad
			{\begin{flalign*}
				\tensor{\tau}{_t} &= E f + \sum\limits_{j \geqslant 2}^\infty e_j x^i & \eqref{eq:k1-taut} \\
				\tensor{\tau}{_t^r} &= \Phi f + \sum\limits_{j \geqslant 2}^\infty \phi_j x^j & \eqref{eq:k1-tautr} \\
				\tensor{\tau}{^r} &= P f + \sum\limits_{j \geqslant 2}^\infty p_j x^i & \eqref{eq:k1-taur}
			\end{flalign*}}
			\\[-4mm]
			Ricci scalar &
			{\begin{flalign*}
				\eR &= R_0 + R_{12} \sqrt{x} + \sum\limits_{j \geqslant 1}^\infty R_j x^j &
			\end{flalign*}}
			\quad & \quad
			{\begin{flalign*}
				\eR &= 2/r_\sg^2 + \sum\limits_{j \geqslant 1}^\infty R_j x^j &
			\end{flalign*}}
			\\
			\bottomrule
		\end{tabular}
	}
	\label{tab:PBHsol}
	}
\end{table}

The three powers $k_a$ should coincide for the effective components $\tau_a$ that are divergent or converging slower than $f$. Then the existence of real-valued series solutions of the Einstein equations \eqref{eq:Gtt}--\eqref{eq:Grr} requires that
\begin{align}
	\tensor{\tau}{_t} \to \tensor{\tau}{^r} \to - \Upsilon^2(t) f^k,
\end{align}
as $r \to r_\sg$, while the sign of $\tensor{\tau}{_t^r}$ remains unconstrained \cite{BMsMT:19,T:20}.

Below we describe the case of $\tensor{\tau}{_t^r}<0$ using transformations between $(t,r)$ and $(v,r)$ coordinates that are regular at the Schwarzschild radius. In the case of $\tensor{\tau}{_t^r}>0$, an analogous derivation employs $(u,r)$ coordinates. In $(v,r)$ coordinates,  the potentially divergent parts of the two scalars are
\begin{align}
	\mathrm{T} &= f \theta_r + 2\theta_{vr}, \label{s1v}\\
	\maT &= \mathrm{T}^2 + 2  (\theta_r\theta_v-\theta_{vr}^2) . \label{s2v}
\end{align}
The convergence of $\rT$ and $\maT$ ensures that the differences between the three components $\tau_a$ are of the order of $f$. Hence Eq.~\eqref{eq:thevr} ensures that $\theta_{vr}$ remains finite as $r \to r_\sg$. As a result,
\begin{align}
	C_+(v,r)=r_+(v)+w_1(v)\big(r-r_+(v)\big)+\ldots , \label{cintro}
\end{align}
where $w_1\defeq -8\pi r_+^2\theta^+_{vr}$, $\theta^+_{vr}=\lim_{r\to r_+}\theta_{vr}$, and higher-order terms have been omitted.  Hence $f=(1-w_1)y/r_++\ldots$, where $y\defeq r-r_+$.

Since $\theta^+_{vr}$ is finite, Eq.~\eqref{s1v} implies that $\theta_r$ can either diverge slower than $1/f$ or converge to a finite value as $r\to r_+$.  Eq.~\eqref{s2v} implies that the product $\theta_r\theta_v$ reaches a finite limit.

If $k<0$, then the component $\theta_v$ diverges according to Eq.~\eqref{eq:thev}, which implies that $\theta_r \to 0$. Using Eq.~\eqref{Grr}, we find that
\begin{align}
	h_+(v,r)=\xi_1(v)y^{1+\epsilon}+\ldots,
\end{align}
for some $\xi_1(v)$ and $\epsilon>0$. Substituting these expressions for $h_+$ and $C_+$ into Eq.~\eqref{eq:Gvv} results in a finite non-zero value of $\theta_v^+$, which is a contradiction.

Consider now $0<k<1$. The convergence of $\rT$ and $\maT$ ensures that  $\theta^+_{vr}$ is finite, and according to Eq.~\eqref{eq:thev} $\theta^+_v=0$, and thus $\theta_{r}$ diverges not faster than $f^{-k}$. Given Eq.~\eqref{cintro}, it follows from Eq.~\eqref{Grr} that $h_+$ is finite. Thus Eq.~\eqref{eq:Gvv} is satisfied at $r=r_+$ only if $w_1\equiv 1$. Moreover, a careful analysis\cite{T:20} shows that to ensure that $\rT$ remains finite, only integer powers of $y$ are allowed for terms of the power $s\leqslant 2$ in the expansion of both $C_+$ and $h_+$. Hence $\theta^+_r$ is finite. To ensure all these relations, the coefficients of the leading ($\propto f^k$) and sub-leading ($\propto f$) terms in the expansion of $\tau_a$ must satisfy a number of relations that cannot be simultaneously satisfied. As a result, the divergent effective EMT components are disallowed, and the only solution with $\tau_a$ converging slower than $f^1$ corresponds to $k=0$.

In a general dynamic case Eq.~\eqref{eq:TwoScalars} does not constrain different powers $k_a \geqslant 1$, and different $\tau_a \propto f^{k_a}$ may in principle converge to zero at different rates. However, consistent solutions must have either $r_\sg=\mathrm{const}$ or have $r_\sg'\neq 0$ throughout the entire evolution. Dynamical solutions with $k_t \geqslant \frac{3}{2}$ require $k_r=k_{tr}=1$, while the solutions with $k_t=1$ and $k_r >1$ and/or $k_{tr}>1$ are inconsistent \cite{T:20}. Verification that the curvature scalars are finite yields $k_t\geqslant2$ and $\rho(r_\sg)=0$, while $p(r_\sg)=-1/(8\pi r_\sg^2)$. As can be observed from Eq.~\eqref{eq:thev}, these solutions belong to the same group with $\theta_v^+=0$ as the $k=1$ solutions that we consider in detail below (see also Sec.~\ref{s:formation}).

\subsubsection{$k=0$ solutions}\label{k=0}
We begin by constructing series solutions without fixing the leading power of $x=r-r_\sg$. We define $\Xi \defeq \lim_{r\to r_g} \tau_t$ and expand the MS mass as
\begin{align}
	C(t,r) = r_\sg + c_\alpha x^\alpha + \ldots .
\end{align}
Eq.~\eqref{eq:Gtt} results in
\begin{align}
	-c_\alpha^2 \alpha x^{2\alpha-1}=8\pi r_\sg^2\Xi,
\end{align}
which determines $\alpha=\half$. The negative sign of $c_\alpha<0$ follows from the definition of the Schwarzschild radius: $C(r,t)<r $ for $r>r_\sg$. To obtain a real solution it is necessary that $\Xi <0$ and thus equals some $-\Upsilon^2(t)$. Substitution into Eq.~\eqref{eq:Grr} yields the leading terms of the function $h$.

Thus the leading terms of the metric functions are \cite{BMsMT:19,T:20,T:19}
\begin{align}
		C &= r_\sg - 4 \sqrt{\pi} r_\sg^{3/2} \Upsilon \sqrt{x} + \mathcal{O}(x) , \label{eq:k0C} \\
		h &= - \frac{1}{2}\ln{\frac{x}{\xi}} + \mathcal{O}(\sqrt{x}) , \label{eq:k0h}
\end{align}
where $\xi(t)$ is determined by the choice of time variable, and the higher-order terms are matched with the higher-order terms in the EMT expansion \cite{BsMT:19}. We list some relationships that involve higher-order terms in \ref{solutions-long}. Eq.~\eqref{eq:Gtr} must then hold identically. Both sides contain terms that diverge as $1/\sqrt{x}$, and their identification results in the consistency condition
\begin{align}
	r'_\sg/\sqrt{\xi} =   4 \epsilon_\pm\sqrt{\pi r_\sg} \, \Upsilon , \label{eq:k0rp}
\end{align}
where $\epsilon_\pm = \pm 1$ corresponds to the expansion and contraction of the Schwarzschild sphere, respectively. Evaluation of the curvature scalars using the metric functions of Eqs.~\eqref{eq:k0C}--\eqref{eq:k0h} results in finite quantities on the apparent horizon once the consistency condition Eq.~\eqref{eq:k0rp} is used \cite{BMsMT:19,T:19}. The values of the non-zero scalars depend on the higher-order terms of the EMT.

The near-horizon geometry is most conveniently expressed \cite{BMsMT:19} in $(v,r)$ coordinates for $\tensor{\tau}{_t^r} \approx - \Upsilon^2$, i.e.\ $r'_\sg<0$, and in $(u,r)$ coordinates for $\tensor{\tau}{_t^r} \approx + \Upsilon^2$, i.e.\ $r'_\sg>0$. The metric functions in these cases are continuous across the horizons, and the expansions of ingoing and outgoing geodesic congruences can be readily evaluated. We find that the case $r_\sg'<0$ corresponds to an evaporating PBH, and $r_\sg'>0$ to an expanding white hole\footnote{This is contrary to erroneous interpretations in Refs.~\citenum{BMsMT:19,T:19,T:20} that misidentified the expanding white hole as an accreting PBH.}.

This solution has a number of remarkable properties that we now describe. The limiting form of the $(tr)$ block of the EMT as $r\to r_\sg$ is
\begin{align}
	\tensor{T}{^a_b} = \begin{pmatrix}
		\Upsilon^2/f & -\epsilon_\pm e^{-h}\Upsilon^2/f^2 \vspace{1mm}\\
		\epsilon_\pm e^h  \Upsilon^2 & -\Upsilon^2/f
	\end{pmatrix},
	\qquad
 	\tensor{T}{_{\hat{a}}_{\hat{b}}} = \frac{\Upsilon^2}{f} \begin{pmatrix}
		-1 & \epsilon_\pm  \vspace{1mm}\\
		\epsilon_\pm   & -1
	\end{pmatrix},
  	\label{tneg}
\end{align}
where the second expression is written in the orthonormal frame.

In the test-field limit \cite{CF:77,BD:82,E:83,V:96a,LO:16} quantum fields propagate on a given gravitational background and the resulting EMT is not permitted to back-react on the geometry via the Einstein equations. It is instructive to compare the tensor of Eq.~\eqref{tneg} with the explicit results that are obtained in the test-field limit.

Out of the three popular choices for the vacuum state (see \ref{vac3} for a summary) only the Unruh vacuum results in an EMT with non-zero $\tensor{T}{_t_r}$ components \cite{CF:77,U:76}. The state itself corresponds to the requirement that no particles impinge on the collapsing object from infinity \cite{U:76}. In the context of a static maximally extended spacetime its counterpart is a state with unpopulated modes at past null infinity and at the white hole horizon \cite{BD:82,FN:98}.

\begin{figure}[!htbp] \centering
	\includegraphics[width=0.95\textwidth]{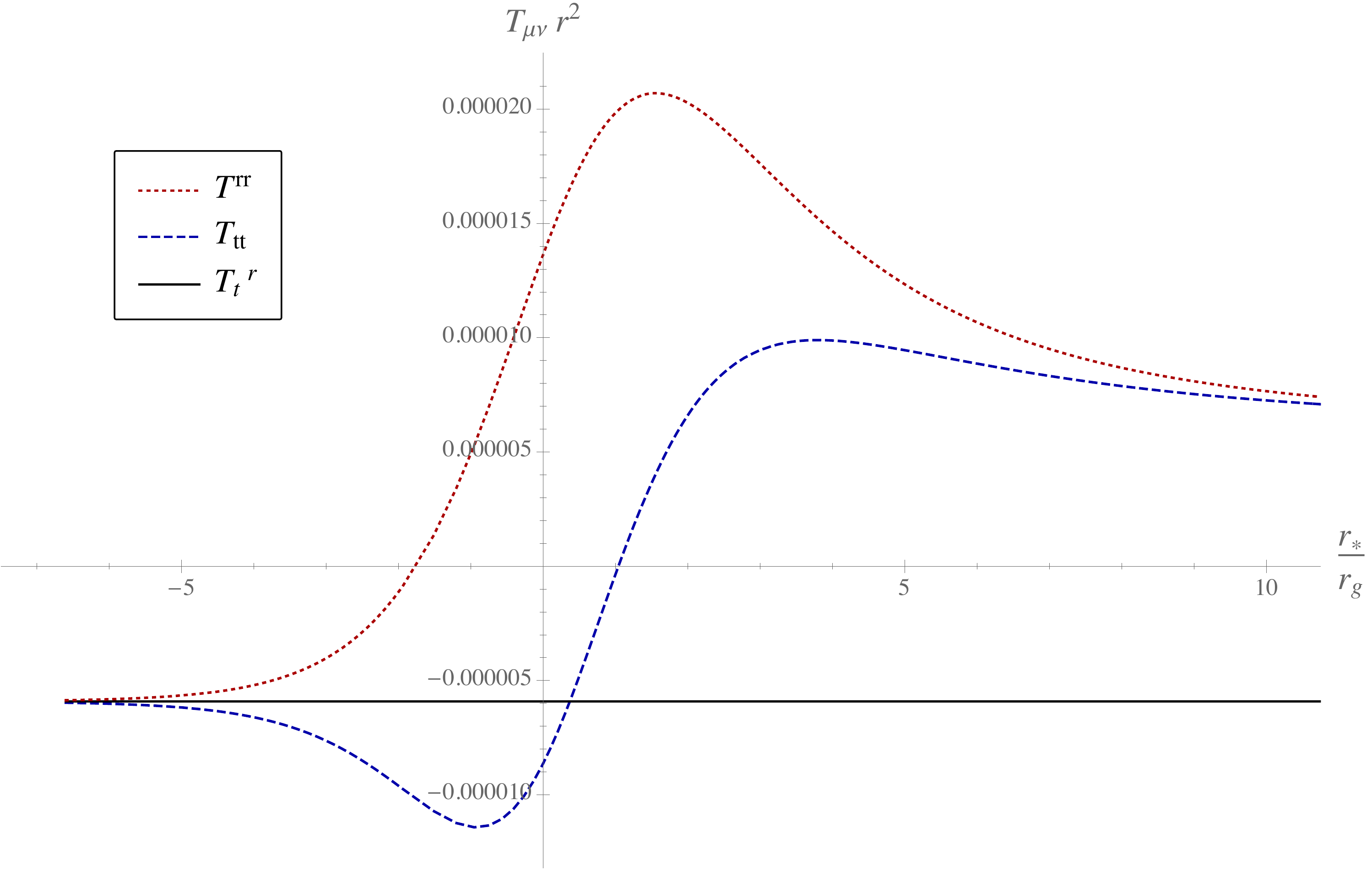}
 	\caption{Expectation values of the renormalized EMT components for the Unruh vacuum of the massless scalar field on the background of a Schwarzschild black hole of mass $M$ as a function of the tortoise coordinate $r_* \defeq r + 2M\ln[r/(2M)-1]$. Values of the EMT components are given in units $\hbar M^{-2}$. On a static background $r^2 \tensor{T}{_t^r} = \mathrm{const}$, resulting in the solid horizontal  line. This figure corresponds to Fig.~1 of Ref.~\citenum{LO:16} by Levi and Ori. The smooth curves were obtained by interpolating the data that was kindly provided by the authors.}
	\label{fig:emt}
\end{figure}

Using various semi-analytic and numerical methods that are based on conformally coupled fields \cite{V:97} and minimally coupled scalar fields \cite{LO:16, L:17}, the expectation values of the renormalized components $\tensor{T}{^r^r}$, $\tensor{T}{_t_t}$, and $\tensor{T}{_t^r}$ have been worked out explicitly. They approach the same negative value as $r \to r_\sg$ (see Fig.~\ref{fig:emt}). An explicit construction of the Wightman function and the renormalized EMT in $(1+1)$-dimensional Vaidya spacetime (see Sec.~\ref{models}) produced by a collapsing null shell revealed that at late times the state approaches the Unruh state \cite{JL:18,Tjoa:2020eqh}.

The NEC \cite{HE:73,exact:03,MV:17,KS:20} is the weakest of all energy conditions and is satisfied by normal classical matter. It posits that for any null vector $k^\mu$ the contraction with the EMT is non-negative, i.e.\ $\tensor{T}{_\mu_\nu} k^\mu k^\nu \geqslant 0$. The assumption of its validity forms the basis of the laws of black hole dynamics and of a majority of the singularity theorems \cite{HE:73,FN:98,SG:15,P:68}. On the other hand, Hawking radiation violates this condition \cite{FN:98,MV:17}, making the laws of black hole mechanics inapplicable in its presence. Violation of the NEC is also one of the ways to circumvent the Buchdahl theorem that limits the compactness of self-gravitating objects by $r_0/r_\sg<9/8$, thus enabling a class of horizonless ECOs with circumferential radius $r_0$ arbitrarily close to $r_\sg$ \cite{CP:19,COY:15}.

The NEC is always violated by the EMT of Eq.~\eqref{tneg}. In the case of an expanding white hole solution ($\epsilon_\pm \to \epsilon_+ = +1$) the inward-pointing null vector $l_{\mathrm{in}}$ leads to its violation, $T_{\mu\nu}l_{\mathrm{in}}^\mu l_{\mathrm{in}}^\nu < 0$, and for an evaporating PBH the NEC is violated by the outward-pointing null vector $l_{\mathrm{out}}$ \cite{BMsMT:19}. The violation of the NEC is caused by the negative sign of the components $\tensor{\tau}{_t}$ and $\tensor{\tau}{^r}$ at the apparent horizon, which in turn is the consequence of requiring real solutions to the Einstein equations at finite time $t_\rS$.

This result should be compared with the conclusions of Sec.~9.2 of Ref.~\refcite{HE:73} that in general asymptotically flat spacetimes with an asymptotically predictable future the trapped surface cannot be visible from future null infinity unless the weak energy condition is violated \cite{FN:98} (note that the rigourous derivation of this result is given in Ref.~\citenum{cC:05}). The above derivation of the NEC violation is more restrictive insofar as we have only considered spherically symmetric settings, but it is more general in the sense that no assumptions about the asymptotic structure of the spacetime were made.

Classification of the EMT according to the Segre\textendash{}Hawking\textendash{}Ellis  scheme \cite{exact:03,HE:73,MV:17} requires transformation to the orthonormal basis. Among the classes \RNum{1}--\RNum{4}, the known classical matter distributions correspond to classes \RNum{1} and \RNum{2}. An expectation value of the renormalized EMT that is so different from classical intuition was advocated in Ref.~\refcite{pR:86} using a two-dimensional model. Ref.~\refcite{APT:19} demonstrated that the EMT for the massless conformal scalar in the Unruh state is of type \RNum{4} everywhere outside of the apparent horizon. For the minimally coupled scalar field the EMT is of type \RNum{4} for \cite{LO:16,L:17} $r \leqslant 1.39 r_\sg$. However, once backreaction is included, in many interesting scenarios the more exotic forms of the EMT (types \RNum{3} and \RNum{4}) are excluded \cite{MV:21}. Our non-perturbative dynamical result shows that if finite-time formation of the apparent horizon is considered (\ref{solutions-long}), this is indeed the case.

The behavior of radial null geodesics reveals an important difference between $k=0$ solutions and classical black hole solutions such as the Schwarzschild metric. For incoming null geodesics crossing the apparent horizon and outgoing ones starting arbitrarily close to it,
\begin{align}
	\lim_{r \to r_\sg}\frac{dt}{dr} = \pm \left.\frac{e^{-h}}{f}\right|_{r=r_\sg}= \pm\frac{1}{|r_\sg'|}, \label{drdt0}
\end{align}
where we used the leading terms in the expansion of the metric functions Eqs.~\eqref{eq:k0C}--\eqref{eq:k0h} and the consistency condition Eq.~\eqref{eq:k0rp}. This result implies that the horizon crossing takes a finite time according to Bob.

Eq.~\eqref{drdt0} allows us to express the transformation law between $(t,r)$ and $(w_\pm,r)$ coordinates in the vicinity of the apparent horizon \cite{T:20, MsMT:21}. The transformation to $(v,r)$ coordinates in the case $r_\sg'<0$ is effected as follows: a point on the apparent horizon has the coordinates  $(v,r_+)$ and $(t,r_\sg)$ in the two coordinate systems, respectively. Moving from $r_+(v)$ along the line of constant $v$ (i.e.\ backwards along the ingoing radial null geodesic) by $\delta r$ leads to the point $(t+\delta t, r_\sg+\delta r)$, where
\begin{align}
	\delta t=-\left.\frac{e^{-h}}{f}\right|_{r=r_\sg} \!\!\!\!\! \!\!\!\!\delta r=\frac{\delta r}{r'_\sg} . \label{dtrv}
\end{align}
This implies that along the ingoing radial null geodesic $t(v,r_++\delta y) = t(r_\sg) - \delta y/|r_\sg'|$, resulting in
\begin{align}
	x(r_++y,v) =r_++y-r_\sg\big(t(v,r_++y)\big)=  -r''_\sg y^2/(2 r_\sg'{}^2)+\cO( y^3). \label{xyrel}
\end{align}
Eq.~\eqref{xyrel} relates the coordinates $x= r-r_\sg(t)$ and $y(r,v)\defeq r-r_+(v)$ in the vicinity of the apparent horizon. These two results are also valid also for $k=1$ solutions \cite{sMT:21,T:20}.

Next, using invariance of the MS mass $C_+(v,r)=C\big(t(v,r),r\big)$ and Eq.~\eqref{xyrel}, we find \cite{MsMT:21}
\begin{align}
	w_1 = 1 - 2 \sqrt{2 \pi r_\sg^3 |r_\sg''|} \frac{\Upsilon}{|r_\sg'|}. \label{w1t}
\end{align}
Eqs.~\eqref{eq:ther} and \eqref{Grr} ensure that in the case of evaporation the limit $\theta_{r}^+$ is finite [its explicit value is given by Eq.~\eqref{therrA}]. Hence, in the case of evaporation, the non-singular expression for the metric in $(v,r)$ coordinates is given by
\begin{align}
	C_+(v,r) &= r_+(v)+w_1(v)y+\cO(y^2), \label{Cpl} \\
	h_+(v,r) &= \chi_1(v) y+\cO(y^2), \label{hpl}
\end{align}
where $w_1\leqslant 1$ due to the definition of the apparent horizon, $\chi_1(v)$ is some function, and we used the freedom of redefining the $v$ coordinate to set $h(v,r_+) \equiv 0$.

A direct calculation of the Lie derivative $\mathcal{L}_{l\mathrm{in}}\vartheta_\mathrm{out}$ at the Schwarzschild radius,
\begin{align}
	l^\mu_\mathrm{in}\pad_\mu \vartheta_\mathrm{out}|_{r_\sg}=-(1-w_1)/r_+^2<0,
\end{align}
indicates that this is indeed a location of the outer horizon\cite{H:94,vF:15,B:05}. Note that this strict inequality is not satisfied by $k=1$ solutions.

For a white hole, Eqs.~\eqref{eq:ther} and \eqref{Grr} imply that the function $h_+$ diverges logarithmically as $y \to 0$. Analogously, $(u,r)$ coordinates are regular at the anti-trapping horizon of a white hole and singular for a PBH.

We conclude by noting that, contrary to the classical solutions, the outer apparent horizon and the anti-trapping horizon are timelike surfaces \cite{BMsMT:19}. Using the terminology of Ref.~\refcite{AK:04}, this makes them timelike membranes, and makes some authors wonder if they should be called horizons at all \cite{B:16}. This is different from the usual black hole models where the expanding apparent horizon is spacelike \cite{FN:98,SAK:20}.

\subsubsection{$k=1$ solutions}\label{k=1}
A static $k=0$ solution is impossible, as in this case the scalar $\mathfrak{T}$ would diverge at the apparent horizon. Consequently, the effective EMT components $\tensor{\tau}{_t}$ and $\tensor{\tau}{^r}$ should converge to zero at least as fast as the metric function $f(r)$ or faster.

Indeed, many models of static black holes with \cite{NQS:96,MB:96,BR:06} and without \cite{B:68,H:06,A:08,M:21,HMP:11,vpF:16} a singularity have finite values of energy density $- \tensor{T}{^t_t} \eqdef \rho$ and pressure $\tensor{T}{^r_r} \eqdef p$ at the Schwarzschild radius $r_\sg$. With respect to the invariants of Eq.~\eqref{eq:TwoScalars}, these are the $k=1$ solutions with
\begin{align}
	\tensor{\tau}{_t} \to E f, \qquad \tensor{\tau}{^r} \to P f, \qquad \tensor{\tau}{_t^r} = 0, \label{k1-stat}
\end{align}
where at the apparent horizon the energy density and pressure are $\rho(r_\sg)=E$ and $p(r_\sg)=P$, respectively. The Reissner--Nordstr\"{o}m black holes \cite{HE:73,MTW,C:92,C-B:09,P:04} have
\begin{align}
	ds^2=-f(r)dt^2+f(r)^{-1}dr^2+r^2d\Omega_2, \qquad f=1-\frac{2M}{r}+\frac{Q^2}{r^2}, \label{m:RN}
\end{align}
where $M$ and $Q$ denote the mass and electric charge, respectively. The two roots of $f(r)=0$,
\begin{align}
	r_\pm^\mathrm{h}=M\pm\sqrt{M^2-Q^2},
\end{align}
are the event horizon $r_\mathrm{eh}\equiv r_\sg=r_+^\mathrm{h}$ and the (inner) Cauchy horizon $r_\rin=r_-^\mathrm{h}$, respectively. The EMT satisfies $\rho=-p$ and the expansion of the function $f$ outside the Schwarzschild radius gives
\begin{align}
	f=\left(1-\frac{Q^2}{r_\sg^2}\right)\frac{x}{r_\sg}+\cO(x^2),
\end{align}
while $E=-P=Q^2/r_\sg^4$. We discuss the inner horizon of Reissner--Nordstr\"{o}m black holes in Sec.~\ref{sc:inner}.

A particularly simple model of a static RBH (see Sec.~\ref{models} for more details) is given by the Hayward metric \cite{H:06}
\begin{align}
	h=0, \qquad f=1-\frac{2mr^2}{r^3+2b^2m}, \label{Hmodel}
\end{align}
where $b$ and $m$ are positive constants. This form of the MS mass can be obtained \cite{PI:88} by assuming that the energy density is proportional to the square of the curvature $C/2r^2$.

For $m<m_*\defeq 3\sqrt{3}b$, the equation $f(r)=0$ has no zeroes and the spacetime has the causal structure of Minkowski spacetime. For $m=m_*$, there is a single double root $r_\sg=r_*\defeq\sqrt{b}$ and the metric describes an extreme black hole with a single marginally trapped surface. For $m>m_*$, the inner and outer Killing horizons of the RBH are located at the two simple roots  of $f(r)=0$. When $m\gg b$   these roots are
\begin{align}
	r_\sg \approx 2m-\frac{b^2}{2m}, \qquad r_\mathrm{in}\approx \frac{5b}{4}-\frac{3b^2}{32m},  \label{horizonsHF}
\end{align}
specifying the locations of the outer and inner horizons, respectively.

From the Einstein equations it follows that in this model $\rho=-p$ everywhere. The causal structure of this spacetime is represented in Fig.~\ref{fig:innercausal}. Moreover, a general property of static $k=1$ models is that \cite{sH:13}
\begin{align}
	\rho(r_\sg) = -p(r_\sg)
\end{align}
and
\begin{align}
	8 \pi r_\sg^2 \rho(r_\sg) \leqslant 1,
\end{align}
with equality corresponding to extreme black holes\cite{NQS:96,MB:96}.

For a dynamical solution, the relations given in Eq.~\eqref{k1-stat} are replaced by
\begin{align}
	\tensor{\tau}{_t} \to E(t) f, \qquad \tensor{\tau}{^r} \to P(t) f, \qquad \tensor{\tau}{_t^r} \to \Phi(t) f.  \label{eq:k1taus}
\end{align}
Any two functions can be expressed algebraically in terms of the third. The energy density satisfies $8 \pi r_\sg ^2 E \leqslant 1$ to ensure that $C(t,r)-r_\sg>0$ for $r>r_\sg$. The explicit form of the solutions is given in \ref{As1}. Again, for an evaporating black hole the expansions of $C_+$ and $h_+$ in $(v,r)$ coordinates include only positive integer powers of $y \defeq r-r_+$.

Only the extreme value of $E=(8\pi r_\sg^2)^{-1}$ corresponds to non-extreme dynamical black/white holes\cite{sMT:21} (i.e.\ those with a non-zero volume of the trapped region and $r_\sg \neq \mathrm{const}$). From Eqs.~\eqref{eq:k1taus} and \eqref{eq:thev} it follows that $\theta^+_v=0$. Eq.~\eqref{the1} shows that $r_+'\neq 0$ is possible only if $w_1 \equiv 1$. On the other hand, $\mathbb{D}_v(r)\defeq C_+(v,r)-r$ is negative for $r>r_+$, and is positive for $r<r_+$. Consequently, the leading terms in the expansion of the MS mass in Eq.~\eqref{Cpl} are $C_+ = r_+ + y + w_3 y^3\equiv r +{ w_3 y^3}$, with $w_3 \leq 0$, and do not contain the even term $y^2$. If $w_3=0$, the non-linear terms begin from a higher odd power.

This expression for the MS mass must coincide with $C\big(t(v,r_++y),r_++y)$. Using Eq.~\eqref{xyrel} leads to
\begin{align}
	C_+= C=r_\sg+y+(1-8\pi r_\sg^2 E)\frac{r''_\sg y^2}{2r'_\sg{}^2}+\mathcal{O}(y^3),
\end{align}
and thus $E\equiv 1/(8\pi r^2_\sg)$ to ensure that $w_2\equiv 0$.  Thus
\begin{align}
	C(t,r)=r+ c_{32}(t)x^{3/2}+\cO(x^2), \label{fk1}
\end{align}
for some coefficient $c_{32}(t)<0$, setting via Eq.~\eqref{eq:k1taus} the scaling of other leading terms in the EMT. Consistency of Eqs.~\eqref{eq:Gtr} and \eqref{eq:Grr} implies $P=-E=-1/(8\pi r^2_\sg)$ and $\Phi=0$. From the next-order expansion, we obtain
\begin{align}
	h=-\tfrac{3}{2}\ln (x/\xi)+\mathcal{O}(\sqrt{x}), \label{hk1}
\end{align}
and the consistency relation
\begin{align}
	r'_\sg=\pm |c_{32}|\xi^{3/2}/r_\sg. \label{rder1}
\end{align}
Details of the calculation are presented in  \ref{As1}. Once again, the case of $r_\sg'<0$ corresponds to an evaporating PBH, and $r_\sg'>0$ to an expanding white hole.

Unlike for $k=0$, the violation of the NEC in this case is more subtle. At $r=r_\sg(t)$ it is marginally satisfied as the $(tr)$ block of the EMT is
\begin{align}
	\tensor{T}{_{\hat{a}}_{\hat{b}}} = \frac{1}{8\pi r_\sg^2}
	\begin{pmatrix}
		1 & 0 \vspace{1mm} \\
		0 & -1
	\end{pmatrix} .
  	\label{tneg1}
\end{align}
However, the NEC is violated for some range $x>0$ (see \ref{As1}). This behavior plays a role in the formation of PBHs (Sec.~\ref{s:formation}).

Solutions with a time-independent apparent horizon or general static solutions do not require $w_1=1$ to satisfy Eqs.~\eqref{eq:thev}--\eqref{eq:ther}. Since $r_+(v)=r_\sg(t)=\mathrm{const}$, it is possible to have non-extreme solutions. Then Eq.~\eqref{eq:Gtr} implies $\Phi=0$, and the identity $E=-P$ follows from Eq.~\eqref{eq:ther}, leading to a regular function $h(t,r)$. However, in this case Eq.~\eqref{drdt0} indicates that the apparent horizon cannot be reached in a finite time $t$ of Bob.

\subsubsection{Inner apparent horizon} \label{sb:in}
Solutions of classical collapse models (i.e.\ those with matter that satisfies at least the NEC), both analytical and numerical, provide several qualitative scenarios for black hole formation \cite{HE:73,FN:98,H:13,JM:11,RZ:13,BS:10}. Generically, the first marginally trapped outer surface forms in the bulk  and subsequently one branch moves inward, while the outer apparent horizon approaches the event horizon. In spherical symmetry, depending on the detailed properties of the model, the apparent horizon may form on the matter-vacuum boundary, while the inner apparent horizon propagates to the center of symmetry, reaching it with the formation of a singularity. Alternatively, the apparent horizon may form at the center and propagate outwards.

To discuss the formation of PBHs (Sec.~\ref{s:formation}, Fig.~\ref{f:formation}) we need to investigate properties of the inner apparent horizon. Assuming that after some $0<t_{\rS}<\infty$ the equation $f(t,r)=0$ has only two roots, we apply the same regularity requirements of Eq.~\eqref{eq:TwoScalars} to the smaller root $r_\rin$. We consider only the $k=0$ case as it is the most relevant \cite{T:19}. The finite times $t>t_\rS$  clearly have an operational meaning  if the trapping horizons are closed, i.e.\ if the object of study is a transient (even if long-lived) RBH (Fig.~\ref{fig:time-g}(b)).

By repeating the analysis of Sec.~\ref{k=0} around $r_\rin$, we find that
\begin{align}
	\lim_{r\to r_\rin} \tensor{\tau}{_t} = \lim_{r \to r_\rin} \tensor{\tau}{^r} = + \Upsilon_\rin^2(t)
\end{align}
for some $\Upsilon_\rin(t)$. On the other hand, analysis of the Einstein equations indicates that
\begin{align}
	&\lim_{r \to r_\rin} \tensor{\tau}{_t^r} = + \Upsilon_\rin^2, \qquad r'_\rin<0,  \\
	&\lim_{r \to r_\rin} \tensor{\tau}{_t^r} = - \Upsilon_\rin^2, \qquad r'_\rin>0.
\end{align}
Outside of the trapped/anti-trapped region, the metric functions are given by
\begin{align}
	C=r_\rin(t)-4\sqrt{\pi} r_\rin^{3/2}\Upsilon_\rin\sqrt{r_\rin-r}, \qquad h=-\frac{1}{2}\ln\frac{r_\rin-r}{ {\xi_\rin}} ,
\end{align}
for some ${\xi_\rin}$, and the consistency condition is
\begin{align}
	r'_\rin /\sqrt{ {\xi_\rin}} = - 4 \epsilon_\pm \sqrt{\pi r_\sg} \, \Upsilon_\rin .
\end{align}
Here $\epsilon_\pm \to \epsilon_+ = +1$ when the inner horizon propagates towards the center.

The $(tr)$ block of the EMT is now
\begin{align}
 	\tensor{T}{_{\hat{a}}_{\hat{b}}} = \frac{\Upsilon_\rin^2}{f}
 	\begin{pmatrix}
		1 & \epsilon_\pm \vspace{1mm}\\
		\epsilon_\pm & 1
	\end{pmatrix}
  \label{tpos}.
\end{align}
Therefore, regardless of whether $r_\rin$ advances towards the origin ($r'_\rin < 0$) or retreats from it ($r'_\rin > 0$), the NEC is satisfied.

Similar to the Schwarzschild radius it is possible to obtain a non-singular expression for the metric in the vicinity of $r_\rin$. From the transformation of the EMT components, it follows that $(v,r)$ coordinates provide such a description when $r'_\rin < 0$, and $(u,r)$ coordinates when $r'_\rin > 0$.  For a PBH with $r_\rin$ propagating towards the origin, we have
\begin{align}
	C(v,r)=r_{<+}(v)+w_{<1}(v)\left(r-r_{<+}(v)\right)+\ldots,
\end{align}
with $r_{<+}(v)\equiv r_\rin(t)$ and $w_{<1}\geqslant 1$. Then $\mathcal{L}_{l\mathrm{in}}\vartheta_\mathrm{out}$ indicates that the locus of points $r_\rin(t)$ is indeed the inner horizon in the sense of Ref.~\refcite{H:94}. Since the region between the two roots of $f(u,r)=0$ describes a white hole, only solutions with $r'_\rin < 0$ are relevant for PBHs, where it describes the inner apparent horizon. As the NEC is satisfied in its vicinity, the inner apparent horizon is either timelike or null \cite{H:94}.

\subsection{Spherical symmetry: relations with popular models} \label{models}
Vaidya metrics
\begin{align}
	ds^2 = - \left(1-\frac{r_\pm(w_\pm)}{r}\right) dw_\pm^2 + 2 \epsilon_\pm {dw_\pm} dr + e^2 d\Omega_2, \label{Vaidya}
\end{align}
where $w_\pm$ are the outgoing (ingoing) null coordinates and $\epsilon_\pm = \pm 1$, are the simplest non-stationary generalizations of the Schwarzschild solution \cite{V:51}. The apparent horizon of these metrics is located at $r_\sg=r_\pm\equiv C_\pm$.  Comparison of the EMT components shows that all Vaidya metrics are $k=0$ solutions. Moreover, $\tau_t=\tau^r=\pm\tau_t^{~r}$ holds as an exact relationship.

The outgoing metric \cite{BBM:62} ($w_-=u$) with decreasing mass $r'_-(u)<0$  was originally proposed to model the radiation of stars by supplying the Einstein equations for the exterior domain with an EMT of the geometric optics form \cite{G:74}. Its interior, however, describes an anti-trapped region. Beyond their pedagogical use \cite{B:16,P:04} in illustrating the differences between various types of masses and horizons, Vaidya metrics are widely applied in studies of stellar dynamics and gravitational collapse \cite{BOS:89,FT:08,PS:20}.

The ingoing metrics ($w_+=v$) model the spacetime geometry in the vicinity of the apparent horizon \cite{VZF:76,H:81,H:87,BMPS:95}, and the ingoing Vaidya metric with decreasing mass \cite{H:87,BMPS:95} is used to model the effect of Hawking radiation at distances $r \gtrsim \cO(3r_\sg)$. Despite their apparent simplicity, these metrics describe a variety of spacetime structures, e.g.\ the formation of a transient trapped region when used as an exterior metric for particular models of collapsing stars, or a singularity (naked or hidden behind the event horizon) \cite{FT:08,BDE:17}.

The EMT has only one non-zero component
\begin{align}
	\tensor{T}{_w_w} = \frac{\epsilon_\pm r'_\pm(w)}{8 \pi r^2},
\end{align}
where the $\pm$ subscripts on $w$ have been omitted to reduce clutter. Hence out of four possibilities only two --- the metrics with $r'_+(v)<0$ and $r'_-(u)>0$ that violate the NEC --- can describe the geometry near an apparent horizon that was formed in time $t$. For the other two metrics, any attempt to construct an explicit transformation to $(t,r)$ coordinates leads to complex-valued functions \cite{BMsMT:19}. The two admissible metrics belong to the $k=0$ class of solutions: all functions $w_i(u)$ and $w_i(v)$ are identically zero. We summarize these properties in Table~\ref{tab:EinEqRealSol}.

\begin{table}[htpb!]
	\tbl{Properties of the four types of Vaidya metrics. The Einstein equations have real solutions at finite time $t>t_\rS$ only if the NEC is violated.}
	{
	\centering
	\begin{tabular}{ccccc} \toprule \\[-2mm]
		$\text{sgn}(\tensor{T}{_t_t})$ & $\text{sgn}(\tensor{T}{_t^r})$  & \begin{tabular}{@{}c@{}} Time-evolution of \\ Vaidya mass function \end{tabular} & \begin{tabular}{@{}c@{}} Black/ \\ White hole \end{tabular} & \begin{tabular} {@{}c@{}} NEC \\ violation \end{tabular} \\ \colrule
		$-$ & $-$ & $C^\prime(v) < 0$ & B & \cmark \\
		$-$ & $+$ & $C^\prime(u) > 0$ & W & \cmark \\
		$+$ & $-$ & $C^\prime(u) < 0$ & W & \xmark \\
		$+$ & $+$ & $C^\prime(v) > 0$ & B & \xmark \\[1mm] \bottomrule
	\end{tabular}
	\label{tab:EinEqRealSol}
	}
\end{table}

One of the simplest dynamical models of a RBH \cite{M:21,F:14,H:06,vpF:14} is modification of the metric of Eq.~\eqref{Hmodel} by adding the dependence of the mass parameter on the advanced time,
\begin{align}
	C_+(v,r) = \frac{2m(v)r^3}{r^3+2b^2m(v)}. \label{hf-m}
\end{align}
The non-zero components of the EMT at the apparent horizon are
\begin{align}
	\tensor{\Theta}{_v_v} \approx \frac{m'(v)}{16 \pi m^2(v)}, \qquad \tensor{\Theta}{_v_r} \approx -\frac{3b^2}{128 \pi m^4}.
\end{align}
This model belongs to the $k=0$ class.  {For $m'(v)<0$} it is consistent with the formation of an apparent horizon in finite time of a distant observer. Several generalizations of this model are described in Ref.~\refcite{BHL:18}.

A different modification of the Vaidya metric that is motivated by renormalization group considerations \cite{BR:06} is based on
\begin{align}
	h_+=0, \qquad C_+=2M(v)G(r),
\end{align}
where $G(r)$ is obtained as a consequence of the running of the Newton constant at different energy scales (e.g.\ in Planck units it is the laboratory value $G_0$ that is set to unity). Such a spacetime exhibits an event horizon, and the static limit of the improved Schwarzschild metric describes a cold (i.e.\ approaching zero Hawking temperature) remnant (a wider context and potential issues are presented in Ref.~\refcite{COY:15}).

However, regardless of any other consideration, models with finite $h_+ $ may be a consistent description of only the evaporation part of the evolution of a PBH. They cannot describe its formation at finite $t_\mathrm{S}$ because the NEC is not violated for $m'(v)>0$. In fact, no model that uses $(v,r)$ coordinates and has an accreting phase (for example the models of Refs.~\refcite{F:14,BR:06,H:06,H:81,SMV:19}) can describe a PBH that is formed at some finite time $t_\rS$. On the other hand, models that contain anti-trapping regions \cite{M:17,CDLV:20,A:20} cannot have a continuous $h_+$ when expressed in $(v,r)$ coordinates, since
\begin{align}
	\frac{\pad_r h_+}{r} = 4 \pi {\Theta}_{rr} \to \frac{16 \pi}{f^2} \Upsilon^2
\end{align}
ensures the divergence of at least $\pad_r h_+$.

Properties of the inner horizon $r_\rin$ impose additional constraints. Even if mass inflation (see Sec.~\ref{sc:inner}) leads to a curvature singularity at some finite value of the evolution parameter, until that point the NEC is satisfied (Sec.~\ref{sb:in}). A direct calculation shows \cite{BHL:18} that in models with $h_+=0$ the violation of the NEC is determined by $\pad_v f$. Hence, in models where the sign of the latter is directly related to the sign of $\pad_v r_+$, as it is in Eq.~\eqref{hf-m}, the NEC is not violated while $\pad_v r_+>0$. Therefore, these models cannot describe the vicinity of the inner horizon during the evaporation stage.

The anti-trapped region is part of black-to-white hole models \cite{M:17}, and in particular loop quantum gravity inspired models of evaporation, where a transition region of strong quantum effects \cite{AB:05,A:20} links the trapped and anti-trapped regions \cite{A:20}.
We discuss additional properties of RBHs in Sec.~\ref{subsec:HorizonFormation}.

In this review we do not discuss wormholes and their role in alternative collapse scenarios. However, we note the Simpson--Visser metric \cite{SV:19}
\begin{align}
	ds^2=-\left(1-\frac{2m}{\sqrt{\eta^2+a^2}}\right)dt^2+\frac{d\eta^2}{1 -\frac{2m}{\sqrt{\eta^2+a^2}} } +(\eta^2+a^2)d\Omega_2, \label{svm}
\end{align}
where $-\infty<\eta<\infty$, and  $a$ is a parameter, interpolates between the Schwarzschild black hole, RBHs, and traversable wormholes  depending on the value of $a$. The MS mass is
$C=2m+a^2(r-2m)/r$. For $a< 2m$ there is a pair of horizons at $\eta_\pm=\pm\sqrt{(2m)^2-a^2}$ (i.e. at $r=r_\sg\equiv 2m$).

\subsection{Axial symmetry} \label{axial}
A general time-dependent axisymmetric metric contains seven functions of three variables (say, $t$, $r$, and $\theta$) that enter the Einstein equations via six independent quantities \cite{C:92}, evading general explicit recipes for locating the apparent horizon \cite{T:07,BS:10}. However, to verify that at least some of the remarkable properties of PBHs are not artifacts of spherical symmetry, a much simpler model is sufficient.

A general stationary axisymmetric metric is described by six independent functions \cite{C:92}. A form that is particularly convenient for studying black holes \cite{KRZ:16} is
\begin{align}
	ds^2=-\frac{N^2-W^2\sin^2\theta}{K^2}dt^2-2Wr\sin^2\theta dtd\phi+K^2r^2\sin^2\theta d\phi^2+S \left(\frac{B^2}{N^2}dr^2+r^2d\theta^2\right),
\end{align}
where $B$, $K$, $N$, $W$, and $S$ are functions of $r$ and $\theta$ and it is possible to set $S=1+a^2\cos^2\theta/r^2$. This metric reduces to the Kerr metric in Boyer--Lindquist coordinates for
\begin{align}
	B=1, \qquad K^2=\frac{(r^2+a^2)^2-a^2\Delta\sin^2\theta}{r^2\rho^2}, \qquad N^2=\frac{\Delta}{\rho^2}, \qquad W=\frac{2 a M}{\rho^2}.
\end{align}
Here and elsewhere
\begin{align}
	\rho^2=r^2+a^2\cos^2\theta, \qquad  \Delta=r^2-2Mr +a^2,
\end{align}
where $a\defeq J/M$, $M$ is the mass, and $J$ the angular momentum.

The Kerr metric can be represented \cite{P:04} using either the ingoing  $dw_+\equiv dv=0$ or the outgoing $dw_-\equiv du=0$ null congruences \cite{C:92},
\begin{align}
	\begin{aligned}
    	ds^2 = & - \bigg(1-\frac{2 M r}{\rho^2}\bigg)dw_\pm^2+2\epsilon_\pm dw_\pm dr-
    	\frac{4 a M r \sin^2\theta}{\rho^2} dw_\pm d\psi_\pm \\
   		& - 2 \epsilon_\pm a \sin^2\theta dr d\psi_\pm  \!  +\rho^2 d\theta^2   +
    	\frac{(r^2+a^2)^2\!-\!a^2 \Delta \sin^2\theta}{\rho^2}\sin^2\theta d\psi_\pm^2,
    \end{aligned}
    \label{kv}
\end{align}
where $\epsilon_\pm = \pm 1$ for the ingoing and outgoing coordinates, respectively. The geodesics propagate at $d\psi_\pm=d\theta=0$, and their interrelations can be found, e.g., in Refs.~\refcite{C:92,P:04}. The simplest formal way to obtain the Kerr metric in this form is to apply the complex-valued Newman--Janis transformation \cite{NJ:65} starting with the Schwartzschild metric written in the ingoing or the outgoing Eddington--Finkelstein coordinates.

The simplest non-stationary metrics \cite{MT:70,CK:77} are obtained by introducing evolving masses $M(v)$ and $M(u)$. The metric of Eq.~\eqref{kv} with variable $M(u)$ is obtained from the outgoing Vaidya metric \cite{GHJ:79}, and its counterpart with variable $M(v)$ from the ingoing Vaidya metric \cite{DT:20}, while other models of regular axially symmetric black holes are obtained by transforming their spherically symmetric counterparts \cite{BM:13}.

\subsubsection{Properties of Kerr--Vaidya metrics}
A schematic form of the EMT in both cases is
\begin{align}
	\tensor{T}{_\mu_\nu} =
	\begin{pmatrix}
		T_{ww} & 0 & T_{w\theta} & T_{w\psi} \\
		0 & 0 & 0 & 0 \\
		T_{w\theta} & 0 & 0 & T_{\theta\psi} \\
		T_{w\psi} & 0 &T_{\theta\psi} & T_{\psi\psi}
	\end{pmatrix},
	\label{emt}
\end{align}
where $w=u,v$. Using the null vector $k^\mu=(0,1,0,0)$ \cite{MT:70}, the EMT can be represented as
\begin{align}
	\tensor{T}{_\mu_\nu} = \tensor{T}{_w_w} k_\mu k_\nu + q_\mu k_\nu + q_\nu k_\mu, \label{MT-emt}
\end{align}
where the components of $\tensor{T}{_\mu_\nu}$ and the auxiliary vector $q_\mu$, $q_\mu k^\mu=0$ are given in \ref{A-KV}. This EMT (for the mass $M(u)$) was identified in Ref.~\refcite{CGK:90} as belonging to type $[(1,3)]$ in the Segre classification scheme \cite{exact:03}, which corresponds to type \RNum{3} of the Hawking--Ellis classification scheme \cite{HE:73,MV:17,MV:21}, indicating that the NEC is violated for any $a \neq 0$.

A detailed investigation reveals some interesting properties of this EMT\cite{DT:20}. We use a tetrad in which the null eigenvector $k^\mu=k^{\ha}e^\mu_\ha$ has the components $k^\ha=(1,1,0,0)$, the third vector $e_{\hat 2}\propto \pad_\theta$, and the remaining vector $e_{\hat{3}}$ is found by completing the basis. Then, the EMT takes the form
\begin{align}
	\tensor{T}{^{\hat{a}}^{\hat{b}}} =
	\left(\begin{tabular}{cc|cc}
 		$\nu$ & $\nu$  & $q_{\hat 2}$& $q_{\hat 3}$\\
 		$\nu$ & $\nu$  & $q_{\hat 2}$& $q_{\hat 3}$\\ \hline \label{emt-t}
 		$q_{\hat 2}$& $q_{\hat 2}$ & 0 & 0 \\
		$q_{\hat 3}$ & $q_{\hat 3}$& 0 & 0
 	\end{tabular}\right)
\end{align}
for all possible types of the metric, irrespective of the sign of $M'(w_\pm)$.

For an arbitrary null vector the NEC  is violated unless $\nu \geqslant 0$ and $q_{\hat 3}=q_{\hat 2}=0$. However, the latter condition holds only for $a=0$, when the metric reduces to its Vaidya counterpart and the EMT becomes a type \RNum{2} tensor.

The EMT of Eq.~\eqref{emt-t} has a single doubly-degenerate eigenvalue $\lambda=0$. The other two eigenvalues
\begin{align}
	\tilde\lambda_{1,2} = \nu \pm \sqrt{2( q_{\hat 2}^2 + q_{\hat 3}^2)+\nu^2}
\end{align}
of $\tensor{T}{^{\hat{a}}^{\hat{b}}}$ are non-zero. As a result, the EMT of Eq.~\eqref{emt-t} cannot be brought into a generic type \RNum{3} form: it corresponds to the special case where two of the three parameters $p=\rho=0$ (see \ref{A-KV} for details).

The apparent horizon of a Kerr black hole coincides with its event horizon. It is located at the largest root of $\Delta=0$,
\begin{align}
	r_0\defeq M+\sqrt{M^2-a^2}.
\end{align}
For the metric of Eq.~\eqref{kv} in $(u,r)$ coordinates with a variable mass $M(u)$ the relation $r_\sg=r_0$ also holds \cite{X:99}. It is not so for the $(v,r)$ version of this metric with variable $M(v)$ \cite{ST:15}. In this case, the expansion of the outgoing null geodesics that are orthogonal to the sphere of constant $v$ and $r$ is
\begin{align}
	\vartheta_{{\mathrm{out}}}=\frac{1}{2\Sigma^{3/2}}\left[\left(\Delta(2 r\rho^2+a^2\sin^2\theta(M+r)\right)+M'(v) r a^2 \sin^2\theta(r^2+a^2+\Sigma)\right],
	\label{varthetout}
\end{align}
where
\begin{align}
	\Sigma=(r^2+a^2)^2-a^2\Delta \sin^2\theta.
\end{align}
As a result $\vartheta_{{\mathrm{out}}}|_{r_0}\propto M'(v)$, and the difference $r_\sg(v, \theta)-r_0$ is of the order $|M_v|a^2/M$. The apparent horizon
\begin{align}
 	r_+(v,\theta)=r_0(M(v),a)-M'(v) a z(a/M,\theta),
\end{align}
where $z \geqslant 0$ and is zero at the poles. It can be obtained numerically and shows  some unusual properties, such as $z' \neq 0$ at the poles \cite{DT:20}.

The static limit --- the boundary of the region where static observers cannot exist --- is given in the Kerr spacetime by
\begin{align}
	r_\mathrm{sl}=M+\sqrt{M^2-a^2\cos^2\theta}.
\end{align}
The same expression with $M \to M(w)$ is valid for Kerr--Vaidya metrics.

It is easy to see that at least for some trajectories the infall time according to a distant observer is finite. Consider the equatorial plane $\theta=\pi/2$ and the null geodesics with constant $\psi$. Under these conditions the equations of motion reduce to the equations of the radial null geodesics in the Vaidya metric. The latter is a special case of the $k=0$ solutions, for which the crossing time according to a distant clock is finite.

\section{Physical Implications} \label{physcs01}
We now turn our attention to exploring properties of PBHs. As it is already apparent from the coordinate transformation Eq.~\eqref{dtrv}, falling into a PBH could take only a finite amount of time according to a distant observer. For a transient object with eventually decreasing $r_\sg(t)$, the alternatives are either crossing in finite time according to distant clocks (and of course comoving ones), or the impossibility thereof. We describe various scenarios in Sec.~\ref{fifi} and find that, in a clear departure from classical black holes, if a particle can fall through the horizon, it will do so in finite time $t$ of a distant clock. In Sec.~\ref{ident}, we match parameters of the PBH solutions with the known semiclassical results.

In Sec.~\ref{firewall}, we briefly review properties of the thermal atmosphere of radiating Schwarzschild black holes, discuss properties of the quantum ergosphere of evaporating black holes, and conclude with a discussion of the experiences of observers entering and exiting a PBH. We find that the apparent horizon and the anti-trapping horizon are hypersurfaces of intermediate singularity, which manifests itself through a firewall (divergent negative energy density experienced by some observers). For a white hole, the energy density in the frame of an infalling observer Alice diverges fast enough to violate quantum energy inequalities, indicating either the breakdown of semiclassical physics or confirming the instability of white hole horizons. The simplest collapse models are discussed in Sec.~\ref{shells}.

Putting all of this together in Sec.~\ref{subsec:HorizonFormation} allows us to outline a unique scenario for the formation of spherically symmetric semiclassical black holes. We also discuss the question of instability of the inner horizon.

In Sec.~\ref{SurfaceGravity}, we review different definitions of surface gravity and the roles they play. We then demonstrate that the principal  generalizations of surface gravity to non-stationary spacetimes do not agree with each other, and cannot agree with the classical static result. The implications of this become clear in Sec.~\ref{sub:loss}, where we discuss the information loss problem.

\subsection{Finite blueshift and finite infall time}\label{fifi}
The outer apparent horizon is a timelike surface \cite{BMsMT:19}, and thus it is possible to  introduce the surface metric
\begin{align}
	ds^2=-d\sigma^2+r^2d\Omega_2,
\end{align}
where $\sigma$ denotes the proper time on it\cite{P:04}. It is most conveniently expressed using $(v,r)$ coordinates as
\begin{align}
	d\sigma^2=2|r'_+|dv^2, \label{ah-met}
\end{align}
while additional useful expressions are given in \ref{As0}.

In contrast to classical black holes, an observer Alice can move with the apparent horizon, and for her the blue-shift of light arriving from infinity is finite. This is easily seen by comparing the time difference between two wave crests at infinity ($\delta\tau_\infty=\delta t_B=\delta v_B$) and at the apparent horizon, resulting in $\omega_\sg/\omega_\infty=1/\sqrt{2|r'_\sg|}$. For macroscopic semiclassical black holes this quantity is expected to be extremely large, but nevertheless it offers a natural cut-off.

Typical crossing times of the apparent horizon according to a distant Bob are also finite, even if they are on the scale of the evaporation time. We describe the geodesic motion of (massive or massless) test particles on a given gravitational background via the Lagrangian
\begin{align}
	\eL = - \half \tensor{\sg}{_\mu_\nu}\dot x^\mu \dot x^\nu,
\end{align}
where the overdot denotes a derivative with respect to a suitable evolution parameter\cite{C:92,FN:98}. It is well-known that a massive particle falling into a Schwarzschild or Kerr black hole takes a finite amount of proper time, but infinite time according to an observer at infinity \cite{C:92,HE:73,FN:98}. Leaving aside the questions of backreaction and radiation by infalling particles \cite{FN:98} (and the dynamics of accretion disks of astrophysical black holes), we focus on the motion of test particles on PBH backgrounds.

On the one hand, it is conceivable that the horizon crossing time according to Bob becomes finite \cite{BDE:17}. On the other hand, if $r_\sg$ recedes faster than the particle approaches it, there will be no crossing at all. We  investigate both of these non-classical possibilities: (i) that Alice's and Bob's times are finite, and (ii) that horizon crossing does not happen. Except for the motion in the Kerr--Vaidya metric presented at the end of this section, we deal with radial motion in spherically symmetric metrics.

Let the particle's motion be described by $x_\mathrm{A}^\mu(\lambda)=(Z,R,0,0)$ in $(z,r)$ coordinates, where $z=t,u,v$, and the parameter $\lambda $ is the proper time $\tau$ for a massive particle or a suitable parameter (such as $-R$) for a massless one. To simultaneously verify horizon crossing and resolve the question of Bob's time, we work in $(t,r)$ coordinates. The particle will cross the apparent horizon when the gap \cite{BMT:18,KMY:13}
\begin{align}
 	X(\lambda)\defeq R(\lambda)-r_\sg\big(T(\lambda)\big)   \label{gap1}
\end{align}
reaches zero. The crossing is clearly prevented if for some $X>0$ and $r_\sg>0$ the gap begins to increase,
\begin{align}
	\dot X=\dot R-r_\sg'\dot T>0,
\end{align}
while the result $\dot X\approx 0$ at leading order requires a more careful treatment.

For massive radially moving particles
\begin{align}
	\dot T=\frac{\sqrt{\dot R^2+F}}{e^HF}, \label{Ttime}
\end{align}
where $F \defeq f(T,R)$ and $H\defeq h(T,R)$. For infalling massless particles parameterized by $R=-\lambda$ (and thus having $\dot R=-1$) the relationship is simply $\dot T=e^{-H}/F$. For massive particles of non-zero radial velocity near the apparent/anti-trapping horizon this expression is approximately true, as
\begin{align}
	\dot T\approx \frac{\sqrt{\dot R^2+F}}{|r_\sg'|}\approx -\frac{\dot R}{|r_\sg'|}, \label{timeap0}
\end{align}
where we used Eq.~\eqref{dtrv}, and the last equality holds for $\dot R^2\gg F$. However, the evolution of $\dot R$ in the geodesic motion depends on whether the test particle moves on the background of a PBH or a white hole \cite{B:16,DMT:21}.

For an expanding white hole $r_\sg'>0$, and since $\dot R < 0$, we have $\dot X<0$. As $\dot T$ is finite, horizon crossing happens at some finite time as measured by Alice and Bob. It is easy to see that the same conclusion is also true in the opposite regime $|\dot R|^2\lesssim F$.

Due to Eq.~\eqref{drdt0}, the case of evaporation requires the use of higher-order terms. Different conclusions ensue for the $k=0$ and $k=1$ solutions, as well as for slow and fast-moving particles.

Consider first $k=0$ solutions. If $\dot R^2\gg F$, then
\begin{align}
	\dot X = \frac{|\dot R|}{2\sqrt{\pi}\Upsilon^2r_\sg^{3/2}}\left(\sqrt{\pi}r_\sg^{3/2}(e_{12}-p_{12})-\Upsilon\right)\sqrt{x}+
\frac{2\sqrt{\pi r_\sg}\Upsilon}{|\dot R|}\sqrt{x} + \cO(x), \label{dXk0l}
\end{align}
where the last term on the right-hand side (rhs) that is proportional to $\sqrt{x}$ is absent for massless particles (\ref{As0}).

The expression in the brackets on the rhs of Eq.~\eqref{dXk0l} equals $(w_1-1)\Upsilon<0$, and thus massless particles always cross the receding apparent horizon in finite time. Fast massive particles do so as well\footnote{The results for $k=0$ solutions were derived in Ref.~\refcite{T:19} for the $w_1=0$ scenario.}: using the results of \ref{AForm}, \ref{As0}, and Sec.~\ref{ident}, we find that the rhs of Eq.~\eqref{dXk0l} is negative so long as
\begin{align}
	2\dot R^2>|r_\sg'|.
\end{align}
Slow particles on the other hand, i.e.\ those that at some $R$ satisfy $|\dot R| \lesssim \sqrt{F}$, do not cross the apparent horizon as in this case the last term on the rhs of Eq.~\eqref{dXk0l} dominates, yielding $\dot X>0$. Many interesting scenarios for evaporating Vaidya black holes (i.e.\ $C_+(r,v)=r_+(v)$) are considered in Ref.~\refcite{PK:21}.

For $k=1$ black hole solutions, where only a single marginally trapped surface exists, this surface cannot be reached by an infalling observer that starts from $r>r_\sg$, regardless of their intial velocity \cite{T:19}. However, the outcome for non-extreme $k=1$ PBHs ($w_2=0$, $E=-P=(8\pi r_\sg^2)^{-1}$) whose metric functions are given by Eqs.~\eqref{fk1}--\eqref{hk1} depends on the specific behavior of the higher-order terms. This lack of clarity is not a serious limitation, as will become apparent in Sec.~\ref{subsec:HorizonFormation}.

Crossing into an expanding white hole is less straightforward. Using $(u,r)$ coordinates, we identify a divergent deceleration for the incoming particles,
\begin{align}
	\ddot R=\frac{2r_-r'_-}{(1-w_1)}\frac{\dot R^2}{Y^2}+\cO(Y^{-1}),
\end{align}
where $Y\defeq R(\tau)-r_-\big(U(\tau)\big)$. Moreover, stopping and reversal of massive test particles happens before the antitrapping horizon is reached, as at $R=r_-$ the condition $u_\mathrm{A}^2=-1$ is satisfied only if $\dot R>0$. Hence the crossing happens only if the expanding horizon overtakes the test particle\cite{DMT:21}.

Consider now Alice on the background of a Kerr--Vaidya metric \cite{DT:20}. For a generic trajectory her four-velocity is written as $u_\rA^\mu=\big(\dot{U}, \dot{R},  \dot{\Theta},\dot \Psi\big)$. Here, we again use capital letters to distinguish the trajectories, $W=w_\pm(\tau)$ and $\Psi=\psi_\pm(\tau)$, and treat the two coordinate systems simultaneously as long as it is feasible.

In spherically symmetric spacetimes Alice's radial trajectory implies zero angular momentum. In axial symmetry the condition to be a zero angular momentum observer (ZAMO) \cite{FN:98,P:04} is $\vK_\psi \cdot u_\mathrm{A}=0$, where the Killing vector $\vK_\psi=\pad_\psi$. In the Kerr--Vaidya metric this implies $\dot\Psi_Z \equiv -(\tensor{\sg}{_u_\psi}\dot U + \tensor{\sg}{_r_\psi}\dot R)/\tensor{\sg}{_\psi_\psi}$, so we consider Alice's trajectories with
\begin{align}
 	u^\mu_\mathrm{A}=\big(\dot{W},  \dot{R},  \dot{\Theta},\dot \Psi_Z\big).
\end{align}
We have $\dot R<0$, and the velocity component $\dot W>0$ is obtained from the normalization condition $u_\rA^2=-1$. Similar to the Vaidya metrics,  the result depends \cite{DT:20} on whether this is a retarded or an advanced metric and on the sign of $M'(w)$. In $(u,r)$ coordinates
\begin{align}
	\dot U=\Big(-\dot{R}(1-\delta)+\sqrt{\dot R^2(1-\delta)^2+\mf(1+\dot\Theta^2\rho^2-\dot R^2 a^2\rho^2\sin^2\theta/\Sigma)}\Big)/\mf, \label{Utime}
\end{align}
where
\begin{align}
	\delta=\frac{2 a^2 M r \sin^2\theta}{\Sigma}, \qquad \mf=1-\frac{2Mr}{\rho^2}(1-\delta).
\end{align}
It is easy to see that as Alice approaches $r_0$, $\dot U$ diverges (for a fixed value of $\dot R$) as
\begin{align}
	\dot U\approx \frac{2|\dot R|\big(1-\delta(r_0)\big)}{\mf'(r_0)Y},  \label{UKerr}
\end{align}
where $Y=R(\tau)-r_-\big(U(\tau)\big)$ is the gap function, and $\mf'(r_0)=\pad \mf/\pad r |_{r=r_0}$, similar to the Vaidya case. On the other hand,
\begin{align}
	\dot V=\Big(\dot{R}(1-\delta)+\sqrt{\dot R^2(1-\delta)^2+\mf(1+\dot\Theta^2\rho^2-\dot R^2 a^2\rho^2\sin^2\theta/\Sigma)}\Big)/\mf \label{Vtime}
\end{align}
remains finite, and as in the previous discussion, for $M'(v)<0$ horizon crossing occurs unless $|\dot R|$ is too small.

\subsection{Identification of metric functions} \label{ident}
The values of $\Upsilon$ and $\xi$ can be obtained from first principles only if one performs a complete analysis of the collapse of some matter distribution and the quantum excitations it generates. Such an analysis would provide a constructive proof of the existence of semiclassical PBHs. In absence of such results we can still establish some useful general relations and then match them with the semiclassical results for evaporation \cite{MsMT:21}.

The apparent horizon of a PBH that was formed at some finite time of Bob is timelike, and the surface metric Eq.~\eqref{ah-met} is defined on its outer branch. To remove the notational ambiguity, we express coordinates of the apparent horizon as functions of its proper time, such as $r_\ah(\sigma)$, $t_\ah(\sigma)$, $v_\ah(\sigma)$. The invariance of the apparent horizon in spherically symmetric foliations means $r_\ah(\sigma)\equiv r_\sg\big(t_\ah(\sigma)\big)$, etc, and its rate of change is given by
\begin{align}
	\frac{dr_\ah}{d\sigma}=r'_\sg\big(t_\ah(\sigma)\big)\dot t_\ah=r'_+\big(v_\ah(\sigma)\big)\dot v_\ah. \label{ah-rate}
\end{align}
If one assumes  that  $r_\sg$ is a monotonically decreasing function of time, one can write
\begin{align}\label{Gamma-rels}
	\dot r_\ah=\Gamma_\ah(r_\ah), \qquad r'_\sg=\Gamma_\sg(r_\sg), \qquad r'_+=\Gamma_+(r_+),
\end{align}
where the relations between the functions $\Gamma_\ah$, $\Gamma_\sg$, and $\Gamma_+$ follow from Eq.~\eqref{ah-rate}. Without assuming any particular relation between $r_\sg'$ and $r_+'$, by using the first expression of Eq.~\eqref{eq:thev} and Eq.~\eqref{the1} with $\tensor{\tau}{_t} = - \Upsilon^2 + \mathcal{O}(\sqrt{x})$, we obtain
\begin{align}
	\Upsilon=\frac{1}{2}\sqrt{\frac{|r_\sg''|}{2\pi r_\sg}}  \frac{|r_+'|} {|r_\sg'|}, \label{Upseq}
\end{align}
and from Eq.~\eqref{eq:k0rp}
\begin{align}
	\xi = \frac{r_\sg'^4}{2|r_\sg''|r_+'^2} .
\end{align}

The semiclassical analysis is based on perturbative backreaction calculations that represent the metric as modified by the Hawking radiation that is produced by a slowly varying sequence of Schwarzschild metrics. It results in \cite{FN:98,jB:81,BMPS:95,P:76,V:97,LO:16,APT:19,FZ:17}  $\Gamma_\sg(r)=\Gamma_+(r)$, and specifically (\ref{Haw-em})
\begin{align}
	\frac{dr_\sg}{dt}=-\frac{\alpha}{r_\sg^2}, \qquad \frac{dr_+}{dv}=-\frac{\alpha}{r_+^2},  \label{paget}
\end{align}
where $\alpha$ is a constant. Using this result, we obtain
\begin{align}
	\Upsilon = \frac{1}{2}\sqrt{\frac{ |r_\sg''|}{  2\pi r_\sg} }=\frac{\alpha}{2\sqrt{\pi}r_\sg^3} , \label{Upseva}
\end{align}
and
\begin{align}
	\xi = \frac{r_\sg'^2}{2 \vert r_\sg'' \vert} = \frac{1}{4}r_\sg, \label{xi0}
\end{align}
where the last equalities on the rhs follow from Eq.~\eqref{paget}. We note that this result agrees within an order of magnitude with the guess of Ref.~\refcite{BMsMT:19}, but as we will see below the assumptions of Ref.~\refcite{BsMT:19} are not fulfilled, and its estimate is in general incorrect.

\subsection{Quanta and firewalls} \label{firewall}

\subsubsection{Thermal atmosphere on the Schwarzschild background}\label{ss:atmo}
The $\tensor{T}{_t^r}$ (or $\tensor{T}{_v^r}$) component of the EMT that is calculated on the background of a Schwarzschild black hole (say, assuming the Unruh vacuum) exhibits a rather simple behavior. The conservation law $\nabla_\mu \tensor{T}{^\mu_\nu}=0$ imposes the relation
\begin{align}
	r^2 \tensor{T}{_t^r} = \mathrm{const} = - L / 4 \pi,
\end{align}
where $L$ denotes the luminosity \cite{FN:98,jB:81,LO:16}. However, the emission field is far from monotonous, and the domain $r < 3r_\sg$ that is known as the thermal atmosphere of the black hole \cite{FN:98,W:01,H:16,TPM:86} is distinguished according to several different criteria. From about this radius, the exterior perturbative metric of an evaporating black hole is well-described by a pure outgoing Vaidya metric [Eq.~\eqref{Vaidya}] with decreasing mass \cite{BMPS:95,jB:81}. From Fig.~\ref{fig:emt} we see that on the fixed Schwarzschild background the products $r^2 \tensor{T}{_t_t}$ and $r^2 \tensor{T}{^r^r}$ evolve from their common negative limit with $r^2 \tensor{T}{_t^r}$ at the horizon, reach their distinct maxima, and begin to descend to their common long-distance limit \cite{LO:16,L:17} within this range, while the NEC is violated for $2.5 r_\sg \gtrsim r $.

Interpreting these results in terms of particles (the populated localizable spacetime modes \cite{FN:98,H:75}) provides a useful related intuition. The Hawking temperature $T_{\mathrm{H}} = \kappa / 2 \pi$ represents the temperature measured by Bob at infinity. For a static observer Eve that is located at some $r$, the relevant temperature is given by the Tolman relation \cite{W:01}
\begin{align}
	T_\mathrm{E} = T_\mathrm{H}\left(1-\frac{r_\sg}{r}\right)^{-1/2}. \label{localT}
\end{align}
This temperature value appears in the analysis of excitation probabilities of model detectors \cite{FN:98} or via the effective temperature function \cite{BBG:11,BBGJ:16}. Hence in the particle picture the thermal atmosphere is full of particles at temperatures $T_\mathrm{atm}\gtrsim 1$, but very few of them escape it.

The argument of Ref.~\refcite{STU:93}, originally made within the stretched horizon paradigm, illustrates this property. At $T_\mathrm{atm} \sim 1$ the number of particles emitted per unit area per unit proper time is of order one in Planck units. If all of these particles made it out to infinity, then according to a distant Bob the emission rate would be equal to the local production rate multiplied by the black hole area and the time dilation factor,
\begin{align}
	dN/dt \sim 1\times d\tau/dt \times  M^2 \sim \frac{\varepsilon}{M}  M^2 \sim M,
\end{align}
where $M=r_\sg/2$ is the black hole mass and $\varepsilon$ the distance of the atmosphere from the horizon, which is generally assumed to be of order unity in Planck units. On the other hand, the actual particle arrival rate at infinity is obtained by dividing the black hole luminosity $L \propto M^{-2}$ by the typical energy of a particle at the Hawking temperature, resulting in
\begin{align}
	dN/dt\sim 1/M.
\end{align}
Two mechanisms prevent the vast majority of particles from escaping. Propagation of perturbative fields (e.g.\ scalars) on the black hole background is governed by the effective gravitational potential. Its maximum strongly rises with the angular momentum of the perturbation, forming a barrier whose peak is located between $r=4r_\sg/3$ for $l=0$ and $r=3r_\sg/2$ in the limit of large $l$ \cite{C:92,FN:98}. Most excitations have too large an angular momentum to have a good chance of escaping \cite{H:16,TPM:86}. Among those (predominantly $s$-wave) excitations that can in principle penetrate the barrier, typical wavelengths are large enough to be strongly affected by the spacetime curvature, mostly within the region $9r_\sg/8 \lesssim r \lesssim 6 r_\sg$, and are driven back towards the horizon \cite{TPM:86}.

To discuss experiences of various observers, we quote some parameters of the EMT for the Unruh vacuum of a conformally coupled scalar field on the Schwarzschild background, adapting the semi-numerical representation of Ref.~\refcite{V:97}. We find that, using the expansion parameter $z \defeq r_\sg/r$, the elements of the EMT in Eq.~\eqref{tspher} are
\begin{align}
	\rho &= \frac{z^2}{1-z}p_\infty\big(\sigma_0+\sum_{k=2}^5\sigma_k z^k\big) , \\
	p &= \frac{z^2}{1-z}p_\infty\big(\sigma_0+\sum_{k=2}^5\tilde{\sigma}_k z^k\big) , \\
	\psi &= \frac{z^2}{1-z}p_\infty\sigma_0 .
\end{align}
The parameter \cite{V:97}
\begin{align}
	p_\infty=\big(90(16\pi)^2r_\sg^4\big)^{-1}
\end{align}
is obtained from the trace anomaly of the renormalized EMT\cite{FN:98,BD:82,PT:09} and is directly related to the luminosity. In the geometric optics approximation the luminosity is $L_\mathrm{geo} = 81 \pi M^2 p_\infty$, while the total luminosity $L$ (obtained by taking into account the effects of wave propagation\cite{LO:16,P:76,E:83}) is $L \approx 7.44/7 L_\mathrm{geo}$. The parameters $\sigma_i$ and $\tilde{\sigma_i}$ satisfy $\sum_{i=2}^5\sigma_i=\sum_{i=2}^5\tilde{\sigma}_i=0$ within numerical precision, and $\sigma=-5.35$. The most recent four-dimensional calculations are reported in Refs.~\refcite{LO:16} and \refcite{L:17}, with the results for Unruh vacuum presented in Fig.~\ref{fig:emt}. Analogous calculations for the Kerr background are reported in Ref.~\refcite{LEOM:17}.

For a static Eve (whose four-velocity $u_\rE$ is just the normalized Killing vector $\vK=\pad_t$), both the local Hawking temperature and the local energy density $\rho_rE=\tensor{T}{_\mu_\nu} u^\mu_\rE u^\nu_\rE$ diverge as her fixed location is placed closer and closer to the horizon. In the context of entropy calculations, Ref.~\refcite{W:01} refers to this as a ``new ultraviolet catastrophe" that can be cured by imposing a cut-off on locally measured frequencies. On the other hand, a freely (or just sufficiently fast) falling Alice does not notice any appreciable effects. As with the calculations of Sec.~\ref{sc:atmo}, the divergent terms in the density $\rho_\rA$ cancel. However, the precision of the coefficients that are given in Ref.~\refcite{V:97} does not allow the determination of its finite value. Finally, in the frame of a radially outward moving observer the energy density diverges \cite{Y:98} near the horizon as $(1-z)^{-2}$.

The late-time response rate of a two-level detector on various stationary orbits outside of a $(1+1)$-dimensional black hole that was formed by a collapsing null shell is consistent with the local Hawking temperature \cite{JL:18,Tjoa:2020eqh}. In four spacetime dimensions the detector response is sensitive not only to the trajectory, but also to the type of vacuum. For a static detector coupled to a massless scalar field on the Schwarzschild background,
the response in the Hartle--Hawking state exhibits the usual thermal form at the local Hawking temperature (as anticipated by the very construction of this state that describes thermal equilibrium \cite{BD:82,TPM:86}), and the response in the Unruh state is thermal at the local Hawking temperature in the limit of a large detector energy gap \cite{HLO:14}.

Properties of the thermal atmosphere played a role in formulation of the generalized second law of thermodynamics that includes the Bekenstein entropy
\begin{align}
	S_\mathrm{B} = \tfrac{1}{4}A,
\end{align}
where $A$ is the horizon area, and in the overall entropy counting \cite{FN:98,W:01,P:05,H:16}. One of the ways to attribute microscopic structure to the black hole entropy is to treat it as an ordinary entropy of the thermal atmosphere \cite{FN:98,W:01,TPM:86}.

Assuming that the thermal atmosphere behaves like a free massless (boson or fermion) gas, its entropy density scales as $T_\rE^3$. The $1/\sqrt{f}$ divergence of this temperature leads to the divergence of the total entropy. It can be cured if one imposes a cut-off on the locally measured frequency of the modes, which is expected to be on the order of the Planck scale. As a result, the thermal atmosphere contributes an entropy on the order of the horizon area. Assuming that the overall state of quantum excitations is pure, this counting evaluates the entropy of the reduced density matrix of Hawking modes outside of the horizon \cite{BKLS:86}. This entropy represents the entanglement between Hawking particles inside and outside of the horizon \cite{W:01} (and more specifically, the standard measure of the bipartite pure entanglement known as the degree of entanglement \cite{PT:04}).

Criticism of this approach centers on the vagueness of the cut-off definition \cite{H:16,P:05}. However, Ref.~\refcite{ABKLU:19} considers a slowly evaporating black hole and uses the difference between the event and the apparent horizon and backreaction of the emitted quanta on the horizon structures to obtain a natural regulator on the order of $1/r_\sg$.

Almost by its very definition the thermal atmosphere is the source of particles that reach distant observers. However, there are different arguments that point to their origin in a thin film near the horizon \cite{VAD:11}, at upper layers of the atmosphere, or just above it \cite{G:16,DLMP:19}.

The thermal atmosphere strongly affects the buoyancy of objects close to the horizon \cite{TPM:86,W:01,P:05,BBGJ:16b}. An assortment of  hypothetical highly reflecting objects (``boxes'') with various content were quasistatically lowered toward the black hole in gedankenexperiments attempting to violate the generalized second law. The temperature gradient in Eq.~\eqref{localT} implies that there is a pressure gradient that produces a buoyancy force on the box. Due to this force the box does not reach the horizon, but stabilizes at the floating point above it, where its weight is supported by the weight of the displaced thermal atmosphere. This maintains the generalized second law \cite{W:01}.

\subsubsection{Quantum ergosphere and its boundaries} \label{sb:ergo}
The apparent horizon of a PBH is a timelike hypersurface. Fig.~\ref{fig:time-g}(a) illustrates the possibility that there is a region that is causally disconnected from future null infinity, i.e.\ a MBH that is bounded by the event horizon. In this case the event horizon is crossed by Alice after she traverses the apparent horizon \cite{BMPS:95,FN:98,jB:81,Y:83}.

Despite the global nature of the event horizon there is an approximate local procedure to identify it \cite{Y:83}. If it forms, the event horizon is a null surface that is generated by the last family of outgoing null geodesics $R(v)$ that do not reach future null infinity. In $(v,r)$ coordinates, the outgoing null geodesics satisfy
\begin{align}
	\frac{dr}{dv}=\frac{1}{2}e^{h(r,v)}f(r,v).
\end{align}
At the apparent horizon both $\vartheta_{{\mathrm{out}}}$ of Eq.~\eqref{varthetout} and $R'(v)$ are zero. Thus the photons are only momentarily at rest and escape to distances $y \gtrsim r_\sg$ in finite (advanced) time \cite{PP:09,BsMT:19}. This implies that the event horizon generators are photons that are ``stuck'', $d^2r/dv^2=0$. Thus the solution of the algebraic equation
\begin{align}
	2\frac{d}{dv}(e^hf)+e^h f\frac{d}{dr}(e^hf)=0 \label{YFsep}
\end{align}
provides a good approximation for the location of the event horizon.

For low luminosity $L \ll 1$, a perturbative analysis establishes that $r_\eh$ is close to $r_\sg$. Using the metric functions Eqs.~\eqref{Cpl}--\eqref{hpl}, we obtain the leading-order expression for $y_\eh \defeq r_\eh(v)-r_+(v)$ as
\begin{align}
	y_\eh\cong \frac{2r_+(1-w_1)r'_+}{ 1+2w_1^2+\big(1+2 r_+(w_2-\chi_1)\big)r'_+ -2\big(1+(1-2\chi_1 r_+)r'_+\big)w_1-2r_+w'_1  }.
\end{align}
Perturbative backreaction calculations\cite{jB:81,Y:83,BMPS:95} lead to an approximately outgoing Vaidya metric with $r'_+=2M'<0$ in the vicinity of the apparent horizon. Then $y_\eh\cong 2r_+r'_+=8MM'$. It is sub-Planckian for $L\propto 1/M^2\ll 1$. This approximation also fits well the exact event horizon in models with a running gravitational constant \cite{BR:06}. The domain between the apparent and the event horizon was named the {\it quantum ergosphere} \cite{Y:83}. Particles can escape from it even before the complete evaporation of the trapped region.

\begin{figure}[!htbp] \centering
	\includegraphics[width=0.925\textwidth]{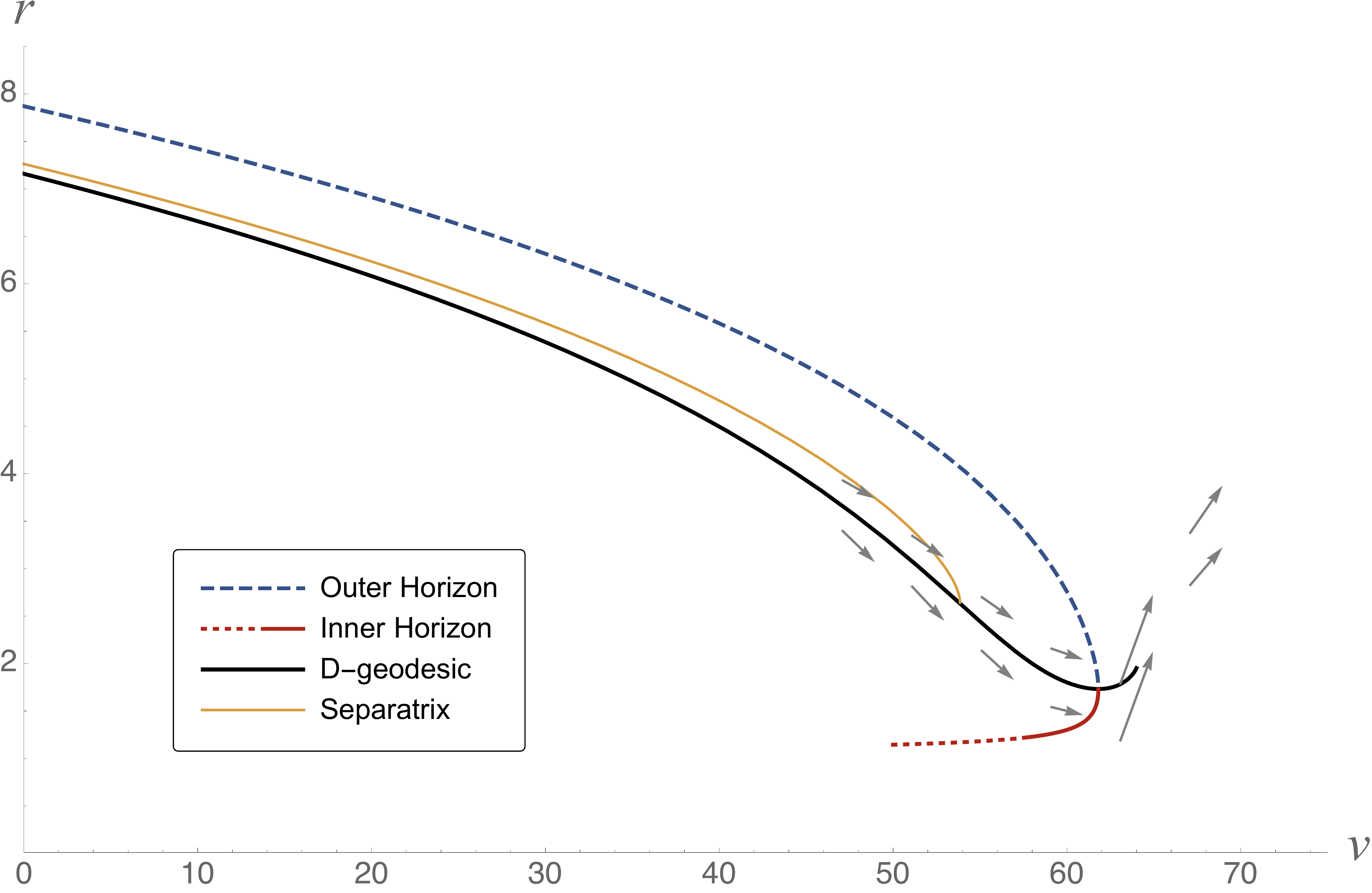}
	\caption{Outer (blue) and inner (red) apparent horizon obtained with $h_+=0$ and $f(v,r)$ of Eq.~\eqref{fHF}. The D-geodesic (black) exits the PBH at their coalescence. We work with the parameters \cite{vpF:16,BHL:18} $b=1$, $m_0=4$, and use the solution $m_1(v) = (64-v)^{1/3}$. The outer apparent horizon $r_+$ in this model is shown only for the evaporation stage $v>0$, as a semiclassical PBH that is formed at some finite time $t_\rS$ can only evaporate. Only a part of the inner apparent horizon is shown to indicate the exit of the D-geodesic. The ``separatrix'' (a segment of the solution of Eq.~\eqref{YFsep} that is closed to the D-geodesic) is shown in orange. Grey arrows indicate the tangents of two selected null geodesics.}
	\label{fig:qergosphere}
\end{figure}

Event horizons are absent in many models of dynamical RBHs. However, there is still a useful distinction between null and timelike geodesics that can leave the trapped region only at the final evaporation event (the two-sphere where the inner and outer horizons join --- see Fig.~\ref{fig:time-g}(b)), and those that can cross the timelike apparent horizon before the evaporation is complete. The boundary of the two domains is a null geodesic (named D-geodesic \cite{BHL:18}). The condition Eq.~\eqref{YFsep} is much easier to implement \cite{vpF:14,vpF:16}, and the resulting curve (the separatrix \cite{BHL:18}) is a good approximation to the D-geodesic \cite{BHL:18}.

Fig.~\ref{fig:qergosphere} illustrates the vicinity of the outer apparent horizon of the RBH model of Hayward \cite{H:06} and Frolov \cite{vpF:14,vpF:16}. Here $C_+(v,r)$ is given by Eq.~\eqref{hf-m}, and thus
\begin{align}
	f(v,r) &= 1 - \frac{2 m(v) r^2}{r^3 + 2 m(v) b^2} , \label{fHF}
\end{align}
and at the evaporation stage the mass evolves according to $(m(v)/b)^3 =(m_0/b)^3 - v/b$, and $h_+\equiv 0$. This is the simplest member of a more general family \cite{BHL:18} with $h_+=0$ and
\begin{align}
 C_+(v,r) = 2m(v)\frac{r^s+\sum_{k=0}^{s-1}a_k(v)r^{k+1}}{r^s+\sum_{k=0}^{s-1}b_k(v)r^k}, \qquad s\geqslant3. \label{genh0}
\end{align}
In such models, the D-geodesic satisfies
\begin{align}
	\frac{dr}{dv} = \frac{f(v,r)	}{2} ,
\end{align}
and  it  passes through the last point of the disappearing trapped region $(v_*,r_*)$ that still satisfies $f(v_*,r_*)=0$ (see Sec.~\ref{k=1}). Using $y=r-r_+(v)$ as an independent variable \cite{ABKLU:19} provides a description of the quantum ergosphere in a convenient way.

\subsubsection{Firewalls and singularities} \label{sc:atmo}

We start by considering the local energy density, pressure, and flux obtained in the frame of a static observer Eve situated just outside of the Schwarzschild sphere. Then we consider two scenarios of Alice crossing the apparent horizon and evaluate these quantities in her local frame: first when she falls through the apparent horizon from the outside, and then when she escapes the quantum ergosphere. Finally, we consider crossing the anti-trapping horizon of a white hole.

In each of the three cases the local time direction is given by the four-velocity $u^\mu$. For   observers outside of $r_\sg$ we take the transverse direction $n^\mu$ to be the unit outward-pointing radial spacelike vector, and inside of the trapped region we choose one of the two possible candidates by requiring consistency with the assignment along the exterior segment of the outgoing trajectory. We present the results only for $k=0$ solutions (as these are the ones that are relevant during the lifetime of trapped regions, as discussed in Sec.~\ref{s:formation}). The exposition largely follows that of Ref.~\refcite{DMT:21}.

For a static Eve, we find that the local quantities are given by the components of the EMT [Eqs.~\eqref{tspher} and \eqref{tneg}]
\begin{align}
	\rho_\rE \defeq \tensor{T}{_\mu_\nu}u^\mu_\rE u^\nu_\rE = \rho, \qquad p_\rE \defeq \tensor{T}{_\mu_\nu} n^\mu_\rE n^\nu_\rE, \qquad \psi_\rE \defeq  \tensor{T}{_\mu_\nu} u^{\mu}_\rE n^\nu_\rE = \psi,
\end{align}
and similar to the EMT of the Schwarzschild background diverge \cite{BMsMT:19} as $1/f$. The experience of a radially-infalling observer Alice moving on the trajectory $x^\mu_\mathrm{A}(\tau)=(T, R,0,0)$ depends on whether the geometry corresponds to an evaporating PBH or an expanding white hole.

We have already seen that the formation of apparent horizons in finite time of Bob is ineluctably connected with a violation of the NEC in their vicinity. Violations of energy conditions in quantum field theory on flat or curved backgrounds are bounded by quantum energy inequalities \cite{KS:20,cjF:17}. While the identification of constants (and thus going beyond the various scalings) is quite a non-trivial task on dynamic curved backgrounds, there is an explicit  inequality Eq.~\eqref{qei-KO} that is valid for spaces of small curvature \cite{KO:15}.

For the expectation value of  the renormalized EMT on an arbitrary Hadamard state \cite{BD:82,PT:09,KS:20} $\omega$, $\tensor{T}{_\mu_\nu} = \6 T^\mathrm{ren}_{\mu\nu}\9_\omega$, and a timelike geodesic $\gamma$ with a tangent four-vector $u^\mu_\tau$, the contraction results in the local density
\begin{align}
	\rho_\tau \equiv \tensor{T}{_\tau_\tau} = \tensor{T}{_\mu_\nu} u^\mu_\tau u^\nu_\tau .
\end{align}
The total integrated energy is obtained with the help of a smearing function of a compact support $\wp(\tau) \geqslant 0$ that can be taken to be $\wp \cong 1$ for an arbitrary large fraction of the domain $\wp>0$. Then
\begin{align}
	\int_\gamma d\tau \wp^2(\tau)\rho(\tau)\geqslant -B(\gamma,\eR,\wp),
	\label{qei-KO}
\end{align}
where $B>0$ is a bounded function that depends on the trajectory, the Ricci scalar, and the sampling function \cite{KO:15}.

For a macroscopic black hole the curvature at the apparent horizon is small and thus the bound specified above is applicable. Therefore, when investigating the experiences of Alice, we focus on geodesic trajectories that admit use of this bound. Given Alice's trajectory, we can choose $\wp=0$ outside of the NEC-violating domain, and choose $\wp \approx 1$ within this domain up to her horizon crossing.

For an evaporating black hole ($r_\sg'<0$), energy density, pressure, and flux in Alice's frame are finite. If the geometry is well-approximated by the Vaidya metric with $r'_+(v)<0$, then at the horizon crossing $R(\tau)=r_+$ the energy density, pressure, and flux are equal to
\begin{align}
	\rho_\rA=p_\rA=\psi_\rA=-\frac{\Upsilon^2}{4\dot R^2}. \label{Veva-in}
\end{align}
In the general case, these quantities are still finite, but the precise values depend on the higher-order terms in the metric. The expressions take their most compact form in $(v,r)$ coordinates, namely
\begin{align}
	\rho_\rA &= \frac{(1-w_1)r'_+}{32\pi r_+^2\dot R ^2}+\frac{w_1}{8\pi r_+^2}+\frac{\dot R^2\chi_1}{4\pi r_+},\\
	p_\rA &= \frac{(1-w_1)r'_+}{32\pi r_+^2\dot R ^2}-\frac{w_1}{8\pi r_+^2}+\frac{\dot R^2\chi_1}{4\pi r_+},\\
	\psi_\rA &= \frac{(1-w_1)r'_+}{32\pi r_+^2\dot R ^2}-\frac{\dot R^2\chi_1}{4\pi r_+}.
\end{align}
By using the explicit expressions for the EMT components in the two frames it is easy to show that the first term above equals the rhs of Eq.~\eqref{Veva-in}.

As Alice exits the quantum ergosphere of a PBH she experiences a divergent energy density \cite{DMT:21}. Working in $(v,r)$ coordinates (see \ref{AForm} for relations between the four-velocity components), we recall that inside of the trapped region $\dot R \leqslant - \sqrt{-F}$ for both ingoing and outgoing trajectories. On the other hand, if the outgoing trajectory crosses the apparent horizon, it does so with $\dot R=0$.

From the geodesic equations, it follows that, as $Y \defeq R - r_+ \to 0_-$, arbitrarily large initial values of the radial velocity $|\dot R|$ are damped down to the minimal possible value $\dot R = - \sqrt{-F}$. Since for the outgoing geodesics
\begin{align}
	\dot V=\frac{\dot R -\sqrt{\dot R^2+F}}{e^HF}, \qquad R \leqslant r_+,
\end{align}
 close to the apparent horizon we have
\begin{align}
	\dot V\approx \sqrt{\frac{r_+}{(1-w_1)|Y|}}, \qquad \dot R\approx -\sqrt{\frac{(1-w_1)|Y|}{r_+}}.
\end{align}
This leads to a whimper singularity
\begin{align}
	\rho_\rA\approx-\frac{\Upsilon^2 r_+}{(1-w_1)|Y|}\approx\frac{r'_+}{8\pi r_+|Y|},
\end{align}
as the energy density in Alice's frame diverges on her exit from the quantum ergosphere whereas the curvature scalars are finite by construction of this solution.
However, taking the gap $Y \defeq R(\tau)-r_+\big(V(\tau)\big)$ as the integration variable, we find
\begin{align}
	d\tau \approx -\frac{\sqrt{Y}}{r'_+}dY,
\end{align}
and the integration of $\sqrt{Y}$ for some $Y<0$ to $0$ results in a finite expression on the lhs of Eq.~\eqref{qei-KO}. As a result, the integrated energy density is not in obvious violation of this bound.

Similarly, if Alice approaches the anti-trapping horizon of an expanding white hole (corresponding to the second row in Table \ref{tab:EinEqRealSol}) she encounters a firewall. In the vicinity of an expanding Schwarzschild radius, the limit $\lim_{r \to r_\sg} \tensor{\tau}{_t^r} = + \Upsilon^2$, and instead of canceling the divergent terms add up, resulting in \cite{T:19,sMT:21}
\begin{align}
	\rho_{\mathrm{A}}=-\frac{4\dot R_\rA^2\Upsilon^2}{F^2}+\cO(F^{-1})=-\frac{  \dot{R}_\mathrm{A}^2}{4\pi r_\sg X} + \mathcal{O}(1/\sqrt{X}), \label{fire}
\end{align}
where $X =R(\tau) - r_\sg \big( T(\tau) \big)$ and $F=f(T,R)$. At leading order in the vicinity of the anti-trapping horizon
\begin{align}
	\rho_\rA = p_\rA = - \psi_\rA.
\end{align}
However, by itself this does not constitute a firewall. Inside of the anti-trapped region $\dot R>0$ for both radial geodesics (\ref{AForm}). Therefore, an ingoing test particle can cross the anti-trapping horizon from the outside only if it has a zero proper radial velocity. In fact, the geodesic equations contain the radial stopping term. Taking into account Eq.~\eqref{udout}, we see that the radial infall is either stopped or even reversed before the expanding anti-trapping horizon overtakes the particle. In any case, the negative energy density on approach to the anti-trapping horizon can diverge at most as $1/\sqrt{R-r_-}$, and hence the integrated energy density remains finite. Nevertheless, $\rho_\rA$ can take on arbitrarily large values, and it remains to be seen if it is compatible with the bounds on the NEC violation\cite{DMT:21}.

The divergence of the energy density in Alice's frame indicates the presence of a matter singularity \cite{DMT:21}: while the curvature scalars are finite, more restrictive regularity conditions are not satisfied. This intermediate singularity (\ref{singu}) can be characterized using the Ricci tensor scalars that are obtained from the Newman--Penrose tetrads (\ref{a:nptet}). Using the two real null vectors of Eq.~\eqref{null-v} with a pair of complex-conjugate  basis vectors, such as given by Eq.~\eqref{mmbar}, we find that the values of all non-zero scalars are finite on the apparent horizon. That is, there is a frame in which all components of the Riemann tensor are bounded.

However, the null vectors can be rescaled as $l^\mu_\mathrm{out}\to Al_\mathrm{out}^\mu$, $l_\mathrm{in}^\mu\to l_\mathrm{in}^\mu/A$, and the values of $\Phi_{00}$ and $\Phi_{22}$ depend on this choice. By choosing $A=f(v,r)$ (this form of the tangent vectors may appear more natural in $(t,r)$ coordinates), we find \cite{DMT:21}
\begin{align}
	\Phi_{00}\propto f, \qquad \Phi_{22}\propto f^{-1},
\end{align}
again demonstrating that the apparent horizon is a surface of intermediately singular behavior. On the other hand, the original basis leads to a divergent $\Phi_{00}$ on the expanding anti-trapping horizon.

We thus observe that if an apparent/anti-trapping horizon forms, it is a surface of intermediate singularity. A negative energy firewall constitutes the counterpart to the arbitrarily large tidal forces that could tear apart an observer falling into such a singularity. Their fate depends on the integrated tidal stress induced by the environment of the singularity \cite{EK:74,FN:98}.

\subsection{Thin shell collapse} \label{shells}
The simplest model of gravitational collapse is the contraction of a spherically symmetric infinitesimally thin massive dust shell that separates a flat interior region from a curved exterior spacetime \cite{P:04}. A thin shell is a mathematical idealization  of a narrow transition region that is modeled as a hypersurface of discontinuity $\Sigma$. Its dynamics is determined by so-called junction conditions, i.e.\ rules for joining the solutions of the Einstein equations on both sides $\Sigma_{\pm}$ of $\Sigma$. The junction conditions \cite{exact:03,P:04,S:18} prescribe continuity of the induced metric and changes in the extrinsic curvature across $\Sigma$. The resulting joined geometry is a solution of the Einstein equations with an additional distributional EMT concentrated on $\Sigma$. The equations of motion of the shell are often referred to as Israel equations \cite{S:18}.

Using thin shells allows one to circumvent some of the controversial issues, such as the structure of the EMT within the collapsing body. They have applications in modeling domain walls, branes, gravitational layers, and impulsive gravitational waves. Due to its inherent simplicity and the ability to model radiative processes of both classical and quantum origin, the thin shell formalism has become a popular resource for analyzing the final stages of gravitational collapse. Nevertheless, in this case the results primarily illustrate their underlying assumptions. Models that use the outgoing Vaidya metric or general metrics that satisfy certain regularity conditions exhibit horizon avoidance \cite{KMY:13,KY:15,BMT:18,MNT:18}, while arguments based on the iterative evaluation of the effects of backreaction indicate that the shell eventually crosses the horizon  \cite{PP:09,CUWY:18}. On the other hand, comprehensive constructions of rigorous Penrose diagrams for various dynamical spacetimes rely on trains of infalling thin shells for a description of the collapse and model the effect of Hawking radiation by emitting shells of negative and positive mass \cite{SA:18,SAK:20}.

The relevant details of the classical thin shell collapse whose trajectory is $\big(T(\tau),R(\tau)\big)$ are summarized in \ref{A-shell}.  The radial motion is governed by a relatively simple equation that we symbolically denote as $\mathcal{D}(R)=0$, where $\mathcal{D}(R)$ is given by Eq.~\eqref{D(R)eq}. It can be integrated in quadratures leading to a finite proper time of horizon crossing, $R(\tau_\rS)=r_\sg$. This is the formation event of the black hole and corresponds to an infinitely distant moment of Bob's time, $T(\tau_\rS)=\infty$.

We illustrate the issues that arise when evaporation is taken into account by considering two models that lead to opposing conclusions. Both are massive thin shells that propagate on  prescribed backgrounds. The flat Minkowski metric describes the spacetime inside of the shells, while the two possible exteriors are given by a pure outgoing and a pure ingoing Vaidya metric, respectively, with the respective prescribed evaporation laws of Sec~\ref{V-out} and Sec.~\ref{V-in}. Without evaporation the shell's rest mass $m=4\pi\sigma R^2$ is conserved, and therefore provides a diagnostic for monitoring the surface density $\sigma$, which is the only non-zero component of the surface EMT.

While the outputs of different thin shell collapse models are seemingly incompatible, there are no contradictions, as their basic assumptions describe different physical regimes. The behavior of the rest mass plays an important role in the interpretation of the results. Nevertheless, models with a prescribed exterior geometry do not have an independent physical meaning \cite{BsMT:19}, but rather simply illustrate their underlying assumptions. If the outgoing Vaidya metric --- corresponding to the impossibility of horizon formation in finite time of Bob --- is used, then the model will predict horizon avoidance. Similarly, if one uses the ingoing Vaidya metric that is associated with the finite-time formation of trapped regions, it will predict horizon crossing.

\subsubsection{Outgoing Vaidya metric} \label{V-out}
Vaidya metrics (Eq.~\eqref{Vaidya}) with $C'_-(u)<0$ satisfy the NEC. The corresponding EMT cannot represent an immediate neighborhood $r \sim r_\sg$ of the trapped region that has formed in finite time of Bob. In fact, it describes the exterior of a contracting white hole. Consequently, the trapped region cannot form in these models \cite{FT:08,BMT:18} by their very design, and the only question is which form the horizon avoidance actually takes.

\begin{figure}[!htbp] \centering
	\includegraphics[width=0.85\textwidth]{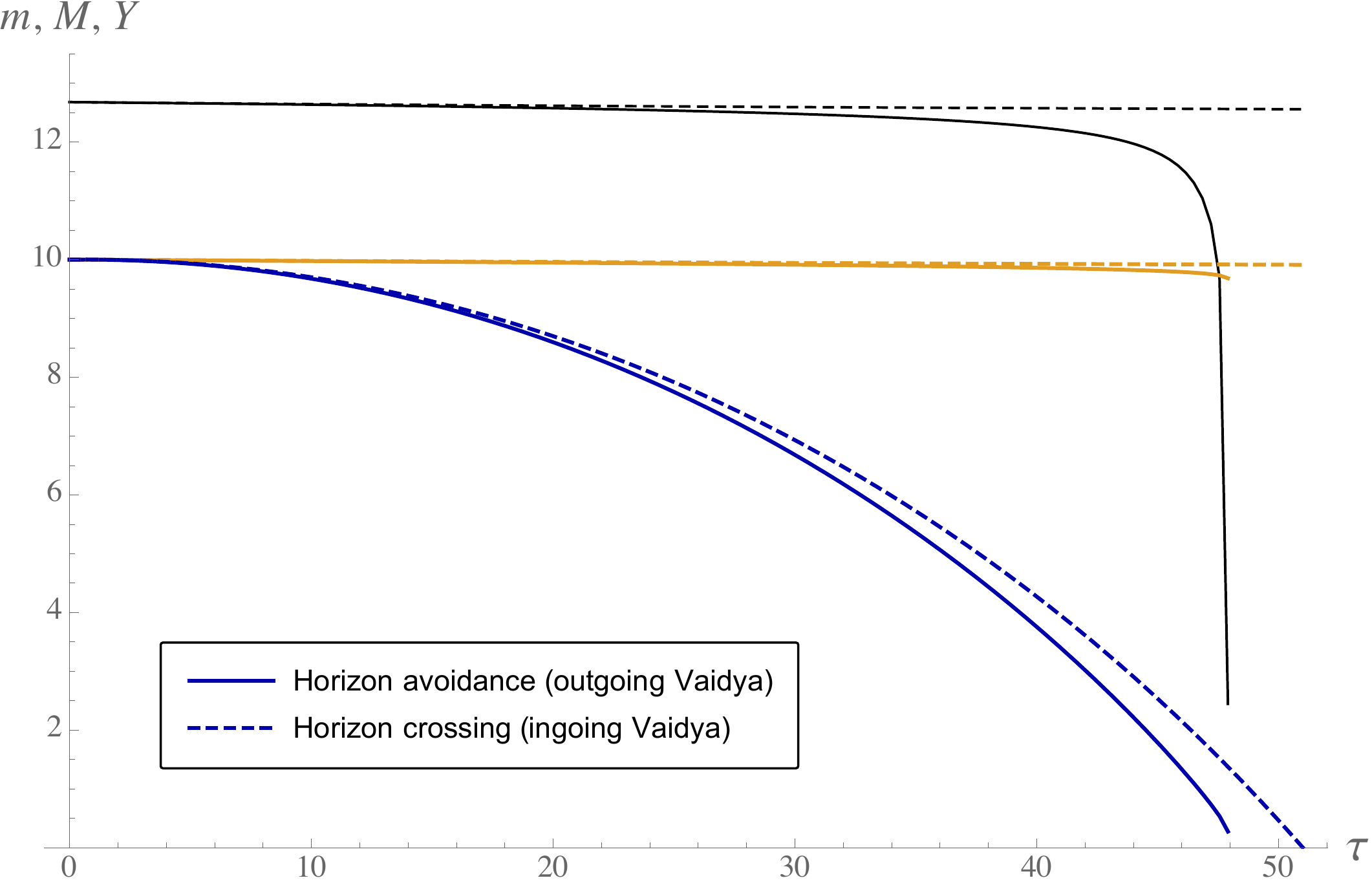}
	\caption{Graviational and rest masses of the shell and the gap $R-2M$. Solid and {dashed} lines correspond to the outgoing and ingoing Vaidya metric, respectively. Evaporation is modeled by a Page-like evaporation law, i.e.\ $dr_-/du = - \alpha /r_-^2$ and $dr_+/dv = - \alpha/ r_+^2$ for the outgoing and ingoing Vaidya metric, respectively, with $\alpha=1$. The black lines represent the rest mass $m(\tau)$, and the orange lines the gravitational mass $M(\tau) = r_\sg(\tau) / 2$ (for both cases $M(0)=10$ and it remains roughly constant throughout the collapse). The initial gap in the two cases is $Y_-(0)=Y(0)=10$. The gap is depicted as a solid blue line for the outgoing Vaidya metric, where $Y_- > 0$ for all times $\tau$ and the transition to the null trajectory occurs at $Y_\infty = 0.250$. For the ingoing Vaidya metric, the gap is illustrated as a dashed blue line that reaches $Y=0$. Horizon crossing occurs at $\tau_\rS = 51.010$.}
	\label{fig:thin-shell-collapse}
\end{figure}

The geometry of such models is described by $h=0$ and
\begin{align}
	F\big(U(\tau),R(\tau)\big)=1-\frac{2M\big(U(\tau)\big)}{R(\tau)},
\end{align}
where $M=C/2$. The shell's equation of motion is then given by \cite{BMT:18}
\begin{align}
	\mathcal{D}(R) + \frac{F_U}{F \sqrt{F + \dot{R}^2}} \left( \frac{1}{2} - \dot{R} \dot{U} \right) = 0,
	\label{eq:EOM-outVaidya}
\end{align}
where $F_U\defeq \pad F/\pad U$.
In the vicinity of the Schwarzschild sphere for non-zero $\dot{R}$, we obtain the asymptotic expression \cite{BsMT:19}
\begin{align}
	\ddot{R} = \frac{16 M M_U \dot{R}^4 }{Y_-^2}+\cO(Y_-^{-1}) ,
	\label{eq:acc-outVaidya}
\end{align}
where $Y_-\defeq R(\tau)-  {r_{-}\big(U(\tau)\big)}$  and $M_U\defeq M'(U)$. We see that evaporation accelerates the collapse of the shell as $M_U<0$. Nonetheless, since
\begin{align}
	\dot Y_-\approx -\dot R\left(1-\frac{2r'_-r_-}{Y_-}\right),
\end{align}
the gap only decreases until   $Y_- \approx \epsilon_\star$, where
\begin{align}
	\epsilon_\star \defeq 2 \frac{dr_\sg}{dU} r_\sg = 8 M \vert M_U \vert ,
\end{align}
and the shell never crosses the ever-shrinking Schwarzschild sphere \cite{BsMT:19} at $r=r_\sg=r_-$ (cf.\ solid blue line in Fig.~\ref{fig:thin-shell-collapse}).

However, the shell sheds its entire rest mass $m$ and becomes null in finite proper time. This happens before reaching the minimal distance from the Schwarzschild sphere (cf.\ solid black line in Fig.~\ref{fig:thin-shell-collapse})\cite{CUWY:18}. This behavior is also true for general sufficiently regular spherically symmetric metrics of the same type \cite{MNT:18}. It is possible to estimate the gap at the timelike-to-null transition \cite{MNT:18} as
\begin{align}
	 Y_\infty = 9 \epsilon_\star / 4.
\end{align}
The shell's subsequent evolution depends on additional assumptions. Once the shell is null, its persistence on a null trajectory can be ensured by several mechanisms. Without changes to its surface EMT, the shell either becomes tachyonic or else ceases radiating\cite{CUWY:18}. If it acquires surface pressure on its transition, it can continue as a null shell and evade the Schwarzschild radius through its complete evaporation. Alternatively, the pressureless collapse may continue if the exterior geometry assumes a more general form. In any case \cite{BsMT:19}, as long as these mechanisms do not lead to a violation of the NEC, the shell will not cross the Schwarzschild radius in finite $t$.

\subsubsection{Ingoing Vaidya metric}\label{V-in}
In the limit $r \to r_\sg$, the near-horizon geometry of an evaporating black hole is described semiclassically by the ingoing Vaidya metric with decreasing mass \cite{jB:81,BMPS:95}. Using it as the exterior metric, the equation of motion of the shell is given by
\begin{align}
	\mathcal{D}(R) - \frac{F_V}{F \sqrt{\dot{R}^2 + F}} \left( \frac{1}{2} + \dot{R} \dot{V} \right) = 0.
\end{align}
In the vicinity of the Schwarzschild sphere for non-zero $\dot{R}$, we obtain
\begin{align}
	\dot V\approx-\frac{1}{2\dot R}+\frac{F}{\dot R^3}.
\end{align}
Using this expression, the radial acceleration is given by \cite{BsMT:19}
\begin{align}
	\ddot{R} \approx - \frac{F'}{2} + \frac{F_V}{F} \approx - \frac{1}{2 r_+} + \frac{\alpha}{r_+^2 Y},
\end{align}
and the rate of change of the gap $Y\defeq R(\tau)-r_+\big(V(\tau)\big)$ is
\begin{align}
	\dot Y\approx \dot R -\frac{1}{2\dot R}.
\end{align}
Thus evaporation prevents collapse only if $Y < \epsilon_\star = 2 \alpha / r_+$ such that $\ddot{R} > 0$. The influence of evaporation on the dynamics of a macroscopic shell is weak (in agreement with the analysis of Ref.~\refcite{bAC.wgU:2018}), and merely causes the formation of the black hole to be slightly delayed (in agreement with the results of Ref.~\refcite{PP:09}).

This model fixes the problem of causal contradiction in the purely classical scenario, in which the crossing time according to Bob is infinite, but the evaporation time is finite. Similarly, a signal that is sent by Alice just before crossing the apparent horizon reaches Bob in finite time, while in the classical case Bob's time diverges as Alice approaches $r_\sg$. However, the shell preserves nearly all of its rest mass when it crosses $r_\sg$ in finite time for both Alice and Bob \cite{BsMT:19}. This is problematic as it implies the existence of a region of positive energy density in the vicinity of the apparent horizon.

\subsection{Formation of the trapped region} \label{subsec:HorizonFormation}

\subsubsection{Inner horizon and its stability}\label{sc:inner}

\begin{figure*}[!htbp]
	\centering
	\hspace*{-6mm}
  	\begin{tabular}{@{\hspace*{0.025\linewidth}}p{0.45\linewidth}@{\hspace*{0.05\linewidth}}p{0.45\linewidth}@{}}
  		\centering
   		\subfigimg[scale=0.475]{(a)}{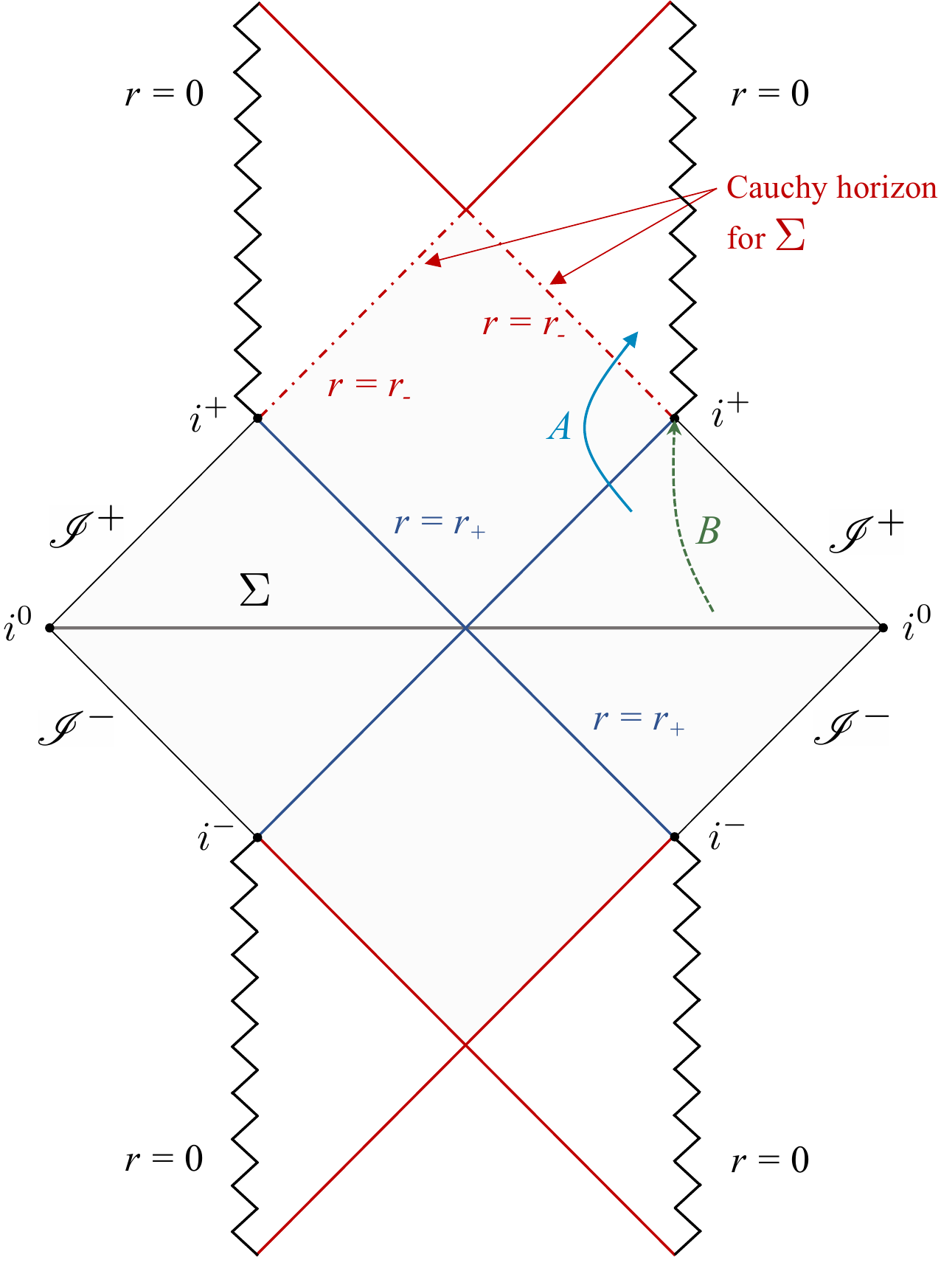} & \hspace*{-1mm}
   		\subfigimg[scale=0.5]{(b)}{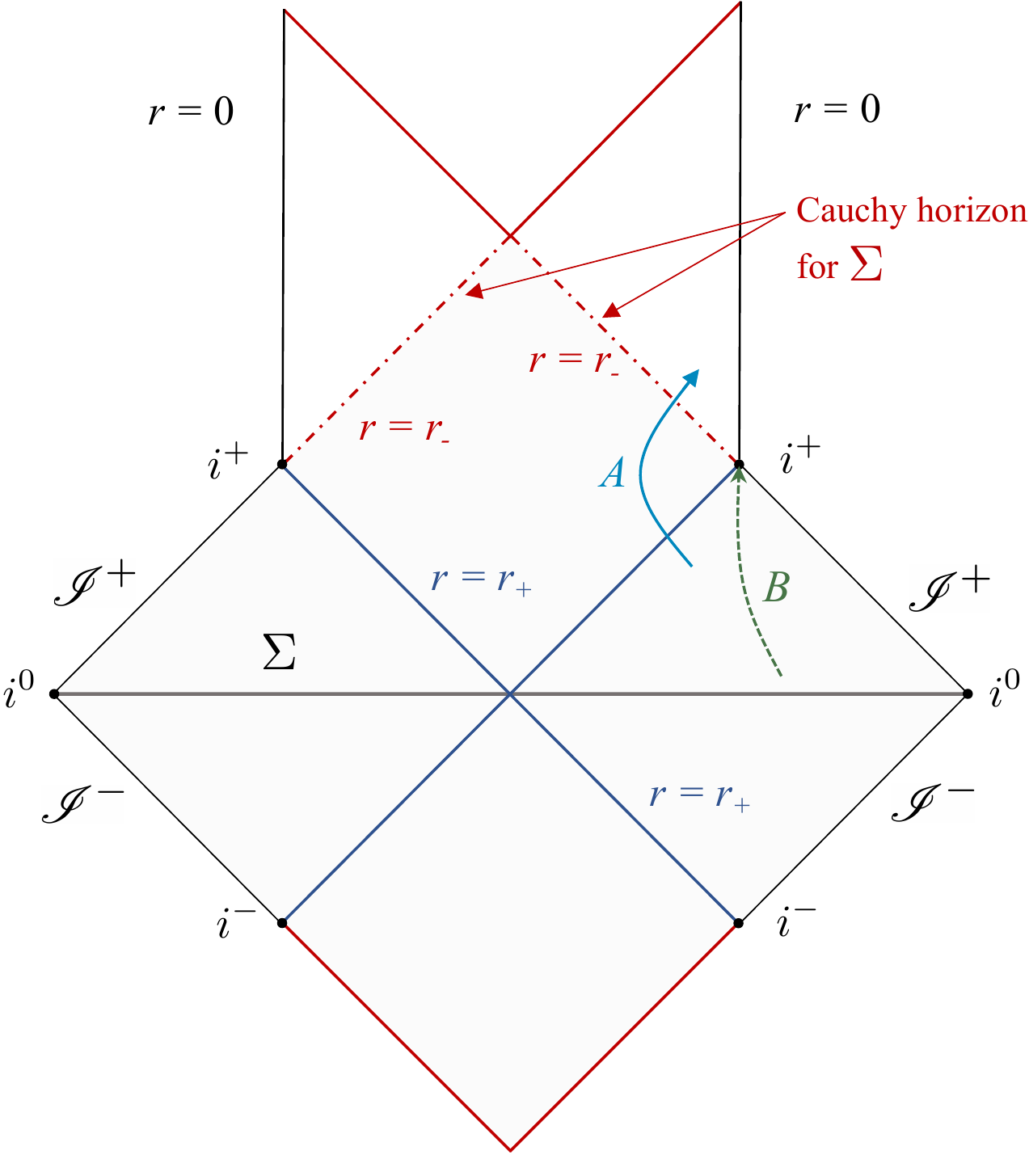}
 	\end{tabular}
  	\caption{Partial internal structure of the maximal causal domain of a (a) subcritical Reissner--Nordstr{\"o}m black hole \cite{HE:73,P:68} (b) regular Hayward black hole \cite{H:06}. Cyan arrows labeled ``A'' indicate the trajectory of an observer Alice crossing the event and Cauchy horizons. The trajectory of a distant observer Bob is indicated by the dashed green arrows labeled ``B''.}
  	\label{fig:innercausal}
\end{figure*}

Multiple horizons are a general property of black hole solutions. The Kerr metric and its generalizations (Sec.~\ref{axial}) have both an outer (event) horizon and an inner (Cauchy) horizon, and their maximal extensions exhibit an intricate causal structure. A spherically symmetric Reissner--Nordstr\"{o}m  metric (Sec.~\ref{k=1}) provides a simple setting for  studying the inner horizon.  Fig.~\ref{fig:innercausal}(a) represents the immediate neighborhood of the domain of dependence of a single Cauchy surface of its extended solution.

The cellular structure of the maximal extension allows for some sci-fi musings: observers deftly navigating their spacecraft away from the timelike curvature singularity at $r=0$ can traverse multiple universes in finite proper times \cite{HE:73}. However, after first crossing the event horizon at $r_\sg$, Alice would see the entire future history of the asymptotically flat region she had left behind within a finite proper time. All signals from that region would become infinitely blue-shifted as their sources approach $\mathscr{I}^+$. This suggests that the Cauchy horizon surface $r=r_\rin$ is unstable \cite{P:68} against perturbations of the initial data on the spacelike surface $\Sigma$.

The emission of gravitational waves allows the collapsar to shed angular momentum and approach the asymptotic solution of Eq.~\eqref{m:RN}. Part of this wave propagates inwards, crosses the horizon in finite advanced time $v$, and is then reflected by the interior gravitational potential \cite{FN:98,HA:10}. It is described as an ingoing flux of positive energy crossing the event horizon with the intensity dropping off as $\sim v^{-p}$ when $v \to +\infty$, and $p>0$ is determined by the multipole moment. Alice encounters this ingoing flux at $r_\rin$. Its intensity is given by $L_1 \propto v^p e^{\kappa_- v}$ and diverges as $v \to +\infty$, where $\kappa_-$ denotes the surface gravity on the inner horizon \cite{PI:89}. While the metric remains regular due to fortunate exact cancellations, taking the outgoing positive energy flux due to backscattering at the black hole interior into account results in the emergence of a real scalar curvature singularity \cite{PI:89,O:91} at the Cauchy horizon $r_\rin$. Moreover, while the infinite curvature occurs only for $v \to \infty$, its exponential growth makes it compatible with the Planck curvature in a short (order of the light crossing) time.

The resulting mass inflation \cite{PI:89,FN:98,HA:10} process is characterized by the exponential growth of the Weyl scalar
\begin{align}
	\Psi_2 \propto v^{-p}e^{\kappa_-v},
\end{align}
indicating an exponential growth of the MS mass at the inner horizon \cite{BKS:21}. Of course, its effects happen beyond the end of time for Bob ($t=\infty$) and are felt only in the ``neighboring universes'' \cite{PI:89,O:91}.

Mass inflation has important consequences for RBH models \cite{M:21}. Static RBHs have a causal structure that is  similar to the Reissner--Nordstr\"{o}m spacetime (Fig.~\ref{fig:innercausal}(b) represents the extension of the Hayward model). This situation is generic: a regular center implies the existence of an untrapped neighborhood in its vicinity, and consequently the existence of an inner (future) trapping horizon \cite{H:06,CDLV:20}. Application of the thin null shells analysis of Refs.~\refcite{PI:89} and \refcite{O:91} leads to the conclusion that static RBHs are generally unstable \cite{CDLPV:18}. This makes it plausible that the inner horizons of dynamic RBHs are also unstable, and there is evidence that this is indeed the case \cite{Brown:2011tv,Brown:2010csa}.

However, the controversy regarding dynamic models, mostly of the types that are discussed in Sec.~\ref{sb:ergo}, is still ongoing \cite{BKS:21,CDLPV:21}. An explicit \textit{ab initio} collapse model of a dust sphere does not demonstrate this instability \cite{HKSW:21}. A class of frozen star models \cite{BM:19,HKY:21} provides an effective description of a highly entropic configuration of fundamental closed strings with the maximal possible negative pressure. Analysis of the resulting Einstein equations demonstrates the stability of static RBH configurations against small time-dependent perturbations \cite{BM:19,BMS:21}.

The self-consistent approach adds another aspect to this discussion. A transient RBH (such as the one in Fig.~\ref{fig:time-g}(b)) forms in finite time according to Bob. Hence the metric in the vicinity of the inner apparent horizon, even during the onset of mass inflation (if it indeed occurs), must satisfy the NEC (see Sec.~\ref{models}). However, the models analyzed in Refs.~\refcite{BKS:21} and \refcite{CDLPV:21} do not have this property. Moreover, even if the energy density in the vicinity of the inner horizon is positive, the very existence of the outer apparent horizon crucially depends on it having a NEC-violating environment. While the infall of positive-energy thin shells onto a self-consistent PBH remains to be investigated, it seems that incursion of a positive energy flux through the apparent horizon is incompatible with its existence. Thus a more detailed investigation is required to understand the stability of the inner horizon of RBHs.

\subsubsection{Formation of physical black holes} \label{s:formation}
The properties of the self-consistent solutions to the Einstein equations described above lead to the identification of a unique scenario for black hole formation \cite{T:20,sMT:21}. Given the results of Sec.~\ref{sc:atmo}, we consider only evaporating ($r'_\sg<0$) PBHs. Let the first marginally trapped surface be denoted by $r_\sg(t_\rS)$. In $(v,r)$ coordinates, it appears at some $v_\mathrm{S}$ at the circumferential radius $r_+(v_\rS)$ that corresponds to Bob's $(t_\rS, r_\sg(t_\rS)=r_+)$.

For $v\leqslant \vS$, the MS mass can be described by modifying Eq.~\eqref{Cpl} as
\begin{align}
	C(v,r) = \Delta(v) + r_*(v) + \sum_{i \geqslant 1}^\infty w_i(v) (r-r_*)^i ,
\end{align}
where $r_*(v)$ corresponds to the maximum of $ \mathbbm{D}_v(r)\defeq C(v,r) - r$, and the deficit function $\Delta(v)\defeq \mathbbm{D}_v(r_*)$. At the advanced time $\vS$, the location of the maximum corresponds to the first marginally trapped surface, $r_*(\vS) = r_+(\vS)$, and $\sigma(\vS)=0$. For $v \geqslant \vS$, the MS mass is described by Eq.~\eqref{Cpl}. For $v \leqslant \vS$, we have $w_1(v) - 1 \equiv 0$ since the (local) maximum of $\mathbbm{D}_v$ is determined by $d\mathbbm{D}_v/dr=0$.

Before the PBH is formed, there are \textit{a priori} no restrictions on the evolution of $r_*$. However, only evaporating black holes, $r'_+(\vS) \leqslant 0$, are consistent with finite formation time of the horizon. Since the trapped region is of finite size for $v>\vS$, the maximum of $C(v,r)$ does not coincide with $r_+(v)$. As a result, $w_1(v)<1$ for $v>\vS$.

\begin{figure}[!htbp]
	\centering
	\includegraphics[scale=0.7]{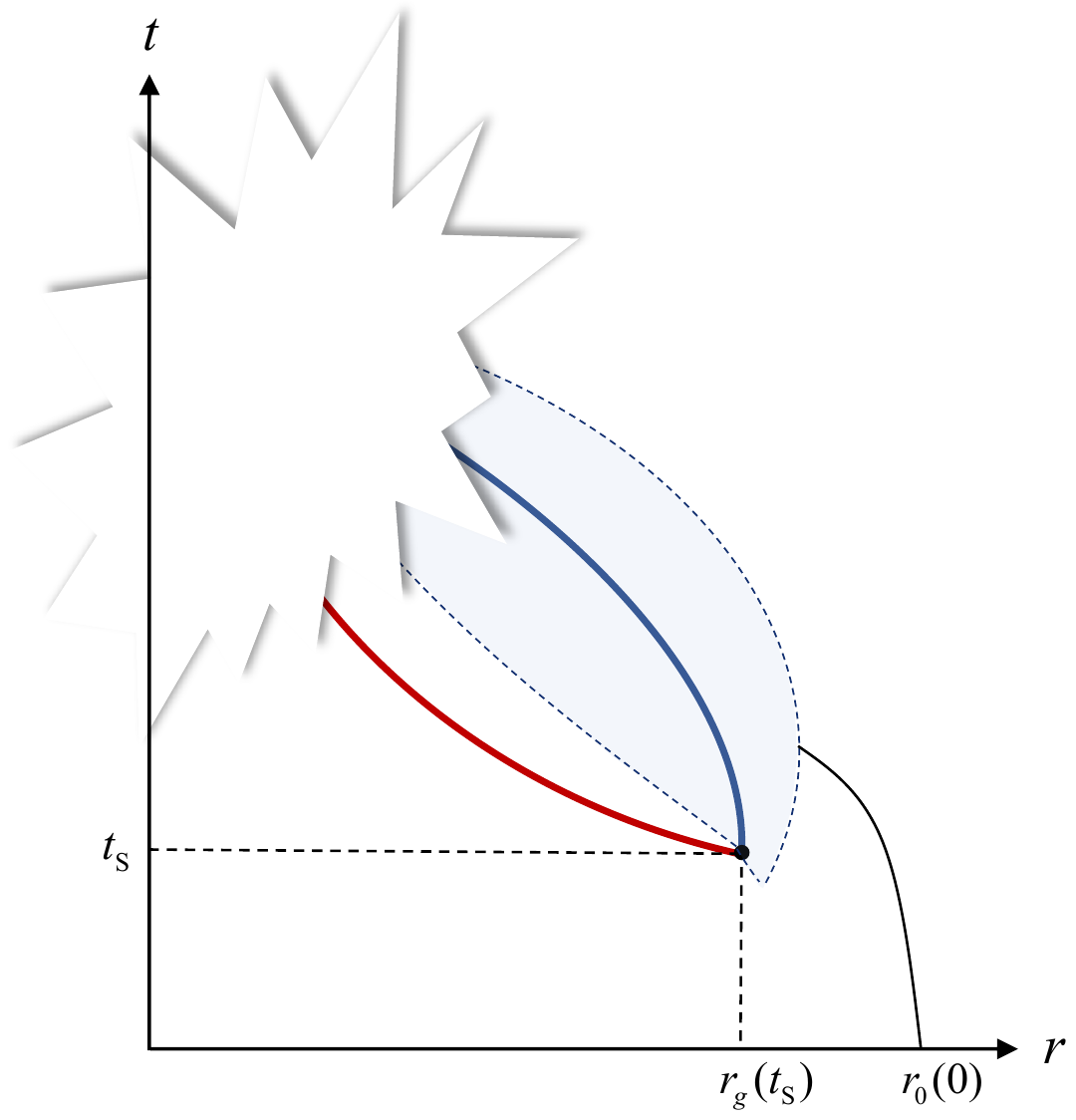}
	\caption{Schematic depiction of the first stages of the evolution of a PBH from the point of view of a distant observer. The dark blue line represents the outer apparent horizon $r_\sg(t)$, and the dark red line the inner horizon $r_\mathrm{in}(t)$. The black dot indicates the first marginally outer trapped surface (MOTS) that is formed at $\big(t_\rS,  r_\sg(t_\rS)\big)$. The NEC-violating region (blue spread, dashed boundary) appears prior to the formation of the first marginally trapped surface $(t_\mathrm{S}, r)$ and covers part of the trapped region. Its outer boundary is not constrained by our considerations. The NEC is not violated at $\big(t_\rS,  r_\sg(t_\rS)\big)$ itself. The thin black line traces the effective boundary of the collapsing $r_0(t)$ body up to the NEC-violating region. The obscured area hides the unknown --- possibly a closing of the trapping horizons (RBH), or a full MBH.}
	\label{f:formation}
\end{figure}

The energy density and pressure at the apparent horizon of a freshly-formed PBH are determined only by its size, as $\rho=-p=(8\pi r_\sg^2)^{-1}$. The NEC is violated in some vicinity of the apparent horizon, but not at $r=r_\sg(t_\rS)$ itself (Sec.~\ref{k=1}).  Analogously, $\mathcal{L}_{l_\rin}\vartheta_\mathrm{out}=0$ at formation, with the apparent horizon branching into the invariantly defined non-spacelike inner and outer horizons. This scenario means that at its formation a PBH is described by a $k=1$ solution. This can be seen from Eq.~\eqref{eq:thev} or Eq.~\eqref{w1t}, as $w_1=1$ implies $\Upsilon=0$. It immediately switches to the $k=0$ solution, with matching decrease in $w_1(v)$ and increase in $\Upsilon \big(t(v,r_+)\big)$. This abrupt transition is only of conceptual importance: metric functions in $(v,r)$ coordinates evolve continuously, and there is no discontinuity according to observers crossing the  $r=r_\sg(t)$ surface.

As the self-consistent approach on its own cannot predict the final state of the collapsing matter, the lines in Fig.~\ref{f:formation} run into the unknown (obscured area). {Due to the non-spacelike nature of the apparent horizon, the formation of a PBH is an event that has an invariant meaning, at least in all coordinate systems that respect spherical symmetry. For a RBH, the same is true also for the final event --- disappearance of the trapped region. Inner and outer apparent horizons correspond to the smallest and largest roots of $f(z,r)=0$ for $z=t,u,r$, respectively.  Moreover, the two segments cannot join smoothly (see Fig.~\ref{f:formation}). This is different from the sketch of Fig.~\ref{fig:time-g}(b).}

Several properties of PBHs are nevertheless clear from the preceding analysis. Since the energy density and pressure are negative in the vicinity of the apparent horizon and positive in the vicinity of the inner horizon (Secs.~\ref{k=0}, \ref{sb:in}), there is a hypersurface at which the NEC is marginally satisfied, and the sphere $r=r_\sg(t_\rS)$ is part of it.

 Hypersurfaces of constant $r$ are timelike outside of the trapped region and spacelike inside of it, while the opposite is true for hypersurfaces of constant $t$. We illustrate how the transition between the two regimes is effected at the apparent horizon\cite{DMT:21} on the hypersurfaces $\Sigma_t$. A hypersurface  can be defined by restricting the coordinates via $\Psi(\Sigma_{t_0}) \eqdef t - t_0 \equiv 0$. Then $\mathfrak{l}_\mu\defeq\Psi_{,\mu}$ is the normal vector field \cite{P:04}, which is timelike for a spacelike segment of the hypersurface and spacelike for a timelike segment. Using $\Psi_{,\mu}$, one can define a normalized vector field that points in the direction of increasing $\Psi$.

Using either $(t,r)$ or $(v,r)$ coordinates, we find that
\begin{align}
	\mathfrak{l}_\mu \mathfrak{l}^\mu = -e^{-2h} f^{-1}.
\end{align}
For both $k=0$ and $k=1$ solutions, $\mathfrak{l}^2\to 0$ as $r\to r_\sg$ (and similarly at the inner apparent horizon). Thus, along $\Sigma_{t_0}$ that passes through a PBH, the normal field changes continuously. Moreover\cite{DMT:21}, at $\big(t,r_\sg(t)\big)$, the vector $\mathfrak{l}^\mu$ is proportional to $l_{\mathrm{out}}^\mu$ of Eq.~\eqref{null-v}. The hypersurface $\Sigma_{t_\rS}$ is spacelike everywhere apart from $\big(t_\rS, r_\sg(t_\rS)\big)$ where it is null (see Fig.~\ref{fig:time-g}).

It is not clear how this scenario can be realized in nature. A thin shell indeed collapses in a finite time as measured by Bob, but this happens on a timescale comparable to that of evaporation \cite{BsMT:19}, and most of its rest mass remains intact. However, the mandatory violation of the NEC requires some mechanism that converts the original matter into the exotic matter present in the vicinity of the forming apparent horizon, whilst maintaining   normal behavior near the inner horizon.

However, the emission of collapse-induced radiation \cite{H:87,BLSV:06,VSK:07} is a nonviolent process that at late times approaches the standard Hawking radiation with Page's evaporation law \cite{FN:98,P:76} $r_\sg' = - \alpha/r_\sg^2$, $\alpha \sim 10^{-3}-10^{-4}$. Even if the necessary NEC violation occurs in nature without requiring new physics, the process may be too slow to transform the UCOs that we observe into PBHs. The timescale of the last stages of infall according to Bob was discussed in Sec.~\ref{shells} and is set by Eq.~\eqref{drdt0}. Assuming that it is applicable for $x\lesssim r_\sg$, we have $t_\mathrm{in}\sim r_\sg/r'_\sg$. For an evaporating macroscopic PBH, this is of the same order of magnitude as the Hawking process decay time $t_\mathrm{evp}\sim 10^3 r_\sg^3$. Such behavior was found in thin shell collapse models, where the exterior geometry is modeled by a pure outgoing Vaidya metric \cite{sMT:21}. For a solar mass black hole this time is about $10^{64}$ yr, indicating that it is simply too early for the horizon to form. It is also conceivable that the conditions are not met before evaporation is complete or before effects of quantum gravity become dominant \cite{FN:98,COY:15}.

Indeed, there are indications that new physics may be required. Explicit perturbative calculations of the backreaction of quantum vacuum polarization in the presence of a gravitational field \cite{FN:98,BD:82,PT:09} lead to a generalization of the classical Tolman--Oppenheimer--Volkoff equation, and eventually predict the formation of horizonless UCOs \cite{C-R:18}.

\subsection{Surface gravity} \label{SurfaceGravity}

\subsubsection{Role of surface gravity}\label{sg-s}
The surface gravity $\kappa$ plays an important role in black hole thermodynamics and in semiclassical gravity \cite{HE:73,FN:98}. For an observer at infinity the Hawking radiation that is produced on the background of a stationary black hole is thermal with its temperature given by $T_\rH = \kappa/2\pi$  (see \ref{a:Haw} for a summary of the relevant formulas).

Stationary asymptotically flat spacetimes admit a Killing vector field $\vK^\mu$ that is timelike at infinity \cite{HE:73,C:92,exact:03,P:04,C-B:09,P:10,FN:98}. A Killing horizon is a hypersurface on which the norm $\sqrt{\vK^\mu\vK_\mu}=0$. While logically this concept is independent of the notion of an event horizon, the two are related: for a black hole that is a solution of the Einstein equations in a stationary asymptotically flat spacetime, the event horizon coincides with the Killing horizon \cite{HE:73,FN:98,W:01}.

A Killing orbit is an integral curve of a Killing vector field. The Killing property $\vK_{(\mu;\nu)}=0$ results in $\vK^\mu\vK_\mu=\mathrm{const}$ on each orbit. The coincidence of Killing and event horizons \cite{HE:73,FN:98} allows one to introduce the surface gravity $\kappa$ as the inaffinity of null Killing geodesics on the event horizon,
\begin{align}
	\vK^\mu_{~;\nu}\vK^\nu \defeq \kappa \vK^\mu. \label{kapKil}
\end{align}
On the other hand, assuming sufficient regularity of the metric, the expansion of null geodesics outside the apparent horizon   establishes the concept of peeling affine gravity \cite{CLV:13,NY:08}
\begin{align}
	\frac{dr}{dt} = \pm 2 \kappa_\mathrm{peel}(t) x + \mathcal{O}(x^2).
	\label{peeldexp}
\end{align}
The two definitions coincide in stationary spacetimes. For the Schwarzschild metric with mass $M$ the surface gravity is $\kappa=1/(4M)=1/(2r_\sg)$.

The surface gravity of a black hole can be interpreted as the force that would be required by an observer at infinity to hold a particle (of unit mass) stationary at the event horizon \cite{FN:98}. We reproduce here the elementary derivation in $(t,r)$ coordinates. Consider an observer Eve at some fixed areal radius $r$. Her four-velocity is $u_\mathrm{E}^\mu = \delta^\mu_0 / \sqrt{-\sg_{00}}$, and her four-acceleration $a_\mathrm{E}^\mu=(0,\Gamma^r_{tt}/\sg_{00},0,0)$. In the Schwarzschild spacetime, we have
\begin{align}
	g \defeq \sqrt{a^\mu_{\mathrm{E}} a_{{\mathrm{E}}\mu}} = \frac{r_\sg}{2r^2\sqrt{1-r_\sg/r}}.
\end{align}
Correcting by the redshift factor $z=-1/\sqrt{\sg_{00}}$ gives the surface gravity on approach to the horizon,
\begin{align}
	\kappa=\lim_{r\to r_\sg}zg=1/(2r_\sg). \label{zkap}
\end{align}
 An equivalent expression is given by
\begin{align}
	\kappa=\left.\frac{1}{2}\frac{\pad f(r)}{\pad r}\right|_{r_\sg}.
\end{align}
Many important results are based on the relationship between Schwarzschild and Rindler metrics, where the role and outcomes of measurements performed by a static Eve near the event horizon are analogous to those of an accelerated Rindler observer, while a free-falling Alice corresponds to an inertial Bob in Minkowski space \cite{FN:98,BD:82,PT:09,tP:05}. Moreover, for a rotating black hole the ZAMO frame provides a preferred frame that is associated with an external observer. In the non-rotating limit it reduces to the standard Schwarzschild frame. In the vicinity of the event horizon these frames coincide with very high accuracy with the Rindler reference frame \cite{TPM:86}.

Using $x=r-r_\sg$ as the radial variable, the Schwarzschild metric near the event horizon takes the form
\begin{align}
	ds^2\approx -\frac{x}{r_\sg}dt^2+\frac{r_\sg}{x}dx^2+r_\sg^2d\Omega_2=-\kappa^2\ell^2dt^2+d\ell^2+dL_\perp^2, \label{SR-0}
\end{align}
where a new independent variable $\ell$ corresponds to the physical distance (\ref{AForm}) and $dL_\perp^2$ is the line element in the transverse space \cite{FN:98,tP:05,tP:10}. (In the vicinity of the event horizon $r \approx r_\sg+\ell^2/4r_\sg$).

Many of the thermal properties are established by analyzing the relevant Euclidean Green functions and their periodicity. The starting point for the exploration of potentially profound relations between gravity and thermodynamics \cite{M:15,tP:05,tP:10} is based on the local gravity-acceleration relationship and the fact that the Hawking temperature is proportional to the acceleration of the Killing orbit \cite{J:95}.

This Schwarzschild--Rindler relationship is important for gedankenexperiments that test the weak equivalence principle \cite{W:14,GF:87}. Classically, a (point-like) detector at rest in a uniform gravitational field and an identical detector in a uniformly accelerated frame in Minkowski spacetime produce indistinguishable results. Matching of appropriate vacua \cite{GF:87,FN:98} leads to the Alice/Bob/Eve equivalences mentioned above: both an accelerating detector in the Rindler vacuum and a fixed detector in Schwarzschild spacetime in the Boulware vacuum do not click. Both a uniformly accelerating detector in Minkowski vacuum and a static detector in the Unruh or Hartle--Hawking vacuum on the Schwarzschild background detect thermal radiation. However, for the same value of acceleration the statistics of the detector in a gravitational field correspond to a higher temperature \cite{SW:11,Ng:2014kha}, and the difference between the two disappears only at the horizon.

The surface gravity for the inner horizon of a Reissner–Nordstr\"{o}m black hole (Sec.~\ref{k=1}) is defined similarly to its counterpart on the event horizon\cite{PI:89,FN:98},
\begin{align}
	\kappa_\rin\defeq\left.-\frac{1}{2}\frac{\pad f(r)}{\pad r}\right|_{r_\rin}=\frac{r_\sg-r_\rin}{2r_\rin^2}.
\end{align}
Its extension to (dynamical) RBH models plays an important role in the debates about stability of  inner horizons\cite{BKS:21,FZ:17,ZLO:20}, as well as their evaporation\cite{FZ:17}.

\subsubsection{Surface gravity of physical black holes} \label{kappaPBH}
Surface gravity is unambiguously defined only in stationary spacetimes. There are several reasonable extensions that are based on different properties  of $\kappa$ in stationary spacetimes, e.g.\ the inaffinity of null geodesics on the horizon, or the peeling off properties of null geodesics near the horizon \cite{vF:15,VAD:11,CLV:13,NY:08}. The latter approach typically uses Schwarzschild or Painlev\'{e}--Gullstrand coordinates \cite{NV:06,VAD:11} (which are non-singular at the horizon) to obtain a generalization of $\kappa_\mathrm{peel}$. In spherically symmetric spacetimes, the Kodama vector $\vKo$ (Sec.~\ref{spher:a}) shares enough properties with the Killing vector to define the surface gravity by constructing an expression analogous to that of Eq.~\eqref{kapKil}.

For slowly evolving horizons of sufficient regularity, different generalizations of surface gravity are practically indistinguishable \cite{VAD:11,CLV:13}. This is important, as the role of the Hawking temperature is captured in various derivations either by the peeling \cite{BLSV:11} or the Kodama \cite{KPV:21} surface gravity, and  to establish the dynamic spacetime counterpart \cite{KSP:07} of the Schwarzschild--Rindler relation of Eq.~\eqref{SR-0}. Indeed, gravitational collapse triggers radiation \cite{H:87,BLSV:06,VSK:07} that for macroscopic black holes at sufficiently late times approaches the standard Hawking radiation. Surface gravity is also related to the observable signatures that distinguish black holes from horizonless UCOs. Classical black holes have zero reflectivity \cite{FN:98,CP:19,BCNS:19}. However, when quantum effects are taken into account \cite{AAOW:20}, its non-zero value strongly depends on the surface gravity \cite{OWA:20}.

The apparent horizon of a PBH is mildly singular, and this affects evaluation of the different versions of surface gravity. First, we note that the red-shifted acceleration of a static observer diverges for $k=0$ solutions as \cite{MsMT:21}
\begin{align}
	z g = \frac{|r'_\sg|}{4x} + \mathcal{O}(x^{-1/2}),
\end{align}
and thus Eq.~\eqref{zkap} results in infinity.

Consider now the peeling surface gravity $\kappa_\mathrm{peel}$. For differentiable $C$ and $h$, it is given by \cite{NV:06,CLV:13}
\begin{align}
	\kappa_\mathrm{peel} = \frac{e^{h(t,r_\sg)} \left( 1 - C'(t,r_\sg) \right)}{2 r_\sg} .
	\label{peeld}
\end{align}
However, this is inapplicable for both the $k=0$ and $k=1$ solutions \cite{T:20,sMT:21}. The metric functions Eqs.~\eqref{eq:k0C}--\eqref{eq:k0h} and Eqs.~\eqref{fk1}--\eqref{hk1} lead to a divergent peeling surface gravity. This happens because Eq.~\eqref{drdt0} ensures that there is a non-zero constant term in the expansion of the geodesics, and instead of Eq.~\eqref{peeldexp} we have
\begin{align}
	\frac{dr}{dt} = \pm r_\sg'+a_{12}(t)\sqrt{x} + \mathcal{O}(x) ,
\end{align}
where $a_{12}$ depends on the higher-order terms of the EMT. The use of Painlev\'{e}--Gullstrand coordinates allows for two different versions of the peeling surface gravity \cite{VAD:11}. One results in zero, and the other depends on the precise form of the Painlev\'{e}--Gullstrand time \cite{MsMT:21}.

The Kodama vector imitates the properties of the Killing vector based surface gravity as \cite{VAD:11,H:98}
\begin{align}
	\frac{1}{2} \vKo^\mu(\nabla_\mu \vKo_\nu-\nabla_\nu \vKo_\mu) \eqdef \kappa_\mathrm{K} \vKo_\nu,
\end{align}
evaluated on the apparent horizon. This Hayward--Kodama surface gravity is explicitly given by \cite{KPV:21,MsMT:21}
\begin{align}
	\kappa_\mathrm{K} = \frac{1}{2} \left.\left(\frac{C_+(v,r)}{r^2} - \frac{\partial_r C_+(v,r)}{r}\right) \right|_{r=r_+} \hspace*{-3mm} = \frac{(1-w_1)}{2r_+}. \label{kappaK}
\end{align}
Thus at the formation of a black hole (i.e.\ of the first MOTS) this version of surface gravity is zero, as $w_1=1$ (Sec.~\ref{s:formation}). At the subsequent evolution stages that correspond to a $k=0$ solution, $\kappa_\mathrm{K}$ is non-zero. However, it can reach the static value $\kappa=1/(4M)$ only if the metric is close to the pure Vaidya metric with $w_1\equiv 0$. Imposing this constraint leads to contradictions \cite{MsMT:21} with the standard semiclassical results for luminosity $L\propto C^{-2}$ and evaporation time $t_\mathrm{e}\propto C^3$.

Constructing the analog of the Schwarzschild--Rindler relation of Sec.~\ref{sg-s} is also not straightforward. Using $x=r-r_\sg$ as an independent variable for a $k=0$ solution with $r'_\sg<0$, we find at leading order
\begin{align}
	ds^2=\sqrt{\frac{\xi}{x}}\left(-2dtdx+\frac{dx^2}{|r'_\sg|}\right)+r_\sg^2d\Omega_2+\cO(x^0), \label{kappaTran}
\end{align}
which is structurally different from Eq.~\eqref{SR-0}.

\subsection{Information loss problem} \label{sub:loss}
For the last four and a half decades, the information loss problem has been one of the most persistent and well-publicized puzzles of theoretical physics. Voluminous literature is devoted to this subject. Refs.~\citenum{BMPS:95,FN:98,W:01,Mathur:2009hf,M:15,COY:15,H:16,M:17,UW:17,info:21,dM:17} provide a comprehensive description of all of its aspects. In its most basic form it can be formulated as follows \cite{W:01}:

\begin{quote}
``an isolated black hole will evaporate completely via the Hawking process within a finite time. If the
correlations between the inside and outside of the black hole are not restored
during the evaporation process, then by the time that the black hole has evaporated
completely, an initial pure state will have evolved to a mixed state, i.e.,
information will have been lost. In a semiclassical analysis of the evaporation
process, such information loss does occur and is ascribable to the propagation of
the quantum correlations into the singularity within the black hole.''
\end{quote}
While there are some minor caveats (such as the impossibility of pure matter states to form a black hole \cite{M:97,HT:10}, revising the initial entropy from zero to ``low''), this quote provides a fair representation of the issues involved. In Sec.~\ref{parel}, we summarize the basic ingredients that are necessary for the formulation of the paradox, and in Sec.~\ref{parPBH} we discuss what these assumptions entail for PBHs.

\subsubsection{Key elements of the paradox} \label{parel}
A crucial assumption underlying black hole radiation is that  quantum states of radiated fields are regular (Hadamard) at the horizon. This assumption is rooted in the equivalence principle, that locally (on sufficiently short time and distance scales) gravity and acceleration are indistinguishable, and so the local behavior of a quantum field in the vicinity of the horizon is the same as it would be in the Minkowski vacuum. This in turn implies that no unusual high-energy behavior should be seen by freely falling observers near the horizon, a situation referred to as the ``no drama" assumption \cite{Almheiri:2012rt}.

This assumption implies that a black hole horizon is `information-free': curved spacetime quantum field theory on the black hole background provides a valid means of describing field modes with wavelengths $l_\rP \ll \lambda \lesssim GM/c^2$. The notion of a particle is contingent on what the vacuum, or `empty space' is defined to be, but the difference between definitions will consist of about 1 quanta for wavelengths as large as the curvature scale $\lambda\sim r_0$, and much less for smaller wavelengths. Since $r_0\sim GM/c^2$ for a black hole, a robust notion of vacuum (i.e.\ empty space) is well-defined for such wavelengths: no modes are present for $l_\rP \ll \lambda < kGM/c^2$, where $k\sim 10^{-1}$.

Black hole radiation is the radiation of quanta, and can be understood as resulting from the time-dependent stretching of spatial slices that foliate both the outside and inside of the black hole. On some smooth initial slice $\Sigma_0$, the matter field that will later form the black hole is in a quantum state $|\Phi(t,r) \rangle$. The entire slice is outside of a horizon (since the black hole has not formed) and therefore no particles (field quanta) are created. Once the black hole forms, this spatial slice necessarily undergoes considerable stretching in order to remain smooth through the horizon. This in turn creates pairs of quanta of short wavelength. The stretching of the slices is localized to a region in the vicinity of the horizon: a field mode in this region gets increasingly stretched to longer wavelengths,   generating pairs of entangled field quanta for as long as the no-drama assumption holds, along with other assumptions such as positivity of energy, smooth evolution, and finiteness of curvature --- a set of assumptions known as `niceness conditions' \cite{Mathur:2009hf,M:15}. One member of the pair remains within the horizon, while the other escapes to an asymptotic region. Since only this latter particle is observable, there is $\ln2$ units worth of entropy created for each particle (each quanta) emitted from the black hole. After $n$ particles are emitted, the entanglement entropy is \cite{PT:04,Mathur:2009hf}
\begin{align}
	S_{\mathrm{ent}} = n\ln2 .
	\label{entent}
\end{align}
This expression is at the core of the information paradox: the entanglement of the radiation state seen by an outside (distant) observer grows without bound. Energy conservation indicates that $n$ can be extremely large. The total mass of the black hole is $M = n E_Q$, where $E_Q$ is the energy per quanta, assuming each quanta contains the same amount of energy. In units of the Planck mass, $E_Q= \sigma (m_\rP/M) m_\rP$; $\sigma$ will be a parameter of order unity. Consequently $n = (M/m_\rP)^2/\sigma$. For a solar mass black hole $n \sim (2\times 10^{30}/( 2\times10^{-8}))^2 = 10^{76}$. There is no upper bound on $n$ since in principle the mass of the black hole can be arbitrarily large, though presumably the largest black hole possible is constrained by the mass of the universe, $M_\mathrm{unv} \sim 10^{52}$ kg, giving $n \leq 10^{120}$.

This is the nub of the information paradox: the entanglement of the radiation state outside of the black hole grows without bound as more pairs are created. Energy conservation implies that this process must eventually terminate, since the radiation cannot contain more energy than the mass $M$ of the black hole (or rather of  the initial quantum state $|\Phi\rangle$ from which the hole formed). Once $M\sim m_\rP$ the niceness conditions will fail to hold, since (for example) the Kretschmann scalar  $\eK= 48 M^2/r^6 \to m_\rP^2 / l_\rP^6 = l^4_\rP$ becomes too large for semiclassical physics to be valid. This situation evidently leaves us with one of three unpleasant choices \cite{Preskill:1992tc}:

\begin{enumerate}[label=\textbf{(\arabic*})]
\item \textbf{Mixedness} \\
The black hole completely evaporates away, with all of its energy contained in the radiation. The radiation will have entanglement entropy $n\ln 2$, but there is no quantum state with which it is entangled. Consequently there is no quantum wavefunction to describe this state; instead we must use a mixed state density matrix $\rho_{\textrm O_n}$. In other words, the evolution begins with a pure state $|\Phi\rangle$ and ends in a mixed state $\rho_{\textrm O_n}$. But there is no unitary matrix that can evolve $\rho_\Phi = |\Phi\9\6\Phi|$ to $ \rho_{\textrm O_n}$ or vice versa, and therefore this option violates unitarity. If one is willing to accept this perspective, then something in the AdS/CFT correspondence conjecture must break down \cite{M:15,info:21}, since unitarity holds in the CFT but will not hold on the AdS (gravity) side once the black hole has evaporated. \\[-2mm]
\item \textbf{Remnants} \mbox{}\\
Once $M \sim M_r \gtrsim m_\rP$ something halts the evaporation process, leaving behind a remnant object of mass $M_r$. A remnant would be a particular kind of ECO, whose number of possible states must be at least as large as the (unbounded) number $n$, since its entanglement with the final state of radiation is $n\ln 2$.   Consequently it must be an $n$-fold degenerate state of finite energy and size, making it a quantum state unlike any normal quantum state we know of \cite{COY:15}. This odd feature is not benign: the remnant must somehow couple to normal matter, and so each of its degenerate states will give a finite loop correction to any scattering process in particle physics. But there are $n$ degenerate states and so the sum over $n$ yields a divergence since $n$ is arbitrarily large. This will occur for any finite matter-remnant coupling, no matter how small. If some other physics yields an upper bound to $n$ (say $n \leq 10^{120}$), then the couplings of all remnant states to normal matter must be kept extremely tiny so as not to significantly modify known scattering processes in particle physics. \\[-2mm]
\item \textbf{Bleaching} \mbox{}\\
The information associated with the state $|\Phi\rangle$ is prevented by some process from entering the black hole or even from forming it in the first place. This begs the question as to what process this might be and as to what object(s) instead of black holes might results. Astrophysical black holes evidently can absorb matter (LIGO/Virgo data indicates that they can absorb neutron stars \cite{LIGOScientific:2021qlt}), so if this option holds then some new kind of physics --- some kind of drama --- must be present at the horizon to either prevent this from happening or to decouple the information in this state from its energy and angular momentum (whatever that means). Alternatively, if black holes never form, then the putative black holes LIGO/Virgo (and the EHT) observe are actually some new form of dark matter whose interactions contain some repulsive effect that counteracts any possible gravitational pull toward collapse.
\end{enumerate}
These three options seem to be the only ones available, given the information paradox. But the paradox itself is contingent on the existence of an event horizon \cite{mV:14,MsMT:21}. The apparent horizon, if it exists, is located within the event horizon (if the NEC is satisfied; see Sec.~\ref{parPBH} for the NEC-violating scenario). However, if there is no apparent horizon, there cannot be an event horizon, and therefore no trapping of information. This would be a manifestation of the last option, which is effectively the ECO option in broad terms.

The last option presents us with choosing between some kind of new physics that either admits horizon avoidance \cite{KMY:13}, leading to the formation of horizonless objects under all possible gravitational collapse scenarios, or that allows the formation of black holes that cannot absorb information. This latter possibility may involve the formation of some new kind of structure (for example, a firewall), or the breakdown of semiclassical physics in some way that eliminates the paradox. It is for these reasons that a better understanding of the near-horizon geometry of PBHs will improve models developed to take full advantage of the new era of multimessenger astronomy \cite{BCNS:19,sbG:17}, using observations not only to learn about the true nature of astrophysical black holes, but also to obtain new insights into fundamental physics.

\subsubsection{Physical black holes and the paradox} \label{parPBH}
In common with the paradoxes of quantum mechanics, the information loss problem combines classical and quantum elements and some counterfactual reasoning. Let us consider the physical and mathematical consequences of having the necessary elements for its formulation realized. We note first that the formulation of the information loss problem involves at least the following:
\begin{enumerate}
\item Formation of a transient trapped region. Such a region either completely disappears or turns into a stable remnant; in either case, this takes place  in finite time as measured by a distant observer Bob. This provides the scattering-like setting to describe the states (and their alleged information content) ``before'' and ``after''.
\item Formation of an event horizon. Its existence is necessary to give an objective, observer--independent significance to tracing out of the black hole degrees of freedom.
\item Thermal or nearly-thermal character of the radiation. It is responsible for the eventual disappearance of the trapped region and for the high entropy of the reduced exterior density operator.
\end{enumerate}
Scrutiny of the technical aspects of these commonly invoked semiclassical notions indicates that they are not realized in the form that mandates the logical contradiction. For example, ``Page time unitarity'' may appear to be violated even if the underlying physics is unitary \cite{SFA:21}. Moreover, there is an indication \cite{BFM:21} that the standard form of the paradox can be consistently rendered only if some new physics begins to play a role before reaching the Planck scale.

Within the purely semiclassical approach, formation of a transient trapped region indicates that it happens at some finite amount of time as measured by Bob (see Fig.~\ref{fig:time-g} and Sec.~\ref{s:shori}). Assuming the absence of a naked singularity makes the PBH model inevitable. In turn, it leads to the problem with matching the Hawking temperature (\ref{a:Haw}) of the radiation \cite{MsMT:21}. Namely, two presumably ``close'' generalizations of surface gravity that underpin different derivations of Hawking radiation on the background of an evolving spacetime are irreconcilable. Disregarding the discordant generalizations of the peeling surface gravity, we focus on the Kodama surface gravity $\kappa_\rK$.

Formation of a PBH (Sec.~\ref{s:formation}) as a $k=1$ solution requires $w_1(t_\mathrm{S})=1$. At the subsequent stages $w_1<1$. This transition is continuous, as  $\Upsilon(t_\mathrm{S}) \equiv 0$ in $k=1$ solutions and it increases thereafter, and thus
\begin{align}
	w_1= 1-r_\sg\frac{r_\sg''}{r_\sg'}  = 1-\frac{2\alpha}{r_+^2}.
\end{align}
Eq.~\eqref{kappaK} implies that for $\kappa_\rK$ to approach the Hawking value $\kappa=1/(2r_\sg)$, it is necessary to have $w_1 \to 0$. However, for the standard evaporation law $w_1 \approx 0$ only when $r_\sg \sim \sqrt{\alpha}$, i.e.\ in the sub-Planckian regime where semiclassical physics indubitably breaks down.

If we try to obtain the evaporation law $\Gamma(r_\sg)$ by requiring $w_1 \equiv 0$, then we must have
\begin{align}
	\Gamma(r_\sg) = r_\sg' =  \ln \frac{r_\sg(t)}{B} ,
\end{align}
where the first equality follows from Eq.~\eqref{Gamma-rels}. Obtaining a negative $r_\sg' < 0$ corresponding to the process of evaporation at times $t > t_\mathrm{S}$ implies  $B=r_\sg(t_\mathrm{S})+\beta>r_\sg(t_\mathrm{S})$. The solution $r_\sg(t)$ can be expressed in terms of the integral logarithm ${\mathrm{li}}(z)=\int_0^zdt/\ln t$. Using its asymptotic form for $\beta\ll 1$, we obtain the evaporation time
\begin{align}
	t_\mathrm{e} \approx r_\sg(t_\mathrm{S}) \ln \frac{r_\sg(t_\mathrm{S})}{\beta},
\end{align}
which is radically different from the standard semiclassical results.

If the Hawking temperature is indeed proportional to a version of the peeling surface gravity, then black holes may freeze or explode at the formation of the trapped region. In either case the semiclassical picture is not valid and it is impossible to formulate the information loss problem. Alternatively, if the Hawking temperature is proportional to the Kodama surface gravity, then it vanishes at the formation of a black hole; although it increases during evaporation, it can never attain the Hawking result without violating the presumed thermality of the radiation. If the Kodama surface gravity reaches the classical value $\kappa_\mathrm{K}=1/(2r_\sg)$, then it cannot be the black hole temperature. Moreover, it is not clear how, given indications to the contrary \cite{C-R:18} (as well as indication that some of the more exotic approaches can actually suppress the Hawking-like emission\cite{K:12}), a process with close to zero flux  can ensure the necessary dominance of quantum effects over normal matter in the vicinity of the outer apparent horizon.

Taken together these results indicate that the circumstances surrounding the formation of black holes do not provide a basis for formulating the information loss paradox. While there is indeed a deep conceptual chasm between quantum mechanics and GR, the intricate questions of the semiclassical black hole analysis do not produce the elements that are necessary for the logical incompatibility that is required for having a paradox.

If the information loss problem cannot be formulated in a way that is self-consistent and physically sound in the framework of semiclassical gravity, we must conclude that speculative resolutions of the ``paradox'' involving new physics are redundant if they do not motivate why new physics should lead to information loss to begin with (as there is currently no evidence to suggest that this is the case).

\section{Black Holes in Modified Theories of Gravity} \label{s:MTG}
The development of various modified theories of gravity (MTG), i.e.\ extensions and or generalizations of GR, is motivated by the prospect of resolving some of its perceived shortcomings (such as the presence of non-spacelike singularities or perturbative non-renormalizability) combined with the possibility of describing additional gravitational degrees of freedom through the inclusion of  additional gravity-related terms in the action functional \cite{CDLNO:09,BLLD:18}. In addition, theoretical considerations encourage us to consider GR as the low-energy regime of some effective theory of quantum gravity \cite{B:04,DH:15}. MTG provide a natural gravitational alternative for dark energy and dark matter, allow for a unified description of early-time inflation and late-time acceleration of the universe, and may even serve as the basis for a unified explanation of dark energy and dark matter \cite{NO:07,S:07}. Compact objects with strong gravitational fields maximally highlight differences in the predictions of GR and alternative theories of gravity \cite{W:14}. Consequently, this is the regime where modifications of the Einstein equations (induced by, e.g., quantum gravitational effects) are expected to be discernible \cite{B:04}. We restrict our attention to the most straightforward generalizations of GR that are obtained by adding higher-order curvature terms in the Lagrangian density.

\subsection{Metric modified theories of gravity} \label{s:mMTG}
We do not make any assumptions about the underlying reason(s) for modifying the gravitational Lagrangian density to include additional curvature-dependent terms, and organize it according to powers of derivatives of the metric as is commonly done in effective field theories \cite{DH:15,B:04,P:10,P:11}. Thus the action $S = \int \eL_\mathrm{g} \sqrt{-\sg}$ is derived from the Lagrangian density
\begin{align}
	\begin{aligned}
		\eL_\text{g}  &= \frac{{m_\rP}^2}{16\pi} \big( \eR + \lambda \eF(\tensor{\sg}{^\mu^\nu}, \tensor{R}{_\mu_\nu_\rho_\sigma}) \big) \\
		&= \frac{{m_\rP}^2}{16\pi} \eR + a_1 \eR^2 + a_2 \tensor{R}{_\mu_\nu} \tensor{R}{^\mu^\nu} + a_3 \tensor{R}{_\mu_\nu_\rho_\sigma} \tensor{R}{^\mu^\nu^\rho^\sigma} + \ldots ,
	\end{aligned}
	\label{eq:gravLagr}
\end{align}
where the cosmological constant term was omitted, $\sg \equiv \det (\tensor{\sg}{_\mu_\nu})$ denotes the determinant of the metric tensor, $m_\rP$ is the Planck mass that we set to one in what follows, and the coefficients $a_1$, $a_2$, $a_3$ are dimensionless. The dimensionless parameter $\lambda$ sets the scale of the perturbative analysis (for details, see Refs.~\citenum{sMT:21b,sM:21}) and is set to unity at the end of the calculations. A prototypical example is the family of $\mathfrak{f}(\eR)$ theories, where $\eL_\mathrm{g}=\mathfrak{f}(\eR)$  is an arbitrary function of the Ricci scalar $\eR$, e.g.\ the Starobinsky model with $\eF=\varsigma \eR^2$, $\varsigma=16 \pi a_1 / m_\rP^2$.

Due to the higher-order curvature terms in the gravitational Lagrangian density Eq.~\eqref{eq:gravLagr}, the resulting modified Einstein equations contain at least fourth-order derivatives of the metric. Variation of the gravitational action results in
\begin{align}
	\tensor{G}{_\mu_\nu} + \lambda \tensor{\EuScript{E}}{_\mu_\nu} = 8 \pi \tensor{T}{_\mu_\nu} , \label{eq:mtgEE}
\end{align}
where the terms $\EuScript{E}_{\mu\nu}$ result from the variation of $\eF(\sg^{\mu\nu}, R_{\mu\nu\rho\sigma})$ \cite{P:11,sMT:21b} and $\tensor{T}{_\mu_\nu} \equiv \langle \tensor{\hat{T}}{_\mu_\nu} \rangle_\omega$ denotes the expectation value of the renormalized EMT as described in Sec.~\ref{s:shori}. In spherical symmetry, the modified Einstein equations take the form
\begin{align}
	& f r^{-2} e^{2h} \partial_r C + \lambda \tensor{\EuScript{E}}{_t_t} = 8 \pi \tensor{T}{_t_t} , \label{eq:mtgEEtt} \\
	& r^{-2} \partial_t C + \lambda \tensor{\EuScript{E}}{_t^r} = 8 \pi \tensor{T}{_t^r} , \label{eq:mtgEEtr} \\
	& 2 f^2 r^{-1} \partial_r h - f r^{-2} \partial_r C + \lambda \tensor{\EuScript{E}}{^r^r} = 8 \pi \tensor{T}{^r^r} . \label{eq:mtgEErr}
\end{align}
We assume that there is a solution of Eq.~\eqref{eq:mtgEE} with the two metric functions
\begin{align}
	& C_\lambda \eqdef \bar{C}(t,r) + \lambda \Sigma(t,r) , \label{eq:pcSigma} \\
	& h_\lambda \eqdef \bar{h}(t,r) + \lambda \Omega(t,r) , \label{eq:pcOmega}
\end{align}
where $\Sigma(t,r)$ and $\Omega(t,r)$ denote the perturbative corrections and the bar labels the unperturbed metric functions that solve the unmodified Einstein equations \eqref{eq:Gtt}--\eqref{eq:Grr}, see Table~\ref{tab:PBHsol}. Artifactual divergences are avoided by using the physical value of $r_\sg(t)$ that corresponds to the perturbed metric $\sg^{(\lambda)}_{\mu\nu} = \tensor{\bar{\sg}}{_\mu_\nu} + \lambda \tensor{\tilde{\sg}}{_\mu_\nu}$, i.e.\ $C_\lambda(t,r_\sg)=r_\sg$. The EMT depends on $\lambda$ through the perturbed metric $\sg^{(\lambda)}_{\mu\nu}$, and potentially also through effective corrections resulting from the modified Einstein equations Eqs.~\eqref{eq:mtgEEtt}--\eqref{eq:mtgEErr}. It is decomposed as
\begin{align}
	\tensor{T}{_\mu_\nu} \eqdef \tensor{\bar{T}}{_\mu_\nu} + \lambda \tensor{\tilde{T}}{_\mu_\nu} ,
	\label{eq:EMTdecomp}
\end{align}
where the bar again labels the unperturbed term $\tensor{\bar{T}}{_\mu_\nu}$.

The perturbations must satisfy the boundary conditions
\begin{align}
	& \hspace*{1.25mm} \Sigma(t,0) = 0 , \\
	& \lim_{r \to r_\sg} \Omega(t,r) / \bar{h}(t,r) = \mathcal{O}(1) , \label{hdiver}
\end{align}
where the former follows from the definition of the Schwarzschild radius, and the latter ensures that perturbations can be treated as small through the requirement that the divergence of $\Omega$ is not stronger than that of $\bar{h}$ on approach to the horizon, i.e.\ as $r \to r_\sg$. Substituting $C_\lambda$ and $h_\lambda$ into Eq.~\eqref{eq:mtgEE} and keeping only first-order terms in $\lambda$ results in
\begin{align}
	\tensor{\bar{G}}{_\mu_\nu} + \lambda \tensor{\tilde{G}}{_\mu_\nu} + \lambda \tensor{\bar{\EuScript{E}}}{_\mu_\nu} = 8 \pi \left( \tensor{\bar{T}}{_\mu_\nu} + \lambda \tensor{\tilde{T}}{_\mu_\nu} \right) , \label{MGscheme}
\end{align}
where $\tensor{\bar{G}}{_\mu_\nu} \equiv \tensor{G}{_\mu_\nu}[\bar{C},\bar{h}]$, $\tensor{\tilde{G}}{_\mu_\nu}$ corresponds to the first-order term in the Taylor expansion in $\lambda$ where each monomial involves either $\Sigma$ or $\Omega$, and $\tensor{\bar{\EuScript{E}}}{_\mu_\nu} \equiv \tensor{\EuScript{E}}{_\mu_\nu}[\bar{C},\bar{h}]$, i.e.\ the modified gravity terms are functions of the unperturbed solutions.

To obtain the explicit form of the equations, we first note that
\begin{align}
	e^{2h}= e^{2 \bar{h}} \left( 1 + 2 \lambda \Omega \right)  +\mathcal{O}(\lambda^2) . \label{hdiver2}
\end{align}
If we adopt the schematic separation of the EMT according to Eq.~\eqref{eq:EMTdecomp} for the effective EMT components defined in Eqs.~\eqref{eq:taus}, i.e.\ $\tensor{\tau}{_a} = \tensor{\bar{\tau}}{_a} + \lambda \tensor{\tilde{\tau}}{_a}$, $a \in \lbrace \tensor{}{_t} , \tensor{}{_t^r} , \tensor{}{^r} \rbrace \equiv \lbrace \tensor{}{_t_t} , \tensor{}{_t^r} , \tensor{}{^r^r} \rbrace$, then the EMT terms of the $tt$ equation can be written as
\begin{align}
	\tensor{\bar{T}}{_t_t} + \lambda \tensor{\tilde{T}}{_t_t} &= e^{2 \bar{h}} \left( 1 + 2 \lambda \Omega \right) \left( \tensor{\bar{\tau}}{_t} + \lambda \tensor{\tilde{\tau}}{_t} \right) \\
	&= e^{2 \bar{h}} \bigl( \tensor{\bar{\tau}}{_t} + \lambda \left( 2 \Omega \tensor{\bar{\tau}}{_t} + \tensor{\tilde{\tau}}{_t} \right) \bigr) + \mathcal{O}(\lambda^2) .
\end{align}
$\tensor{T}{_t^r}$ and $\tensor{T}{^r^r}$ are expanded analogously. The requirement that the scalar curvature invariants $\mathrm{T}$ and $\mathfrak{T}$ of Eq.~\eqref{eq:TwoScalars} be regular implies that the perturbative terms $\tensor{\tilde{\tau}}{_a}$ should either have the same behavior as their $\tensor{\bar{\tau}}{_a}$ counterparts when $r \to r_\sg$, or go to zero faster. Consequently, the schematic of Eq.~\eqref{MGscheme} implies
\begin{align}
	&	\tensor{\bar{G}}{_t_t} = \frac{e^{2 \bar{h}}}{r^3} \left( r - \bar{C} \right) \partial_r \bar{C} , \\
	& \tensor{\tilde{G}}{_t_t} = \frac{e^{2 \bar{h}}}{r^3} \left[ - \Sigma \partial_r \bar{C} + \left( r - \bar{C} \right)
 \left( 2 \Omega \partial_r \bar{C} + \partial_r \Sigma \right) \right] ,
\end{align}
and thus the explicit form of Eq.~\eqref{eq:mtgEEtt} is
\begin{align}
	- \Sigma \partial_r \bar{C} + \left( r - \bar{C} \right) \partial_r \Sigma + r^3 e^{- 2 \bar{h}} \tensor{\bar{\EuScript{E}}}{_t_t} = 8 \pi r^3 \tensor{\tilde{\tau}}{_t} . \label{eq:mGravEFEtt}
\end{align}
Similarly, Eqs.~\eqref{eq:mtgEEtr} and \eqref{eq:mtgEErr} can be written explicitly as
\begin{align}
	& \partial_t \Sigma + r^2 \tensor{\bar{\EuScript{E}}}{_t^r} = 8 \pi r^2 e^{\bar{h}} ( \Omega \tensor{\bar{\tau}}{_t^r} + \tensor{\tilde{\tau}}{_t^r}) , \label{eq:mGravEFEtr} \\
	& \Sigma \partial_r \bar{C} -  ( r - \bar{C}) (4\Sigma \partial_r \bar{h} + \partial_r \Sigma) + 2( r - \bar{C})^2 \partial_r \Omega + r^3 \tensor{\bar{\EuScript{E}}}{^r^r} = 8 \pi r^3 \tensor{\tilde{\tau}}{^r} . \label{eq:mGravEFErr}
\end{align}
The most general spherically symmetric metric is still given by Eq.~\eqref{eq:metric}, and the requirements of finiteness of $\mathrm{T}$ and $\mathfrak{T}$ are still meaningful.  On the one hand,
 they are no longer directly related to the finiteness of the curvature scalars. For example, in $\mathfrak{f}(\eR)$ theories \cite{SF:10,DT:10}, where $\eL_\mathrm{g} = \mathfrak{f}(\eR)$, the trace of the field equations is given by
\begin{align}
	\maf'(\eR) \eR  {- 2 \maf(\eR)} + 3 \square \maf'(\eR) = 8 \pi \mathrm{T},
	\label{eq:f(R)EOMtrace}
\end{align}
and the finiteness of $\tensor{T}{^\theta_\theta}$ is not guaranteed \textit{a priori}. It is conceivable that the metric is such that the curvature invariants are finite, but $\square R$ and thus $\mathrm{T}$ diverge at the apparent horizon. On the other hand, it is possible to incorporate all terms not accounted for by the Einstein tensor into the effective EMT
\begin{align}
	T^\mathrm{eff}_{\mu\nu}\defeq T_{\mu\nu}-\lambda\EuScript{E}_{\mu\nu}/8\pi,
\end{align}
and to define $\mathrm{T}$ and $\mathfrak{T}$ accordingly \cite{MT:21mg}. Then the two classes of solutions described in Sec.~\ref{s:Ein} are still the only possible solutions in spherical symmetry, even if (as we see below) not all solutions are perturbative expansions of their GR counterparts.

Nevertheless, the two classes of solutions described in Sec.~\ref{s:Ein} are  the starting point for the perturbative expansion in $\lambda$. While their existence must be established separately for each MTG, it is clear that divergences stronger than those allowed in GR are not permitted at any order of $T_{\mu\nu} = \bar{T}_{\mu\nu} + \lambda T^{(1)}_{\mu\nu} + \ldots$, as such terms would contribute stronger singularities to the metric functions $C_\lambda$ and $h_\lambda$, and thus invalidate the perturbative expansion close to the apparent horizon.

To describe perturbative PBH solutions in MTG, the equations must satisfy the same consistency relations as their GR counterparts. Taking the GR solutions as the zeroth-order approximation, we express functions using the perturbed metric $\sg^{(\lambda)}_{\mu\nu}$ and represent the modified Einstein equations as series in integer and half-integer powers of the coordinate distance $x \defeq r - r_\sg$ from the apparent horizon. Their order-by-order solution results in formal expressions for the perturbative corrections $\Sigma(t,r)$ and $\Omega(t,r)$ of Eqs.~\eqref{eq:pcSigma}--\eqref{eq:pcOmega}.

To be compatible with the two semiclassical PBH solutions summarized in Table~\ref{tab:PBHsol}, any given MTG must satisfy several constraints that arise because power expansions in various expressions have to match up to allow for self-consistent solutions of the modified Einstein equations \eqref{eq:mGravEFEtt}--\eqref{eq:mGravEFErr} \cite{sMT:21b}. They are summarized in Table~\ref{tab:MTGcc}, and manifest themselves in two ways: first, the MTG terms $\tensor{\bar{\eE}}{_\mu_\nu}$ must conform to the expansion structures prescribed by Eqs.~\eqref{teq:k0-Ett}--\eqref{teq:k0-Err} and Eqs.~\eqref{teq:k1-Ett}--\eqref{teq:k1-Err} for the $k=0$ and $k=1$ solutions, respectively. Second, their coefficients must satisfy the three identities Eqs.~\eqref{teq:k0-cc1cc2}--\eqref{teq:k0-cc3} for $k=0$, and the two identities Eqs.~\eqref{teq:k1-cc1}--\eqref{teq:k1-cc2} for $k=1$. Moreover, the relations between the EMT components that are given by Eqs.~\eqref{eq:thev}--\eqref{eq:ther} must hold separately for both the unperturbed terms and the perturbations.

\begin{table}[!htpb]
	\tbl{Necessary conditions for the existence of semiclassical PBHs in arbitrary metric MTG as perturbations of the PBHs in GR. To be compatible with semiclassical PBHs of the $k=0$ ($k=1$) type, the higher-order terms of arbitrary metric MTG must have the divergence structures prescribed by Eqs.~\eqref{teq:k0-Ett}--\eqref{teq:k0-Err} [Eqs.~\eqref{teq:k1-Ett}--\eqref{teq:k1-Err}] when expanded in terms of the coordinate distance $x \defeq r - r_\sg$ from the apparent horizon. Additionally, their lowest-order coefficients $\ae_j(t)$, $\oe_j(t)$, and $\o_j(t)$ must satisfy the three (two) identities given by Eqs.~\eqref{teq:k0-cc1cc2}--\eqref{teq:k0-cc3} [Eqs.~\eqref{teq:k1-cc1}--\eqref{teq:k1-cc2}]. \vspace*{2mm}}
	{
	\centering
	\resizebox{\textwidth}{!}{
		\begin{tabular}{ >{\raggedright\arraybackslash}m{0.135\linewidth} | @{\hskip 0.02\linewidth} >{\raggedright\arraybackslash}m{0.36\linewidth} @{\hskip 0.02\linewidth} | @{\hskip 0.02\linewidth} >{\raggedright\arraybackslash}m{0.445\linewidth}}
			& $k=0$ solutions & $k=1$ solution
			\\ \toprule
			Decomposition of MTG terms
			&
			{\begin{flalign}
				\tensor{\bar{\eE}}{_t_t} & = \frac{\ae_{\bar{1}}}{x} + \frac{\ae_{\overbar{12}}}{\sqrt{x}} + \ae_0 + \sumj \ae_j x^j &
				\label{teq:k0-Ett} \tag{0.1} \\
				\tensor{\bar{\eE}}{_t^r} & = \frac{\oe_{\overbar{12}}}{\sqrt{x}} + \oe_0 + \sumj \oe_j x^j &
				\label{teq:k0-Etr} \tag{0.2} \\
				\tensor{\bar{\eE}}{^r^r} & = \o_0 + \sumj \o_j x^j &
				\label{teq:k0-Err} \tag{0.3}
			\end{flalign}}
			\quad & \quad
			{\begin{flalign}
				\tensor{\bar{\eE}}{_t_t} & = \frac{\ae_{\overbar{32}}}{x^{3/2}} + \frac{\ae_{\bar{1}}}{x} + \frac{\ae_{\overbar{12}}}{\sqrt{x}}  + \ae_0 + \sumj \ae_j x^j &
				\label{teq:k1-Ett} \tag{1.1} \\
				\tensor{\bar{\eE}}{_t^r} & = \oe_0 + \sumj \oe_j x^j &
				\label{teq:k1-Etr} \tag{1.2} \\
				\tensor{\bar{\eE}}{^r^r} & = \sum\limits_{j \geqslant \frac{3}{2}}^\infty \o_j x^j &
				\label{teq:k1-Err} \tag{1.3}
			\end{flalign}}
			\\[-3mm]
			Relations between MTG coefficients
			&
			{\begin{flalign}
				\ae_{\bar{1}} &= \sqrt{\bar{\xi}} \oe_{\overbar{12}} = \bar{\xi} \o_0 &
				\label{teq:k0-cc1cc2} \tag{0.4} \\
				\ae_{\overbar{12}} &= 2 \sqrt{\bar{\xi}} \oe_0 - \bar{\xi} \o_{12} &
				\label{teq:k0-cc3} \tag{0.5}
			\end{flalign}}
			\quad & \quad
			{\begin{flalign}
				\ae_{\overbar{32}} &= 2 \bar{\xi}^{3/2} \oe_0 - \bar{\xi}^3 \o_{32} \vphantom{\sqrt{\bar{\xi}}} &
				\label{teq:k1-cc1} \tag{1.4} \\
				\ae_{\bar{1}} &= 2 \bar{\xi}^{3/2} \left( h_{12} \oe_0 + \oe_{12} \right) - \bar{\xi}^3 \left( 2 h_{12} \o_{32} + \o_2 \right) \vphantom{\sqrt{\bar{\xi}}} &
				\label{teq:k1-cc2} \tag{1.5}
			\end{flalign}}
			\\
			\bottomrule
		\end{tabular}
	}
	\label{tab:MTGcc}
	}
\end{table}

There is \textit{a priori} no reason to assume that $\bar \sg_{\mu\nu} \gg\lambda \tilde \sg_{\mu\nu}$ should hold in some boundary layer around $r_\sg$ \cite{MVC:20,C:18}. If this condition is not satisfied, then the classification scheme of the GR solutions and a mandatory violation of the NEC are not necessarily true. Properties of solutions without a well-defined GR limit are discussed in Ref.~\refcite{sMT:21b}, together with a derivation of the necessary conditions for their existence.

Similarly, there is \textit{a priori} no reason to believe that the constraints should or should not be satisfied in any particular MTG. If the constraints are not satisfied, the MTG in question may still possess solutions corresponding to PBHs, albeit their mathematical structure must then be fundamentally different from those of semiclassical gravity described in Sec.~\ref{s:Ein}, which may or may not give rise to observationally distinguishable features.

For a given MTG, the constraints may lead to several different outcomes: first, it is possible that some of the higher-order terms in the gravitational Lagrangian density $\eL_\mathrm{g}$ contribute terms to $\tensor{\bar{\eE}}{_\mu_\nu}$ such that their expansions around $x=0$ lead to terms that diverge stronger than any other term in Eqs.~\eqref{eq:mGravEFEtt}--\eqref{eq:mGravEFErr}. If only one higher-order curvature term is responsible for such behavior, then such a theory cannot produce perturbative PBH solutions, and only non-perturbative solutions may be possible, or the corresponding coefficient $a_i$ of that term [cf.\ Eq.~\eqref{eq:gravLagr}] is zero. If the divergences originate from several terms, they can either cancel if a particular relationship exists between their coefficients $a_i, a_{i'},\ldots$, or not. In the former case the existence of perturbative PBH solutions imposes a constraint, not on the form of the available terms, but on the relationships between their coefficients.

It is also possible that the divergences of the terms $\tensor{\bar{\eE}}{_\mu_\nu}$ match the divergences of the GR terms. The constraints can then be satisfied (i) identically, i.e.\ without any additional requirements, thus providing no additional information; (ii) only for a particular combination of the coefficients $a_i$, thereby constraining the possible classes of MTG; (iii) only in the presence of particular higher-order terms, irrespective of the coefficients, and only for certain unperturbed solutions. In the last scenario, where only certain GR solutions are consistent with a small perturbation, this should be interpreted as an argument against the presence of that particular term in the gravitational Lagrangian density $\eL_\mathrm{g}$ of Eq.~\eqref{eq:gravLagr}.

Ref.~\citenum{sM:21} demonstrates that all of the constraints summarized in Table~\ref{tab:MTGcc} are satisfied identically in generic MTG with up to fourth-order derivatives of the metric (which includes $\mathfrak{f}(\eR)$ theories as a special subclass). Consequently, these models are compatible with the semiclassical PBH solutions, which can be regarded as zeroth-order terms in their perturbative solutions.

\subsection{Torsion gravity}
The Einstein\textendash{}Cartan theory of gravity is a modification of GR in which spacetime can have torsion in addition to curvature \cite{CDL:11,H:02,HHKN:76}. Torsion is often presumed to arise from intrinsic spin, although it has been demonstrated that it can also derive from the gradient of a scalar potential \cite{H:02}, e.g.\ the Higgs field \cite{SW:80}. Here, we make no assumptions about the origin of torsion, and simply elaborate the consequences of the results presented in the previous sections for gravitational theories with torsion.

The torsion tensor is expressed as the antisymmetric part of the connection $Q^\mu_{\nu\eta}=\half(\Gamma^\mu_{\nu\eta}-\Gamma^\mu_{\eta\nu})$. Despite having a non-metric part of the connection, it is still assumed that $\nabla \tensor{\sg}{_\mu_\nu} = 0$. The full set of equations therefore consists of the equations for $\tensor{G}{_\mu_\nu}$ that are related to the EMT, and the equations for $Q^\mu_{\nu\eta}$ that relate the torsion to the density of intrinsic angular momentum (or, depending on the origin of torsion, an analogous quantity).

However, it is possible to represent this system by a single set of Einstein equations with an effective EMT on the rhs, i.e.\
\begin{align}
	\mathring{G}_{\mu\nu} = 8 \pi T^\text{eff}_{\mu\nu},
\end{align}
where $\mathring{G}_{\mu\nu}$ is derived from the metric alone and the effective EMT includes terms that are quadratic in spin \cite{H:02,HHKN:76,T:06}. Requiring now that $\tensor{\mathring{R}}{^\mu_\nu}$ and $\tensor{\mathring{R}}{^\mu^\nu}\tensor{\mathring{R}}{_\mu_\nu}$ are finite at the apparent horizon $r=r_\sg$ leads to the same $k=0,1$ types of PBH solutions.

\section{Discussion}
The goal of this review is to survey the consequences of an affirmative answer to the question ``Do trapped regions form in finite time as measured by distant observers?''. While several widely accepted scenarios (see Fig.~\ref{fig:time-g}) implicitly assume this, there is \textit{a priori} no reason to assume that this is in fact the case, and scenarios (1) and/or (2) of Sec.~\ref{sec:bh} are not realized. In any case, since classical black hole solutions are used as a benchmark for interpreting observations, it is natural to investigate if the differences in the near-horizon geometry of PBHs and MBHs translate into potentially observable signatures.

Basing our approach on two reasonable (and often implicit) assumptions of semiclassical gravity --- cosmic censorship and horizon formation in finite asymptotic time --- we find a number of startling implications. First, we find that only two types of dynamic solutions, both describing (depending on the sign of $r_\sg^\prime$) evaporating PBHs and expanding white holes, are admissible. Second, if a particle can fall through the horizon of a PBH, it will do so in finite asymptotic time (the time measured by a distant clock), in a clear departure from classical black holes. Third, we find that the apparent and anti-trapping horizons are hypersurfaces of intermediately singular behavior, which manifests itself via regions of divergent negative energy density experienced by some observers. It would seem appropriate to refer to this phenomenon as a firewall. It remains to be seen if this firewall violates any of the bounds that constrain the violation of the NEC.

The usual practice of modeling black holes (as distinct from ECOs) following scenario (1) as Schwarzschild or Kerr--Newman solutions, perhaps slightly modified by the effects of quantum gravity, is closely related to the no-hair theorems. This group of uniqueness results is one of the most intriguing properties of MBHs \cite{HE:73,C-B:09,P:04,CCH:12}. From a practical perspective, they lead to the expectation that the stationary post-collapse/merger black hole states are parameterized by their mass, angular momentum, and a set of charges. On the other hand, there are black hole solutions with Skyrme, Higgs, dilaton, or non-Abelian fields, with and without supersymmetry, that have hair of various `softness' \cite{CCH:12}, while supertranslation symmetries require (via an infinite number of conservation laws for all gravitational theories in asymptotically Minkowskian spacetimes) the existence of ``a large amount of soft (i.e.\ zero-energy) supertranslation hair'' \cite{HPS:16}. The no-hair property of conventional GR black holes can be tested, largely in a model-independent way, in future gravitational wave experiments \cite{vdB:14}. The basic underlying logic is that after the dramatic evolutionary stages (such as collapse or merger), the quiescent stage of their evolution is well-described by the classic solutions of GR.

The $k=0$ solutions exhibit strong hair without postulating exotic fields or symmetries. They contain a non-trivial function $h(t,r)$ whose characteristic length scale $\xi=r'_\sg{}^2/(2|r_\sg''|)$  (Sec.~\ref{ident}) is another important parameter that characterizes the geometry [see Eq.~\eqref{kappaTran}]. If the evaporation law is well-approximated by the relation $r_\sg'=-\alpha/r_\sg^2$, then $\xi \sim r_\sg$.

PBHs must have an energy density and pressure that is negative in the vicinity of their apparent horizons, but positive in the vicinity of their inner horizons. Consequently, there is a hypersurface where the NEC is marginally satisfied. However, whether or not this kind of object can form in nature is not at all clear. Although a thin shell indeed collapses in a finite asymptotic time, the mandatory violation of the NEC noted above necessitates some mechanism for converting the original matter of the shell into the exotic matter near the forming apparent horizon. From a modeling perspective \cite{RZ:13,BS:10}, an apparent horizon can be obtained in finite time of Bob only if the EMT that is used is different from that of a standard perfect fluid. Apart from violating the NEC, the matter is at least a two-component fluid, where one component is null. The role of fluxes is as important as that of density and pressure.

These results in turn have implications for the information loss paradox, since the circumstances surrounding the formation of PBHs do not provide a basis for formulating it \cite{MsMT:21}. The necessary ingredients for the logical incompatibility required to have a paradox are absent. This in turn suggests that speculative resolutions of the ``paradox'' involving new physics are redundant.

Conventional effects of quantum fluctuations \cite{FN:98} lead to the fuzziness (of the event horizon)
\begin{align}
  \delta r_\sg \sim l^2_\rP/r_\sg.
\end{align}
The very existence of this effect qualitatively changes the classical description of the collapse and infall of a body into a black hole as seen by Bob, as now the final stages of the infall take only a finite amount of time
\begin{align}
	\Delta t \sim r_\sg \ln (r_\sg/l_\rP), \label{delTqg}
\end{align}
where the Schwarzschild metric provides the classical picture. It is interesting to investigate how these considerations play out with the apparent horizon and $k=0$ metric [and the physical distance Eq.~\eqref{lphys0}].

The above estimate is based on modeling the near-horizon geometry with the Schwarzschild metric. However, properties of the $k=0$ solutions suggest a different scaling. Replacing $r_\sg$ with $r_\sg+\delta r_\sg$ as the lower bound of integration, that in the Schwarzschild metric leads \cite{FN:98} to Eq.~\eqref{delTqg}, changes the finite infall time only by $\delta t \sim l^2_\rP/(r_\sg|r'_\sg|)$ [cf.\ Eqs.~\eqref{drdt0} and \eqref{kappaTran}]. Even when taking the new relationship Eq.~\eqref{lphys0} between the physical and coordinate distance into account, this is insignificant.

We also emphasize that our considerations are relevant on timescales much shorter than the age of the universe, since we have no \textit{a priori} restriction on the size of the PBHs we consider. It is logically possible for PBHs to be planet-sized or even human-sized, well within the regime of semiclassical physics, but requiring considerations of much shorter timescales. For example, the standard evaporation rate (\ref{Haw-em}) implies that a black hole of $\sim10^8$kg will evaporate on the timescale of a year. There is no logical reason why such objects cannot form. Primordial black holes \cite{K:10,DlFPR:20,E:22} may be produced even in the absence of mechanisms that allow for the formation of PBHs as the end stage of the collapse. In particular, it is conceivable that the necessary EMT content is locally created as a result of density perturbations during the inflationary era. A different near-horizon geometry and, in particular, the need to maintain the NEC-violating region, may  also affect the subsequent accretion of matter into primordial black holes, and thus their mass distribution.

From an observational perspective the PBH models (as opposed to the Kerr-type paradigm) may or may not be the right description of the observed astrophysical black holes. The point that we would like to emphasize is that PBHs and MBHs are conceptually different objects. A particularly promising direction is the investigation of the response of PBH models to perturbations. As we have seen, there are substantial differences among the geometries at $r\sim r_\sg$ between these scenarios. Differences in the near-horizon geometry and the behavior of generalizations of surface gravity indicate that both the equations that govern quasi-normal modes and the boundary conditions that are imposed on them \cite{C:92,CP:19} are modified, while the effective reflectivity that depends on the surface gravity \cite{AAOW:20} also distinguishes PBHs from other models. All of these indicate the potential for obtaining observable differences between the three possible end results of gravitational collapse. Of course, reliable conclusions can be drawn only after non-spherically-symmetric configurations are thoroughly investigated.

Another avenue of investigation is the response of physical objects (detectors) to quantum effects in the vicinity of a PBH. Recent work \cite{HMZ:21} has shown that the response of an Unruh--DeWitt detector (an atom with 2 energy levels) held static near an ECO is notably distinct from the corresponding situation outside of a black hole \cite{HLO:14,Ng:2014kha}, even when the ECO boundary is perfectly absorbing. Studies of detector entanglement outside of PBHs and ECOs should likewise prove interesting, and can be expected to differ considerably from the scenario for black holes \cite{HHMSZ:18,RHM:21}.

Each possibility will allow important conclusions to be drawn. For example, if astrophysical black holes are described by Kerr black holes, then  \textit{now} these are actually horizonless UCOs. There is no need to introduce ECOs, as the paradoxical inner life of black holes begins only after crossing the horizon, which (under these circumstances) is known not to form. A particularly exciting possibility is the detection of an expanding apparent horizon. Since accretion into a semiclassical PBH is impossible, if such a process is observed, it would constitute a direct (even if not specific regarding the underlying theory) observation of a full quantum gravitational effect, as well as confirming their importance at the horizon \cite{AB:05,B:14,M:17,HKY:21}.

Whatever the outcome, it is clear that investigations of the confluence of horizon formation with (semiclassical) quantum physics still have much to teach us.

\section{Acknowledgments}
Our teachers, students, collaborators and colleagues with whom we discussed, worked on, and argued over black holes and their properties are gratefully acknowledged. We thank Niayesh Afshordi, Valentina Baccetti, Jacob Bekenstein, Ivan Booth, Pisin Chen, Wan Cong, Jolien Creighton, Pravin Dahal, Orsola DiMarco, Josh Foo, Luis Garay, Laura Henderson, Robie Hennigar, Sabine Hossenfelder, Viqar Husain, Bernard Kay, Eleni Kontou, David Kubiz\v{n}\'{a}k, Adam Levi, Etera Livine, Jorma Louko, Swayamsiddha Maharana, Eduardo Mart\'{\i}n-Mart\'{\i}nez, Keith Ng, Amos Ori, Paddy Padmanabhan, Don Page, Asher Peres, Matthew Robbins, Jos\'{e} Senovilla, Joe Schindler, Fil Simovic, Alexander Smith, Lee Smolin, Ioannis Soranidis, Erickson Tjoa, Bill Unruh, Matt Visser, Shao Wen Wei, Mark Wardle, Yuki Yokokura, Dieter Zeh, Chen Zhang, Jialin Zhang, and Magdalena Zych. We also thank several anonymous referees that have kept us on the straight and narrow.

RBM was supported in part by the Natural Sciences and Engineering Research Council of Canada and by the Asian Office of Aerospace Research and Development Grant No.\ FA2386-19-1-4077. SM was supported by an International Macquarie University Research Excellence Scholarship and a Sydney Quantum Academy Scholarship. The work of DRT was in part supported by the ARC Discovery project grant DP210101279, the US Air Force AOARD grants FA2386-1-7-4015 and FA2386‐20‐1‐4016, and the Southern University of Science and Technology, Shenzhen, Guandong, P.\ R.\ China.

\appendix

\section{Classical Results} \label{app:classRes}

\subsection{Horizons} \label{a:hor}
Here we collect definitions of various horizons and list some of their properties. There is  a certain amount of terminological ambiguity that may cause confusion, as the same or similar names are used for related, but subtly different concepts\cite{S:11,vF:15,B:16,B:05}. We describe the nomenclature that is used in this review and comment on the naming discrepancies between the sources we cite.

A MBH $\mb$ can be defined  without making assumptions about the structure of infinity \cite{C-B:09} as (a single connected component of) the complement of the past of the set covered by the null geodesics that have an infinite future canonical (affine) parameter. The \textit{(future) event horizon} $\mh$ (or $\mh^+$) is the boundary $\pad\mb$ of the MBH. This is an observer-independent concept, and for stationary black holes the horizon area $A$ (area of the spacelike two-surface $\mh\cap\Sigma$, where $\Sigma$ is an observer-dependent spacelike hypersurface) is invariant \cite{HE:73,FN:98,P:04}. The two-surface $\mh\cap\Sigma$ is also often referred to as the event horizon \cite{vF:15}.

The definition of a PBH involves several concepts. All quasi-local notions of a horizon use the local concept of a marginally trapped surface \cite{B:16,S:11}. A two-dimensional spacelike submanifold $\ms$ (that is taken to be closed\cite{HE:73,P:68} to eliminate trivial examples) allows one to decompose the tangent space at any spacetime point of $\eM$ into a two-dimensional space of vectors tangent to $\ms$, and a two-dimensional transverse space, which is a $1+1$ Minkowski space that can be spanned by two null vectors $l$ and $n$, $l \cdot n=-1$. The vectors are taken to be outward and inward pointing, respectively, as well as future-directed. The mean curvature vector $K$ is then given by
\begin{align}
	K=-\vartheta_{(n)}l-\vartheta_{(l)}n ,
\end{align}
where $\vartheta_{(n)}$ and $\vartheta_{(l)}$ are the expansions of the inward-pointing and outward-pointing vector fields on $\ms$.

On a \textit{future trapped surface} both expansions are negative and $K$ is timelike future-directed. On a \textit{marginally future trapped surface} (MTS) $\sH'_\Sigma$  $\vartheta_{(l)}=0$ \textit{and} $\vartheta_{(n)}<0$, making $K$ future-directed and null. As there are examples with non-negative expansion of the field $n$, a \textit{marginally outer trapped surface} (MOTS) $\sH_\Sigma$ is considered \cite{K:14}. It is defined solely by requiring $\vartheta_{(l)}=0$. It is \textit{stable} if it becomes untrapped when deformed outward, and an inward deformation results in $\vartheta_{(l)}<0$.

The (4D) \textit{trapped region} $\mt$ consists of spacetime points that lie on closed future trapped surfaces. Its boundary $\pad\mt$ is called the \textit{trapping boundary}. The (3D) trapped region\cite{HE:73,K:14} $\mt^\Sigma$ is the set of points on a spacelike surface $\Sigma$ that lie on closed future-trapped surfaces contained entirely in $\Sigma$. The intersection $\mt\cap\Sigma$ ($\Sigma$ is usually a Cauchy surface) results in what is colloquially referred to as the interior of a PBH at the given instant. Each connected component of the boundary of $\mt^\Sigma$ (labeled as $\pad\mt^\Sigma\defeq\pad(\mt\cap\Sigma)$) that is contained in the intersection of $\pad\mt\cap\Sigma$ is called a (2D) \textit{apparent horizon} \cite{HE:73,K:14}. The outermost such boundary on $\Sigma$ is the outer (2D) apparent horizon. In principle this object (generically a two-surface) may not correspond to any MOTS that lies in the slice $\Sigma$, but in fact it is the unique outermost MOTS \cite{A:09}.

A dynamical \textit{(3D) apparent horizon} $\sH$ (we use an adjective to distinguish this (3D) entity from the (2D) apparent horizon $\sH_\Sigma$) can be thought of as an evolving (2D) apparent horizon that, however, is taken either as a MTS $\sH'_\Sigma$ (Ref.~\citenum{vF:15}) or as a MOTS $\sH_\Sigma$ (Ref.~\citenum{K:14}). In the former case, the dynamical apparent horizon $\sH$ is a \textit{marginally trapped tube} (MTT) \cite{AK:04,K:14}, and in the latter a \textit{marginally outer trapped tube} (MOTT) \cite{S:11,vF:15}. More formally, a MOTT is the closure of the hypersurface that is foliated by MOTSs on the smoothly evolving family of hypersurfaces $\Sigma_T$, where $T$ is a suitable evolution parameter.

Not imposing the requirement $\vartheta_{(n)}<0$ results in the MOTS $\sH_\Sigma$ that we take as the apparent horizon on a time slice $\Sigma$, and thus by a (3D) apparent horizon we mean a MOTT. It is often referred to simply as ``apparent horizon'' \cite{B:16, H:94,vF:15}, and we follow this usage when it does not lead to confusion. It should be noted that in general $\pad\mt$ is not a MTT \cite{S:11,BS:10}.

The location of the apparent horizon $\sH$ is computed via well-defined strategies in numerical relativity \cite{T:07,BS:10,RZ:13}. Unlike the invariantly defined $\mt$ and $\pad\mt$, both $\sH_\Sigma$ and $\sH$ depend on the foliations \cite{S:11,vF:15,FEFHM:17,K:14}. {In particular\cite{WI:91}, even in Schwarzschild spacetimes there are families of non-symmetric hypersurfaces where parts of the hypersurface $\Sigma$ approach the future singularity arbitrarily closely, but a part of it is still outside of the MBH. As a result, the intersection $\Sigma\cap\mh$ is not a complete sphere, and thus there are no marginally trapped surfaces on $\Sigma$.

Numerical simulations also indicate that $\sH_{\Sigma_T}$ may evolve discontinuously; indeed the evolution of MOTSs for binary black hole mergers can be quite complicated \cite{Pook-Kolb:2021gsh}. Starting from the same $\Sigma_0$ and the same MOTS $\sH_{\Sigma_0}$, but with slightly different subsequent slices, $\Sigma_{T'}$ leads (under certain conditions) to a smoothly evolving MOTT $\sH'$ that, however, does not necessarily coincide with $\sH$. Thus the jumps \cite{A:09,BS:10} that are observed in numerical simulations ``are because of the outermost condition: while an individual MTT continues to evolve smoothly, a new MTT can appear further outward'' \cite{K:14}\footnote{Note that in the terminology of Ref.~\citenum{K:14}, a hypersurface foliated by MOTSs is called MTT.}. It should be noted that many smoothness results \cite{vF:15,K:14}, as well as the key property that the apparent horizon is contained inside the event horizon \cite{HE:73,RZ:13}, depend on the matter satisfying at least the NEC.

There are a number of additional useful horizon notions, such as  trapping horizons \cite{H:94,NV:06,B:05} and dynamical horizons \cite{AK:04} (or isolated horizons in the equilibrium case). For practical purposes these are apparent horizons $\sH$ that are subject to additional restrictions \cite{vF:15}.

While the trapped region $\mt$ is invariantly defined, its identification is a non-trivial task even in simple spherically symmetric models, and also raises some conceptual issues \cite{B:16,S:11,BS:11}. Therefore, we define a PBH $\sB$ as a generally foliation-dependent (4D) domain that is enclosed by a (3D) MOTT $\sH$.  A coordinate-independent distinction between inner and outer horizons (for the subtle distinction in foliations by MTS and MOTS see Refs.~\refcite{vF:15} and \refcite{B:05}) identifies the \textit{outer} (\textit{inner}) trapping horizons via
\begin{align}
	\mathcal{L}_n\vartheta_{(l)}=n^\mu\pad_\mu\vartheta_{(l)}<0, \qquad \mathcal{L}_n\vartheta_{(l)}>0, \label{in-out}
\end{align}
respectively.

Several other horizons are important in our discussion. A Killing horizon \cite{HE:73,FN:98,P:04,vF:15} $\mh_\vK$ is a null hypersurface that is everywhere tangent to a Killing vector field $\vK$ that becomes null on it. This Killing vector field is timelike, $\vK^\mu\vK_\mu<0$, in a spacetime region outside of $\mh_\vK$, and spacelike inside of it. Stationary event horizons in GR are usually Killing horizons for suitably chosen Killing vectors \cite{HE:73,FN:98,vF:15}.  For example, a non-static asymptotically flat stationary spacetime must be axisymmetric. An event horizon is a Killing horizon for the Killing field
\begin{align}
	\vK_\phi=\pad_t+\omega_\mh\pad_\phi,
\end{align}
where $\omega_\mh$ is the angular velocity at the horizon (this result requires the predictability of spacetime from a particular Cauchy surface, hyperbolic equations of motion for matter fields that satisfy the dominant energy condition \cite{HE:73}).

A bifurcate Killing horizon consists of a pair of Killing horizons,  both with respect to the same Killing field, that intersect on a spacelike two-surface. This bifurcation surface plays an important role in black hole thermodynamics and in derivations of Unruh and Hawking radiation \cite{BD:82,FN:98,W:01,PT:09}.

A Cauchy horizon\cite{HE:73,C-B:09,W:20} is the boundary of the domain of dependence of an achronal surface (a surface for which no pair of its points can be joined by a timelike curve). More precisely, the domain of dependence $\eD_\Sigma$ of the (co-dimension one) hypersurface $\Sigma$ consists of all points $p$ for which every inextendible causal curve drawn through them meets $\Sigma$. This hypersurface divides $\eD_\Sigma$ into the future $\eD^+_\Sigma$ and past $\eD^-_\Sigma$ domains of dependence. The boundary of the closure of $\eD_\Sigma$ (or $\eD^\pm_\Sigma$)  is the Cauchy horizon $\eH_C$, or its future/past version. The domain of dependence $\eD_\Sigma$ is the largest spacetime region in which deterministic physics can be predicted from the knowledge of initial conditions on $\Sigma$.

White holes can be thought of as time-reversed black holes \cite{FN:98,C-B:09,K:14,vF:15}. For the Schwarzschild solution that is described using the outgoing Eddington--Finkelstein coordinate $u=t-r_*$, it is the spacetime domain $r<r_\sg$ that corresponds to region IV in Fig.~\ref{fig:SPen}. In general, the relevant definitions mimic those of black holes up to the sign reversal. Thus on an \textit{anti-trapped surface}\cite{A:20} both $\vartheta_{(n)}$ and $\vartheta_{(l)}$ [cf.\ Eq.~\eqref{eq:ur-tangent-expansions}] are positive, and on a \textit{marginally outer anti-trapped surface}  $\vartheta_{(n)} = 0$. The (4D) \textit{anti-trapped region} consists of spacetime points that lie on closed anti-trapped surfaces, and it is possible to define various anti-trapping horizons \cite{vF:15} (such as the past inner trapping horizon).

\begin{figure}[!htbp] \centering
	\includegraphics[width=0.7\textwidth]{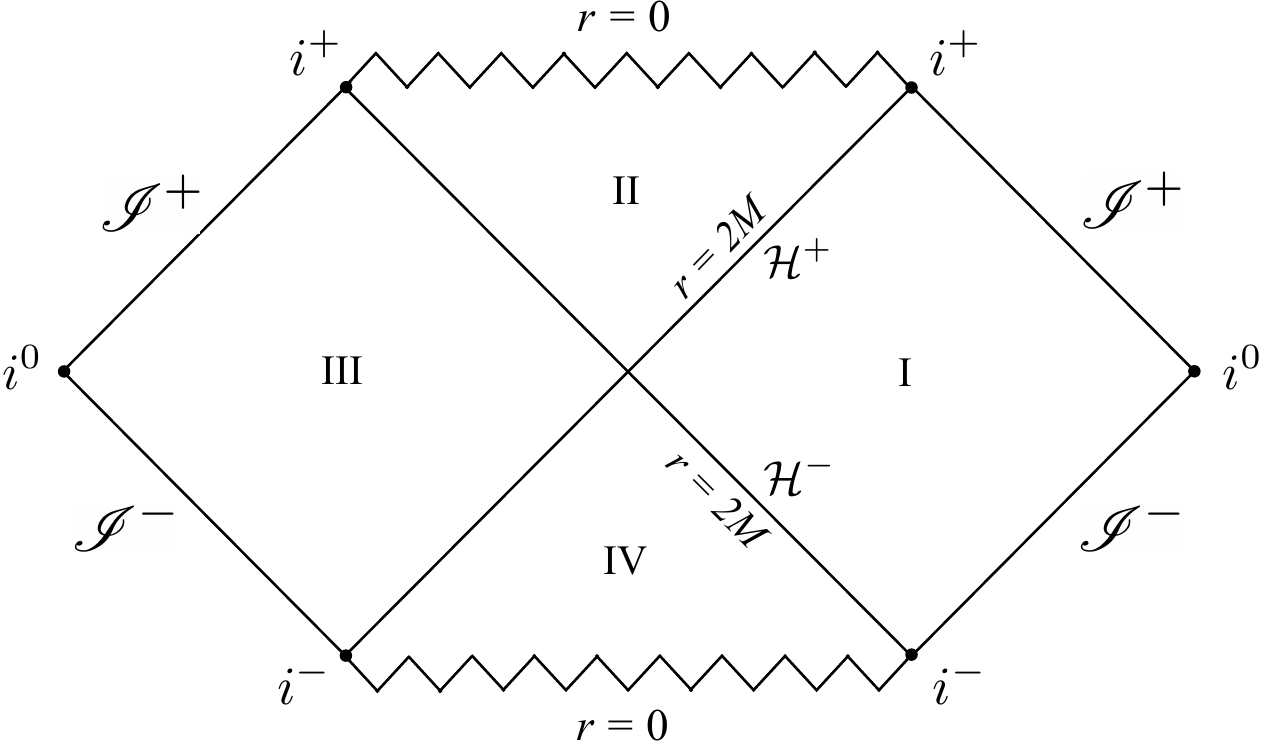}
	\caption{Carter--Penrose diagram of the maximally extended Schwarzschild solution. The spacetime regions \RNum{1} and \RNum{2} correspond to the black hole, and \RNum{3} and \RNum{4} to the white hole. Both black and white hole singularities are spacelike. No particle coming from spacelike infinity or past null infinity can penetrate the white hole, and no particle can leave the black hole.}
	\label{fig:SPen}
\end{figure}

\subsection{Singularities} \label{singu}
An intuitive view of real singularities (as opposed to coordinate singularities or singularities of congruences of curves) is that of spacetime locations where matter and energy densities take on arbitrarily high values, and arbitrarily strong tidal forces crush any object. The Einstein equations and the geodesic deviation equation translate this intuition into the divergence of components of curvature tensors and curvature scalars. However, this concept requires several technical refinements before it can be usefully employed.\cite{P:68,HE:73,TCE:80,C-B:09,J:14}

Spacetime is modeled by a sufficiently well-behaved manifold $\eM$ with the metric satisfying some minimal smoothness conditions.\cite{HE:73,TCE:80,C-B:09} Hence the singular points are excised from the spacetime manifold and ought to be treated as boundary points attached to it \cite{J:14}. In turn, one deduces the existence of a singularity from the behavior of the points contained within the manifold. To uncover a curvature singularity, we can follow, e.g., the evolution of certain curvature scalars along various families of curves. This picture is complicated by the existence of non-curvature singularities. Geodesics  that are  incomplete in the original spacetime indicate that the spacetime is possibly extendible, and if so possibly as a non-globally hyperbolic spacetime \cite{C-B:09}. The simplest example is the Schwarzschild spacetime, in which the range of the radial coordinate is restricted to $r>2M$.

Finally, because the spacetime metric has signature $(-,+,+,+)$, the simplest criterion for locating singular points is geodesic (in)completeness. Unlike the spaces with positive-definite metric, where the Cauchy sequences or a distance function can be used to uncover the missing limit points, the existence of null rays with zero arc length necessitates a more complicated procedure \cite{TCE:80}. We begin by listing the relevant definitions.

A manifold $\eM$ with a metric $\sg$ is \textit{extendible} if it can be imbedded in a manifold $\eM'$ with a metric $\sg'$ that coincides with $\sg$ on the image of $\eM$. Otherwise the manifold is \textit{inextendible}\cite{HE:73}. The Schwarzschild spacetime is extendible, with the corresponding Kruskal extension being the maximal (i.e.\ inextendible) \cite{HE:73,FN:98,C-B:09} manifold $\eM'$.

For an \textit{inextendible curve} $\gamma\in \eM$ there is no smooth path $\gamma'$ for which $\gamma$ is its proper subset. In more detail, a point $p$ is a \textit{future endpoint} of a future-directed non-spacelike (also known as causal) curve $\gamma(\sigma)\in\eM$ if for every neighborhood $\eV$ of $p$ there is a parameter $\sigma$ such that $\gamma(\sigma_1)\in\eV$ for every $\sigma_1 \geqslant \sigma$. A causal curve is \textit{future-inextendible} (respectively, \textit{future-inextendible in a set} $\eS$) if it has no future endpoint (respectively, no future endpoint in $\eS$).\cite{HE:73} Thus the endpoint is associated with the value $\sigma_+$ of the generalized affine parameter, with, e.g., the future-inextendible curve corresponding to the range $\sigma\in [0,\sigma_+)$.

It is generally agreed that a singular spacetime is one that has incomplete causal geodesics. This \textit{geodesic incompleteness} means that some of its inextendible timelike or null geodesics, future or past directed, have a finite proper length or a finite canonical parameter \cite{HE:73,C-B:09,J:14,W:20}. (A refinement \cite{L:21,SG:15}, introduced to eliminate trivial examples of incompleteness, also requires the absence of the spacetime manifold extension $\eM\to\eM'$ that produces an extendible image $\gamma'\in\eM$).

This definition may not cover all types of possible singular behavior, but it does cover an obvious pathological behavior where a timelike observer or a photon can disappear from the spacetime after a finite amount of generalized affine parameter \cite{J:14}. It is the basis of various singularity theorems \cite{HE:73,C-B:09,SG:15,W:20,S:21}. The theorems themselves do not establish the existence of physically relevant curvature singularites \cite{J:14,L:21}. Moreover, a curvature singularity does not imply geodesic incompleteness \cite{C-B:09,CK:85}. Nevertheless, coupled with additional input, such as various forms of the cosmic censorship conjecture, the end results are physically relevant singularities \cite{J:14,SG:15,W:20}.

The singularity theorems follow a generic pattern \cite{SG:15,S:21}. It is assumed that a spacetime of sufficient differentiability satisfies (i) a condition on the curvature; (ii) a causality condition; (iii) an appropriate initial and/or boundary condition. Then there are null or timelike inextensible incomplete geodesics. The curvature condition (i) is translated via the Einstein equations into one of the energy conditions. An important research direction is the establishment of singularities even if energy conditions, including the NEC, are violated \cite{KS:20}.

After describing the general framework, we focus on varieties of curvature singularities and the relationships between them. A distinction between the so-called p.p.\ and s.p.\ curvature singularities is particularly useful \cite{HE:73,TCE:80}.

A \textit{scalar polynomial (s.p.) singularity} is the end point of at least one causal curve on which at least one of the scalar polynomials that are built from the metric and the Riemann tensor is unbounded. If some components of the Riemann tensor or its contractions diverge in some orthonormal basis that is parallelly propagated along a causal curve, then its end point is a \textit{p.p.\ curvature singularity}. An s.p.\ curvature singularity implies a p.p.\ curvature singularity. However, the divergence of any of the components of $\tensor{\sg}{_\mu_\nu}$ is not required. For example \cite{O:00}, a conformally flat metric
\begin{align}
	ds^2=e^\zeta(-dt^2+dx^2+dy^2+dz^2), \qquad \zeta=t\big(e^{-x^2/t^4}-e^{-x^4/t^4}\big),
\end{align}
is everywhere nondegenerate and continuous, but $\eR=6/t^3$ as $x\to 0$.

The behavior of the components of $R^\lambda_{~\mu\nu\varsigma}$ allows the introduction of more refined categories of p.p.\ singularities \cite{EK:74}. If  \textit{all} of the components of $R^\lambda_{~\mu\nu\varsigma}$ tend to finite limits as $\sigma\to\sigma_+$ for \textit{some} orthonormal basis which is parallelly propagated along $\gamma(\sigma)$, the singularity is defined as \textit{a locally extendible} or \textit{quasi-regular} singularity. In this case there always exists an open neighborhood of $\gamma(\sigma)$ that can be extended in such a way that this curve can be continued beyond $p$ in this local extension \cite{TCE:80}. If \textit{some} component $R^\lambda_{~\mu\nu\varsigma}$ with respect to a parallelly propagated basis does not converge to a finite limit as $\sigma\to\sigma_+$, but there is \textit{some other} orthonormal basis along $\gamma(\sigma)$ in which they all tend to finite limits, then the singularity
is an \textit{intermediate singularity}.

The definition of a \textit{curvature singularity} is similarly tied to the behavior of the Riemann tensor components \cite{EK:74} as the absence of \textit{any} orthonormal basis along $\gamma(\sigma)$ such that \textit{all} of the components of $R^\lambda_{~\mu\nu\varsigma}$ tend to finite limits as $\sigma\to\sigma_+$. However, the s.p.\ singularities are precisely those for which the Riemann tensor components are unbounded in all tetrads (not just parallel propagated ones) \cite{TCE:80}.

Another singularity defintion captures its meaning as a locus of total destruction\cite{TCE:80,O:00}. A causal geodesic $\gamma(\sigma)$ terminates in a \textit{strong curvature singularity} at $\sigma_+$ if  a 3-form (for a timelike case) or a 2-form (for a null case) $\omega(\sigma)$ on the normal space to $\gamma(\sigma)$ determined by  linearly independent vorticity-free Jacobi fields approaches zero volume,
\begin{align}
	\lim_{\sigma\to\sigma_+}\|\omega(\sigma)\|=0.
\end{align}
To produce a strong curvature singularity, some component of the Riemann tensor must diverge. Belonging to the s.p.\ class is not required, even if the explicit examples have divergent curvature scalars \cite{TCE:80,O:00}. Both necessary and sufficient conditions for its existence involve not only the divergence of tetrad components of the Riemann, Ricci, or Weyl tensors, but also the divergence of their integrals along causal geodesics terminating in the singularity \cite{CK:85}.

\subsection{Useful formulas for spherically symmetric spacetimes} \label{AForm}
Here we summarize some useful expressions in the case of spherical symmetry. A useful collection of results in axial symmetry can be found, e.g., in Refs.~\refcite{FN:98} and \refcite{C:92}.

The EMT components of Eq.~\eqref{tspher} and the effective components $\tau_a$ are related by
\begin{align}
	\rho = \tensor{\tau}{_t}/f, \qquad p = \tensor{\tau}{^r}/f, \qquad \psi = \tensor{\tau}{_t^r}/f.
\end{align}
In the Schwarzschild metric the physical distance in the vicinity of the event horizon is given by
\begin{align}
	\ell_\rS(x)=\int_0^{x(r)}\frac{dx}{\sqrt{f(x)}}= 2\sqrt{r_\sg x}+\cO(x^{3/2}),
\end{align}
where $r_\sg=2M$.

In $(v,r)$ coordinates outside of the apparent horizon, the relationship between four-velocity components of the timelike trajectory is
\begin{align}
	\dot{V}=\frac{\dot R+\sqrt{\dot R^2+F}}{e^H F}\label{vdin}
\end{align}
for both ingoing ($\dot R<0$) and outgoing ($\dot R>0$) test particles, where $F=f\big(V(\tau),R(\tau)\big)$, and $H=h_+\big(V(\tau),R(\tau)\big)$. On the other hand, inside of the trapped region $f<0$ and thus to maintain the timelike character of the trajectory
\begin{align}
	\dot R\leqslant-\sqrt{ -F}.
\end{align}
Velocity components of ingoing particles still satisfy Eq.~\eqref{vdin}, with the ingoing null geodesics $\dot V=0$ being their ultra-relativistic limit, while the outgoing particles satisfy
\begin{align}
	\dot{V}=\frac{\dot R-\sqrt{\dot R^2+F}}{e^H F}>0.
\end{align}
The limiting values of the EMT components in $(v,r)$ coordinates, $\theta^+_{\mu\nu} \defeq \lim_{r\to r_+} \theta_{\mu\nu}$, are
\begin{align}
	& \theta_v^+= (1-w_1)\frac{r_+'}{8\pi r_+^2}, \label{the1} \\
  	& \theta_{vr}^+=- \frac{w_1}{8\pi r_+^2}, \label{the2} \\
  	& \theta_r^+=\frac{\chi_1}{4\pi r_+}. \label{the3}
\end{align}

In $(u,r)$ coordinates outside of the anti-trapping horizon, the relationship between the components of the four-velocity is
\begin{align}
	\dot{U}=\frac{-\dot R+\sqrt{\dot R^2+F}}{e^H F}\label{udout}
\end{align}
for both ingoing ($\dot R<0$) and outgoing ($\dot R>0$) particles, where $F=f\big(U(\tau),R(\tau)\big)$, and $H=h_-\big(U(\tau),R(\tau)\big)$. On the other hand, inside of the anti-trapped region $f<0$ and thus to maintain the timelike character of the trajectory
\begin{align}
	\dot R\geqslant\sqrt{ -F}
\end{align}
must hold. Velocity components of outgoing particles still satisfy Eq.~\eqref{udin}, with the outgoing null geodesics $\dot U=0$ being their ultra-relativistic limit, while the ingoing particles satisfy
\begin{align}
	\dot{U}=-\frac{\dot R+\sqrt{\dot R^2+F}}{e^H F} . \label{udin}
\end{align}
The EMT components are given by
\begin{align}
	&	\theta_u \defeq e^{-2h_-} \Theta_{uu} = \tensor{\tau}{_t} , \label{eq:theu} \\
	&	\theta_{ur} \defeq e^{-h_+} \Theta_{ur} = \left( \tensor{\tau}{_t^r} + \tensor{\tau}{_t} \right) / f , \label{eq:theur} \\
	&	\theta_r \defeq \Theta_{rr} = \left( \tensor{\tau}{^r} + \tensor{\tau}{_t} + 2 \tensor{\tau}{_t^r} \right) / f^2 . \label{eq:therru}
\end{align}
 The Einstein equations are
\begin{align}
	&- e^{-h_-} \pad_u C_- + f \pad_r C_- = 8 \pi r^2 \tensor{\theta}{_u} , \label{eq:Guu}\\
	& \pad_r C_- = - 8 \pi r^2 \tensor{\theta}{_u_r} , \\
	& \pad_r h_- = 4 \pi r \tensor{\theta}{_r} . \label{Grr-u}
\end{align}
The limiting values of the EMT components in $(u,r)$ coordinates, $\theta^-_{\mu\nu} \defeq \lim_{r\to r_-} \theta_{\mu\nu}$, are
\begin{align}
	& \theta_u^-= -(1-w_1)\frac{r_-'}{8\pi r_-^2}, \label{the1u} \\
  	& \theta_{ur}^-= \frac{w_1}{8\pi r_-^2}, \label{the2u} \\
  	& \theta_r^-=\frac{\chi_1}{4\pi r_-}. \label{theu3}
\end{align}
In these coordinates, a convenient choice for the tangents to the congruences of ingoing and outgoing radial null geodesics is
\begin{align}
	n^\mu = (1,-\tfrac{1}{2} f(u,r)e^{h_-(u,r)},0,0), \qquad l^\mu = (0,e^{-h_-(u,r)},0,0),
\end{align}
respectively. The tangents are normalized by $n \cdot l =-1$, and their expansions are given by
\begin{align}
	\vartheta_{(n)} = - \frac{e^{h_-}f}{r}, \qquad \vartheta_{(l)}=\frac{2e^{-h_-}}{r} .
	\label{eq:ur-tangent-expansions}
\end{align}

\subsection{Newman--Penrose tetrad} \label{a:nptet}
The Newman--Penrose null tetrad \cite{C:92,FN:98,exact:03,K:14} is formed by two real null vectors $l^\mu$, $n^\mu$ that satisfy $l\cdot n=-1$, and a pair of complex-conjugate null vectors $m^\mu$, $\bar m^\mu\defeq m^{\mu*}$ that are orthogonal to the real pair and satisfy $m\cdot \bar{m}=1$. It is particularly well-suited for studies of null surfaces. In this basis, the metric is expressed as
\begin{align}
	\tensor{\sg}{_\mu_\nu} = - l_{(\mu}n_{\nu)} + m_{(\mu}\bar m_{\nu)}.
\end{align}
Four classes of transformations preserve these properties of the null tetrad. They form a representation of the proper orthochronous Lorentz group and are parameterized by two real parameters $A$ and $\psi$, as well as two complex parameters $\alpha$ and $\beta$. These are (i) boosts
\begin{align}
	l\to Al, \qquad n\to A^{-1}n, \qquad m\to m,
\end{align}
(ii) spin rotations in the $(m\bar m)$ plane,
\begin{align}
	m\to e^{i\psi}m, \qquad l\to l, \qquad n\to n,
\end{align}
(iii) null rotations around $l$ (type \RNum{1} rotations),
\begin{align}
	l\to l, \qquad m\to m+\alpha l, \qquad n\to n+\alpha^*m+\alpha\bar m+|\alpha|^2l,
\end{align}
and  (iv) null rotations around $n$ (type \RNum{2} rotations)
\begin{align}
	n\to n, \qquad m\to \beta m, \qquad l\to l+\beta^*m+\beta\bar m+|\beta|^2 n.
\end{align}

The ten independent components of the Weyl tensor $C^\alpha_{~\beta\gamma\delta}$ are expressed with the help of five complex   scalars
\begin{align}
	\Psi_0&\defeq C_{\alpha\beta\gamma\delta}l^\alpha m^\beta l^\gamma m^\delta, \qquad \Psi_1\defeq C_{\alpha\beta\gamma\delta}l^\alpha m^\beta l^\gamma m^\delta, \\
	\Psi_2&\defeq C_{\alpha\beta\gamma\delta}l^\alpha m^\beta \bar{m}^\gamma n^\delta=C_{\alpha\beta\gamma\delta}l^\alpha n^\beta(l^\gamma n^\delta-m^\gamma m^\delta),\\
	\Psi_3&\defeq C_{\alpha\beta\gamma\delta}l^\alpha n^\beta \bar{m}^\gamma n^\delta, \qquad \Psi_4\defeq C_{\alpha\beta\gamma\delta}\bar{m}^\alpha n^\beta \bar{m}^\gamma n^\delta.
\end{align}
Similarly, the components of the Ricci tensor are recovered from one curvature scalar
\begin{align}
	\Lambda\defeq\eR/24,
\end{align}
and three real
\begin{align}
\Phi_{00}&\defeq \half R_{\mu\nu}l^\mu l^\nu,   \\
\Phi_{11}&\defeq \tfrac{1}{4}R_{\mu\nu}(l^\mu n^\nu+m^\mu\bar{m}^\nu),\\
\Phi_{22}&\defeq \half R_{\mu\nu}n^\mu n^\nu,
\end{align}
and three complex scalars
\begin{align}
\Phi_{01}&\defeq \half R_{\mu\nu}l^\mu m^\nu,   \\
\Phi_{02}&\defeq \half R_{\mu\nu}m^\mu m^\nu,   \\
\Phi_{12}&\defeq \half R_{\mu\nu}m^\mu n^\nu,
\end{align}
where $\bar{\Phi}_{ij}\defeq \Phi_{ij}^*=\Phi_{ji}$.

Spherical symmetry leads to considerable simplifications. We use $(v,r)$ coordinates, $(l\equiv l_{\mathrm{out}},n\equiv l_\rin,m,\bar{m})$ that are given in Eq.~\eqref{null-v}, and the pair of complex-conjugate null vectors $m^\mu$, $\bar m^\mu=m^{\mu*}$ is given by
\begin{align}
	m=\frac{1}{\sqrt{2}r}\pad_\theta+\frac{i}{\sqrt{2}r\sin\theta}\pad_\phi, \qquad m\cdot \bar m=1. \label{mmbar}
\end{align}
In this basis, in addition to the Ricci scalar
\begin{align}
	\begin{aligned}
	\eR &= \frac{2 \pad_r C_+ + C_+ \pad_r h_+}{r^2} + \frac{\pad^2_r C_+-4\pad_r h_++3\pad_r C_+\pad_r h_+}{r} \\
	& \qquad - 2 \big( f (\pad_r^2 h_+ + (\pad_r h_+)^2 ) + e^{-h_+} \pad_r \pad_v h_+ \big) ,
 	\end{aligned}
\end{align}
the quantities
\begin{align}
	\Phi_{00} &= \frac{e^{h_+} \pad_v C_+}{2r^2} + \frac{e^{2h_+} f^2 \pad_r  h_+ }{4 r} , \\
	\begin{split}
		\Phi_{11} &= \frac{2 \pad_r C_+ + 3 C_+\pad_r h_+}{16 r^2} - \frac{\pad_r^2 C_+ - 3 \pad_r C_+ \pad_r h_+}{16 r} \\
		& \qquad + \frac{1}{8} e^{-h_+} \pad_r \pad_v h_+ +\frac{1}{8} f \big( \pad_r^2 h_+ + (\pad_r h_+)^2 \big) ,
	\end{split}
	\\
	\Phi_{22} &= \frac{e^{-2h_+} \pad_r h_+}{r} ,
\end{align}
are the only independent non-zero Ricci tensor components. The only non-zero Weyl tensor component is
\begin{align}
	\begin{aligned}
		\Psi_2 =& -\frac{C_+}{2 r^3} +\frac{4\pad_r C_++3C_+\pad_r h_+}{12r^2}
 -\frac{\pad_r^2 C_++\pad_r h_+(3\pad_r C_++2f)}{12r} \\ 
 		& +\frac{1}{6}e^{-h_+}\pad_r\pad_v h_++\frac{1}{6}f\big(\pad_r^2 h_++(\pad_r h_+)^2\big).
	\end{aligned}
\end{align}
Additional explicit expressions can be found in Ref.~\citenum{CM:18}.

\section{Semiclassical Results} \label{a:Haw}
\subsection{Emission formulas} \label{Haw-em}
The standard results on black hole evaporation are collected in Refs.~\citenum{FN:98,P:05} and \refcite{PT:09}. For a Kerr--Newman black hole of mass $M$, charge $Q$ and angular momentum $a$, the event horizon radius, horizon area, surface gravity, angular velocity of the horizon, and the electrostatic potential are
\begin{align}
	&r_\mathrm{eh}=M+\sqrt{M^2-a^2-Q^2}, \\
	&A=4\pi(r_\mathrm{eh}^2+a^2),\\
	&\kappa= \frac{(r_\mathrm{eh}-M)}{(r_\mathrm{eh}^2+a^2)},\\
	&\omega_\mh=\frac{a}{(r_\mathrm{eh}^2+a^2)},\\
    &\varphi=\frac{4\pi Q r_\mathrm{eh}}{A},
\end{align}
respectively, and the Hawking temperature is
\begin{align}
	T_\rH=\frac{\kappa}{2\pi}.
\end{align}
For a Schwarzschild black hole, assuming emission according to the Stefan--Boltzmann law at the Hawking temperature with $\mathcal{N}$ degrees of freedom of massless particles,
\begin{align}
	\frac{dM}{dt}=-\frac{1}{15360 \pi}\left(\frac{m_\rP}{M}\right)^2\left(\frac{m_\rP}{t_\rP}\right)\mathcal{N}.
\end{align}
After restoring the constants, the Hawking temperature is given by
\begin{align}
	T_\rH=\frac{\hbar c^3}{8\pi Gk_{\rB}M}\sim 6\times 10^{-8}\frac{M_\odot}{M} {} \textrm{K},
\end{align}
and the evaporation rate is expressed as
\begin{align}\label{evaprate}
	-\frac{dM}{dt}=\frac{k_\rB^4}{\hbar^3c^2}T^4\sigma_\mathrm{eff}\sim 10^{-30}\mathcal{N}\frac{M_\odot^2}{M^2} \mathrm{J/s} ,
\end{align}
where the effective cross-section is $\mathcal{N} G^2M^2/c^4$.

More detailed emission formulas are given in, e.g., Ref.~\citenum{P:05}. The black hole emission rate is in principle observable by gravitational wave detectors. While current bounds are quite weak, they are expected to tighten as the sensitivity of the detectors improves. Refs.~\citenum{PYB:21} and \citenum{YBC:21} describe the underlying (4 PPN order) physics, bounds, and perspectives on their improvements.  Evidence for  stimulated Hawking radiation was discovered in the analysis of GW190521 data\cite{AMA:22}.

\subsection{Vacuum types} \label{vac3}
Referring to Fig.~\ref{fig:SPen}, there are three distinguished vacuum states that are invariant under the timelike Killing vector $\vK$ outside of a static (eternal) black hole \cite{BD:82,FN:98,PT:09}:
\begin{enumerate}
    \item Boulware vacuum $\ket{0_B}$: this state is defined in Region I and has modes that are positive and negative frequency with respect to the timelike Killing field $\partial_t$ (restriction of $\vK$ to Region I). It is a state with no radiation, and is
 considered unphysical as it is not regular on both future and past horizons $\mathcal{H}^\pm$.  In Rindler space, it
corresponds to the Rindler vacuum.
 \item Hartle--Hawking--Israel (HHI) vacuum $\ket{0_H}$: this state is defined on the full Kruskal--Szekeres extension and has modes that are positive frequency with respect to both past and future horizon generators $\partial_U$ and $\partial_V$. This is a state representing a black hole in thermal equilibrium with a radiation bath, such that the restriction of the state to Region I is KMS at the Hawking temperature $T_\mathrm{H} = (8\pi M)^{-1}$. Note that $T_H$ is the temperature measured by an observer at infinity (see Section~\ref{ss:atmo}).
    \item Unruh vacuum $\ket{0_U}$: this state is defined in Region I and II and has modes that are positive frequency on the Cauchy surface $\Sigma=\mathscr{I}^-\cup \mathcal{H}^-$, the union of past null infinity and the past horizon. Physically, it is a state in which the black hole radiates at temperature $T$, but its surroundings have zero temperature. As such, it corresponds to the vacuum state of an astrophysical
black hole formed by gravitational collapse. The positive frequency modes on the past horizon $\mathcal{H}^-$ are obtained with respect to the null generator $\partial_U$ of $\mathcal{H}^-$  ($U$ being the null affine parameter along $\mathcal{H}^-$); the positive frequency modes on the past null infinity $\mathscr{I}^-$ are obtained with respect to the null generator $\partial_V$ of $\mathscr{I}^-$.
\end{enumerate}
The EMTs of the Boulware and HHI vacuua are diagonal,
\begin{align}
 	\tensor{T}{_{\hat{\mu}}_{\hat{\nu}}} =
 	\begin{pmatrix}
		\rho$ $\,~ & 0$ $~ & 0$ $~ & 0$ $~ \vspace{1mm} \\
		0 & p & 0 & 0 \vspace{1mm} \\
		0 & 0 & { \mathfrak{p}} & 0 \vspace{1mm}\\
		0 & 0 & 0 & { \mathfrak{p}} \vspace{1mm}\\
	\end{pmatrix},
\end{align}
and a generic EMT of the Unruh vacuum is given by Eq.~\eqref{tspher}.

\subsection{EMT classification} \label{emt-cl}
EMTs are characterized by their Lorentz-invariant eigenvalues \cite{HE:73,C-B:09,exact:03,MV:17}. These are the eigenvalues of the matrix $\tensor{T}{^{\hat{\alpha}}_{\hat{\beta}}}$, i.e.\ the roots of the equation
\begin{align}
	\det \left( \tensor{T}{^{\hat{\alpha}}^{\hat{\beta}}} - \lambda \tensor{\eta}{^{\hat{\alpha}}^{\hat{\beta}}} \right) = 0, \qquad \tensor{\eta}{^{\hat{\alpha}}^{\hat{\beta}}} = \mathrm{diag}(-1,1,1,1).
\end{align}
While a more refined classification scheme exists \cite{exact:03}, the standard separation into four classes \cite{HE:73,MV:17} suffices for our purposes. Using an orthonormal basis, the standard matrix forms of the four different types along with their eigenvalues are given by
\begin{alignat}{2}
	\text{Type \RNum{1}:} & \qquad &&
	\tensor{T}{^{\hat{\alpha}}^{\hat{\beta}}} =
	\left(\begin{tabular}{c|ccc}
		$\varrho$ & $0$ & $0$ & $0$ \\ \hline
		$0$ & $p_1$ & $0$ & $0$ \\
		$0$ & $0$ & $p_2$ & $0$ \\
		$0$ & $0$ & $0$ & $p_3$
	\end{tabular}\right) ,
	\quad \lbrace - \varrho , p_1 , p_2 , p_3 \rbrace ,
	\\
	\text{Type \RNum{2}:} & \qquad &&
	\tensor{T}{^{\hat{\alpha}}^{\hat{\beta}}} =
	\left(\begin{tabular}{cc|cc}
		$\mu + \varphi$ & $\varphi$ & $0$ & $0$ \\
		$\varphi$ & $- \mu + \varphi$ & $0$ & $0$ \\ \hline
		$0$ & $0$ & $p_2$ & $0$ \\
		$0$ & $0$ & $0$ & $p_3$
	\end{tabular}\right) ,
	\quad \lbrace - \mu , - \mu , p_2 , p_3 \rbrace ,
	\\
	\text{Type \RNum{3}:} & \qquad &&
	\tensor{T}{^{\hat{\alpha}}^{\hat{\beta}}} =
	\left(\begin{tabular}{ccc|c}
		$\varrho$ & $0$ & $\varphi$ & $0$\\
		$0$ & $-\varrho$ & $\varphi$ & $0$\\
		$\varphi$ & $\varphi$ & $-\varrho$ & 0 \\ \hline
		$0$ & $0$ & $0$ & $p$
	\end{tabular}\right) ,
	\quad \lbrace - \varrho , - \varrho , - \varrho , p \rbrace ,
	\label{IIIf}
	\\
	 \text{Type \RNum{4}:} & \qquad &&
	\tensor{T}{^{\hat{\alpha}}^{\hat{\beta}}} =
	\left(\begin{tabular}{cc|cc}
		$\varrho$ & $\varphi$ & $0$ & $0$ \\
		$\varphi$ & $-\varrho$ & $0$ & $0$ \\ \hline
		$0$ & $0$ & $p_1$ & $0$ \\
		$0$ & $0$ & $0$ & $p_2$
	\end{tabular}\right) ,
	\quad \lbrace - \varrho + i \varphi , - \varrho - i \varphi, p_1 , p_2 \rbrace ,
\end{alignat}
with the corresponding Lorentz-invariant eigenvalues of $\tensor{T}{^{\hat{\alpha}}_{\hat{\beta}}}$ quoted in curly brackets.

Conventional classical matter belongs to types \RNum{1} and \RNum{2}. For both types, the NEC is satisfied if $\rho+p_i\geqslant 0$, and for type \RNum{2} $\varphi>0$ is required as well.

\section{Properties of Spherically Symmetric PBHs} \label{solutions-long}

\subsection{$k=0$}\label{As0}
Here we summarize some higher-order terms for the evaporating ($\tensor{\tau}{_t^r}<0$) PBH  $k=0$ solutions. A convenient way to represent the effective components of the EMT is
\begin{align}
	& \tensor{\tau}{_t} = - \Upsilon^2 + e_{12}\sqrt{x} + e_1 x + \cO(x^{3/2}) , \label{eq:k0taut} \\
	& \tensor{\tau}{_t^r} = \pm \Upsilon^2 + \phi_{12} \sqrt{x} + \phi_1 x + \cO(x^{3/2}) , \label{eq:k0tautr} \\
	& \tensor{\tau}{^r} = - \Upsilon^2 + p_{12} \sqrt{x} + p_1 x + \cO(x^{3/2}) . \label{eq:k0taur}
\end{align}
Since $f\propto \sqrt{x}$ and the rhs of Eq.~\eqref{eq:Grr} results in a finite limit, we have
\begin{align}
	\phi_{12}=\half (e_{12} + p_{12}).
\end{align}
The metric functions are then given by
\begin{align}
C &=r_\sg-4\sqrt{\pi} r_\sg^{3/2} \Upsilon \sqrt{x}+\left(\frac{1}{3}+\frac{4 \sqrt{\pi} e_{12}r_\sg^{3/2}}{3\Upsilon}\right) x + \cO(x^{3/2}),\\
h&= -\frac{1}{2}\ln\frac{x}{\xi}+\left(\frac{1}{3\sqrt{\pi}r_\sg^{3/2}\Upsilon}-\frac{e_{12} - 3 p_{12}}{6\Upsilon^2}\right)\sqrt{x}+\cO(x).
\end{align}
By substitution into Eq.~\eqref{eq:thevr}, we find that the limiting values of the effective EMT components in $(v,r)$ coordinates are given by
\begin{align}
	\theta_v^+ & = - \Upsilon^2, \\
	\theta_{vr}^+ &= \frac{p_{12} - e_{12}}{8\sqrt{\pi r_\sg}\Upsilon}, \label{thevrA} \\
	\theta_r^+ &= \frac{e_1 - 2 \phi_1 + p_1}{16\pi r_\sg\Upsilon^2}. \label{therrA}
\end{align}
The EMT in this orthonormal basis has the form
\begin{align}
	\tensor{T}{_{\hat{\alpha}}_{\hat{\beta}}} =
	\left(\begin{tabular}{cc|cc}
 		$q+\mu_1$ & $q+\mu_2$  & $0$& $0$\\
		$q+\mu_2$ & $q+\mu_3$  & $0$& $0$ \\ \hline \label{IVP}
		$0$ & $0$ & $\mathfrak{p}$ & 0 \\
		$0$ & $0$ & 0 & $\mathfrak{p}$
	 \end{tabular}\right),
\end{align}
which is different from the usual standard forms of the EMT \cite{HE:73,MV:17}. The components in the $(tr)$ block are sums of the divergent quantity
\begin{align}
	q=-\frac{\Upsilon}{4\sqrt{\pi r_\sg x}},
\end{align}
and additional finite terms. In particular, $\mathfrak{p}= \tensor{T}{_{\hat\theta}_{\hat\theta}}>0$, and
\begin{align}
\mu_1&=\frac{4\sqrt{\pi} e_{12} r_\sg^{3/2}+\Upsilon}{24\pi r_\sg^2\Upsilon} +\cO(\sqrt{x}),\\
\mu_2&=\frac{ \sqrt{\pi}r_\sg^{3/2}(e_{12} + 3 p_{12})+\Upsilon}{24\pi r_\sg^2\Upsilon}+\cO(\sqrt{x}),\\
2 \mu_2&=\mu_1+\mu_3+\cO(\sqrt{x}).
\end{align}
Two of the Lorentz-invariant eigenvalues that are important for the EMT classification are
\begin{align}
	\mathfrak{t}_{1,2} = \half \left(\mu_3-\mu_1\pm\sqrt{(\mu_1-2\mu_2+\mu_3)(\mu_1+2\mu_2+\mu_3+q)} \right) \; .
\end{align}
They are real in the vicinity of the apparent horizon if $e_{1} + p_{1} - 2 \phi_1 < 0$. Using Eqs.~\eqref{Grr} and \eqref{therrA}, we find that this is the case if the hair $h_+(v,r)$ is a decreasing function of $r$.

The apparent horizon is a timelike hypersurface. In the case of evaporation we can easily express the proper time $\sigma$ in $(v,r)$ coordinates as
\begin{align}
	d\sigma^2=2|r_+'|dv^2.
\end{align}
The analogous calculation in $(t,r)$ coordinates leads to
\begin{align}
	d\sigma^2=\frac{4\xi}{r_\sg}(1-w_1)dt^2,
\end{align}
where we have used Eqs.~\eqref{the2} and \eqref{thevrA}, while
\begin{align}
	\left|\frac{dv}{dt}\right|_{r_\sg}=\frac{\sqrt{\xi}}{2\sqrt{\pi r_\sg^3}\Upsilon}(1-w_1) .
\end{align}
For dynamic metrics of PBHs the spatial metric changes with time, and the finite values of distance are meaningful only if the rate of change of its parameters is sufficiently small, i.e.\ $r'_\sg x\ll r_\sg$, which is satisfied for $x \sim r_\sg$. For $k=0$ the near-horizon physical distance is given by
\begin{align}
	\ell_0(x)=\frac{2}{3\pi^{1/4}}\frac{x^{3/4}}{(r_\sg\Upsilon^2)^{1/4}}+\cO(x^{5/4}).
\end{align}
Using Eq.~\eqref{eq:k0rp}, the leading term results in
\begin{align}
	\ell_0(x)\approx\frac{4}{3}\frac{x^{3/4}\xi^{1/4}}{\sqrt{|r'_\sg|}}=\frac{\sqrt{2}}{3}\frac{x^{3/4}r_\sg^{1/4}}{\sqrt{|r'_\sg|}}, \label{lphys0}
\end{align}
where the last equality holds if Eq.~\eqref{xi0} is true.

\subsection{$k=1$}\label{As1}
For the class of $k=1$ solutions, the effective EMT components are given by
\begin{align}
	\tensor{\tau}{_t} &= E f + e_2 x^2 + \mathcal{O}(x^{5/2}),
	\label{eq:k1-taut} \\
	\tensor{\tau}{_t^r} &= \Phi f + \phi_2 x^2 + \mathcal{O}(x^{5/2}) ,
	\label{eq:k1-tautr} \\
	\tensor{\tau}{^r} &= P f + p_2 x^2 + \mathcal{O}(x^{5/2}) .
	\label{eq:k1-taur}
\end{align}
The flux $\Phi$ and pressure $P$ at the apparent horizon/anti-trapping horizon can be written as \cite{sMT:21}
\begin{align}
	\Phi = \pm \frac{1 - 8 \pi r_\sg^2 E}{8 \pi r_\sg^2} , \qquad P = \frac{-1 + 4 \pi r_\sg^2 E}{4 \pi r_\sg^2} .
\end{align}
For an evolving apparent horizon $r_\sg^\prime < 0$ of a non-extreme black hole only a single solution with the extreme-valued energy density $E=1/(8 \pi r_\sg^2)$ at the apparent horizon $r_\sg(t)$ is consistent.
The metric functions are
\begin{align}
	C &= r - 4 r_\sg^{3/2} \sqrt{- \pi e_2 / 3} \; x^{3/2} + \frac{4}{7 r_\sg} \left( 1 + \frac{r_\sg^{5/2} \sqrt{3 \pi} e_{52}}{\sqrt{-e_2}} \right) x^2 + \mathcal{O}(x^{5/2}) , \\
	h &= - \frac{3}{2} \ln \frac{x}{\xi} + \frac{3}{14 e_2} \left( \frac{4 \sqrt{- 3 e_2 / \pi}}{r_\sg^{5/2}} + 5 e_{52} - 7 p_{52} \right) \sqrt{x} + \mathcal{O}(x) ,
\end{align}
where $8 \pi r_\sg^2 E \leqslant 1$ due to the definition $C(t,r_\sg) = r_\sg$, and $f>0$ for $r>r_\sg$. Note that
\begin{align}
	c_{32}= - 4 r_\sg^{3/2} \sqrt{- \pi e_2 / 3}<0. \label{c32-1}
\end{align}
The limit of Eq.~\eqref{eq:thevr} results in
\begin{align}
	- \frac{w_1}{8 \pi r_+^2} = \Phi - E ,
\end{align}
which implies $\Phi=0$, and by virtue of Eq.~\eqref{eq:ther} also $P = - E = - 1/(8 \pi r_\sg^2)$. Moreover, Eqs.~\eqref{eq:thevr}--\eqref{eq:ther} impose the conditions
\begin{align}
	e_2 = p_2 = \phi_2 , \qquad e_{52} + p_{52} = 2 \phi_{52}
\end{align}
on the coefficients of the effective EMT expansion Eqs.~\eqref{eq:k1-taut}--\eqref{eq:k1-taur}. Consistency of the Einstein equations \eqref{eq:Gtt}--\eqref{eq:Grr} requires Eq.~\eqref{rder1} and
\begin{align}
	p_{52} &= \frac{2 \sqrt{- e_2}}{\sqrt{3 \pi} r_\sg^{5/2}} + e_{52} .
	\label{eq:k1p52}
\end{align}
Substitution of Eq.~\eqref{eq:k1p52} into
\begin{align}
	h_{12} &= \frac{3}{14 e_2} \left( \frac{4 \sqrt{- 3 e_2 / \pi}}{r_\sg^{5/2}} + 5 e_{52} - 7 p_{52} \right)
\end{align}
leads to the identity $c_2 = c_{32} h_{12}$ between the lowest-order coefficients of $C$ and $h$ \cite{sMT:21b}, which leads to many simplifying cancellations.

At leading order the limiting form of the $(tr)$ block of the EMT as $r \to r_\sg$ is given in the orthonormal basis by
\begin{align}
 	\tensor{T}{_{\hat{a}}_{\hat{b}}} = \frac{1}{8\pi r_g^2}
 	\begin{pmatrix}
		1 +\tfrac{3}{2}c_{32}\sqrt{x} & \pm\tfrac{3}{2}c_{32}\sqrt{x} \vspace{1mm} \\
		\pm \tfrac{3}{2}c_{32}\sqrt{x}  & -1+\tfrac{3}{2}c_{32}\sqrt{x}
	\end{pmatrix},
  \label{tneg1f}
\end{align}
where the upper (lower) sign corresponds to PBH evaporation (white hole expansion).

\section{Kerr--Vaidya metric} \label{A-KV}
In advanced coordinates, the decomposition Eq.~\eqref{MT-emt} of the EMT is obtained with the vectors
\begin{align}
	k_\mu=(1,0,0,-a\sin^2\theta),
\end{align}
and
\begin{align}
	q_\mu=\bigg(0,\; 0,\; T_{v \theta},\; -a\sin^2\theta\frac{r^2-a^2\cos^2\theta}{\rho^4}M_v\bigg).
\end{align}
The orthonormal tetrad is chosen in such a way that $k^\mu=e_{\hat 1}^\mu+e_{\hat 0}^\mu$,
\begin{align}
	e_{\hat 0}^\mu &= \left( -1, \frac{rM}{\rho^2,}0,0 \right), \\
	e_{\hat 1}^\mu &= \left( 1, \frac{1-rM}{\rho^2},0,0 \right),  \\
	e_{\hat 2}^\mu &= \left( 0,0, \frac{1}{\rho},0 \right), \\
	e_{\hat 3}^\mu &= \frac{1}{\rho}\big(a\sin\theta,a\sin\theta,0,\csc\theta\big).
\end{align}
In this basis, the EMT has the form of Eq.~\eqref{emt-t},
\begin{align}
	\tensor{T}{^{\hat{a}}^{\hat{b}}} =
	\left(\begin{tabular}{cc|cc}
 		$\nu$ & $\nu$ & $q_{\hat 2}$ & $q_{\hat 3}$ \\
 		$\nu$ & $\nu$ & $q_{\hat 2}$ & $q_{\hat 3}$ \\ \hline
 		$q_{\hat 2}$ & $q_{\hat 2}$ & 0 & 0 \\
		$q_{\hat 3}$ & $q_{\hat 3}$ & 0 & 0
 	\end{tabular}\right),
 \end{align}
with $\nu=T_{vv}$, and $q^\mu=q^{\hat 2} e_{\hat 2}^\mu+q^{\hat 3} e_{\hat 3}^\mu$ with
\begin{align}
	q^2 &= -\frac{a^2 r M_v}{8\pi \rho^5} \sin2\theta, \\
	q^3 &= -\frac{a(r^2 - a^2\cos^2\theta) M_v}{8\pi \rho^5}\sin\theta.
\end{align}

All four Lorentz-invariant eigenvalues of $T^{\hat{a}}_{\hat{b}}$ are zero. On the other hand, the two non-zero eignevalues of the matrix ${T}{^{\hat{a}\hat{b}}}$ are
\begin{align}
\tilde\lambda_{1,2}=\pm\sqrt{2}\varphi.
\end{align}
As a result we see that Kerr-Vaidya metrics are a special case of type III,  as the Lorentz-invariant eigenvalues of Eq.~\eqref{IIIf} are zero if and only if $\varrho=p=0$. Then the EMT tensor \eqref{emt-t} cannot be brought to a generic type III form by an arbitrary similarity transformation unless $\tilde\lambda_{1}=-\tilde\lambda_{2}$, which is impossible for an evolving mass\cite{DT:20}.

\section{Classical thin shell collapse}\label{A-shell}
The first junction condition asserts that the induced metric is the same on both sides of a $D$-dimensional shell, and is expressed mathematically by
\begin{align}
	ds_{\Sigma}^2 &= \tensor{h}{_a_b} \tensor{dy}{^a} \tensor{dy}{^b} = - d \tau^2 + R^2 d\Omega_{D-1} ,	
\end{align}
where $\tensor{y}{^a} \equiv \left( \tau, \Theta \defeq \theta \vert_\Sigma , \Phi \defeq \phi \vert_\Sigma \right)$ labels the hypersurface coordinates, $\tau$ is a future-directed time coordinate defined on the hypersurface, and upper case letters denote quantities on the shell $\Sigma$, e.g.\ $R \defeq r \vert_\Sigma$. It implies $R_+ \equiv R_- \eqdef R(\tau)$. The surface EMT of a massive thin shell is given by
\begin{align}
	\tensor{S}{^a^b} = \sigma \tensor{v}{^a} \tensor{v}{^b} = \sigma \delta^a_\tau \delta^b_\tau ,
\end{align}
where $\sigma$ denotes the surface density. The rest mass of the shell is $m = 4 \pi \sigma R^2$. The second junction condition relates the jump in extrinsic curvature
\begin{align}
	\tensor{K}{_a_b} \defeq \tensor{\hat{n}}{_{\mu;}_\nu} \tensor{e}{^\mu_a} \tensor{e}{^\nu_b}
\end{align}
to the surface EMT
\begin{align}
	\tensor{S}{_a_b} = - \left( \left[ \tensor{K}{_a_b} \right] - \left[ K \right] \tensor{h}{_a_b} \right) / 8 \pi ,
\end{align}
where $K \defeq \tensor{K}{^a_a}$, and $\left[ K \right] \defeq K \vert_{\Sigma_+} - K \vert_{\Sigma_-}$ is the discontinuity of the extrinsic curvature scalar $K$ across the two sides of the hypersurface $\Sigma_{\pm}$. The trajectory of $\Sigma$ is timelike, and hence the proper time derivatives of time coordinates ($T$, $U$, or $V$) are related to $\dot R$ via Eqs.~\eqref{Ttime}, \eqref{Utime}, and \eqref{Vtime} (with $h=0$, $a=0$).

A straightforward way to determine the fate of the collapsing shell is to monitor the evolution of the gap $X =Y=R(\tau) - r_\sg$, i.e.\ the coordinate distance between the shell and the Schwarzschild radius. The equation of motion for the shell is given by \cite{P:04,BsMT:19}
\begin{align}\label{D(R)eq}
	\mathcal{D}(R) \defeq \frac{2 \ddot{R} + F'}{2 \sqrt{F + \dot{R}^2}} - \frac{\ddot{R}}{\sqrt{1+\dot{R}^2}} + \frac{\sqrt{F + \dot{R}^2} - \sqrt{1 + \dot{R}^2}}{R} = 0 ,
\end{align}
where the last term
\begin{align}
	- 4 \pi \sigma = \frac{\sqrt{F + \dot{R}^2} - \sqrt{1 + \dot{R}^2}}{R}
\end{align}
directly describes the evolution of the surface density.

This equation is simple enough to have an analytic solution $\tau(R)$ \cite{P:04,BMT:18}, leading to the finite proper crossing time $\tau(r_\sg)$ that corresponds to an infinite time according to Bob.

\end{document}